\documentclass[a4paper,11pt]{book}
\usepackage{bbold}
\usepackage{amssymb}
\usepackage{amsfonts}
\usepackage{amsthm}
\usepackage{physics}
\usepackage{amsmath}
\usepackage{graphicx}
\usepackage{subcaption}
\usepackage{float}
\usepackage{etoolbox}
\usepackage{hyperref}
\usepackage{amstext}
\usepackage{amsxtra}
\usepackage{tikz}
\usepackage{setspace}
\usepackage{listings}
\usepackage[titletoc]{appendix}
\usepackage{xcolor}
\usepackage{fancyhdr}
\usepackage{url}
\usepackage{mathrsfs, enumerate}
\usepackage{tikz}
\usepackage{setspace}
\usepackage{longtable}
\usepackage{vmargin}
\setmarginsrb     { 1.25in}  % left margin
                        { 1.0in}  % top margin
                        { 1.0in}  % right margin
                        { 1.0in}  % bottom margin
                        {  20pt}  % head height
                        {0.25in}  % head sep
                        {   9pt}  % foot height
                        { 0.3in}  % foot sep
\begin{document}
% Title Page--------------------------------------------------------------------------------------------------------------
\begin{titlepage}
\begin{center}
{ \Huge \bfseries Monte Carlo Simulations of BFSS \\[0.5cm]and IKKT Matrix Models}\\[3cm]

\textbf{\LARGE Nikhil Tanwar}\\[2.0cm]

{\large \sl A dissertation submitted for the partial fulfillment \\ 
of BS-MS dual degree in Science}\\[4.0cm]
\begin{figure}[ht]
\centering
\includegraphics[width=0.25\textwidth]{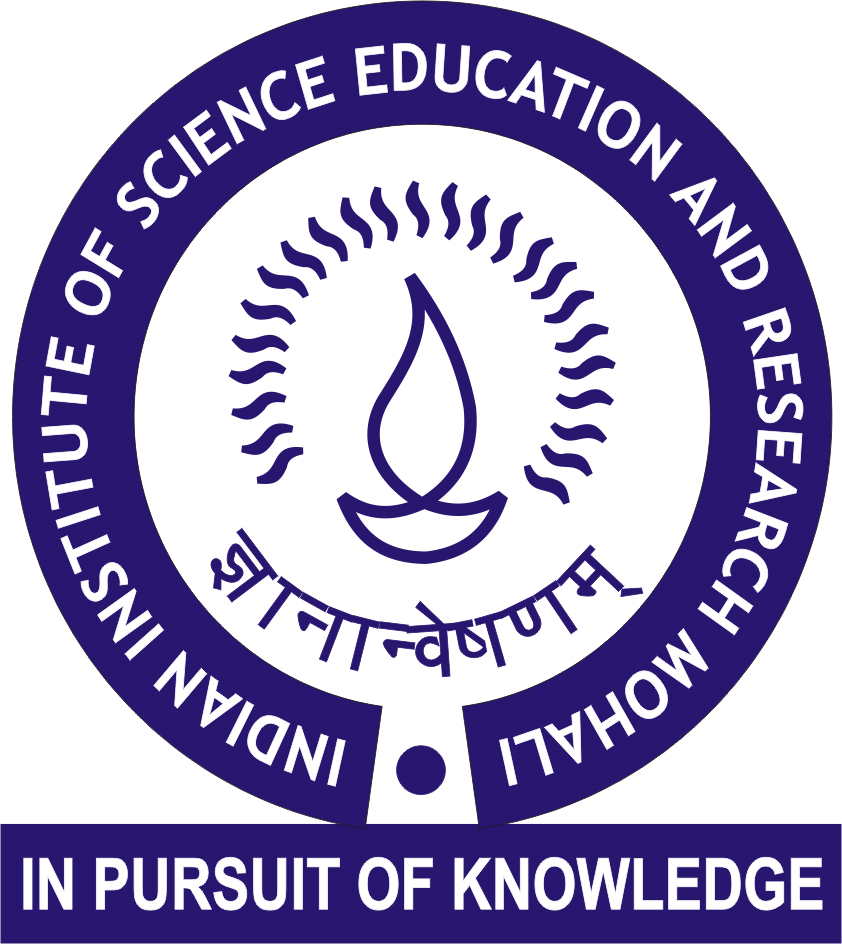}\\[1cm]     
\end{figure}

\textbf{\large Indian Institute of Science Education and Research Mohali\\June 2020}
\end{center}
\date{\nodate}
\end{titlepage}
\clearpage

\graphicspath{{Images/}}

%---------------------------------------------------------------------------------------------------------------------------------------------

%---------------------------------------------------------------------------------------------------------------------------
%\makeindex

\thispagestyle{empty}
%\begin{flushleft}
% Acknowledgement ----------------------------------------------------------------------------------------------------------------------------
\parbox{5in}{
\begin{center}
\large \textbf{Certificate of Examination}
\end{center}
\small{ 
\par

\noindent This is to certify that the dissertation titled 
\textbf{Monte Carlo Simulations of BFSS and IKKT Matrix Models}
submitted by \textbf{Nikhil Tanwar} (MS15111) for the partial fulfillment of BS-MS dual degree 
programme of the Institute, has been examined by the thesis committee duly appointed by the
Institute. The committee finds the work done by the candidate satisfactory and recommends that the
report be accepted.
}
\vspace{1.5cm}
\begin{flushright}
 Dr. Kinjalk Lochan \hspace{2cm} Dr. Sanjib Dey \hspace{2cm} Dr. Anosh Joseph
 
 (Supervisor)
\end{flushright}
\vspace{1.5cm}
\begin{flushright}
Dated: \today 
\end{flushright}
}

\clearpage
\thispagestyle{empty}
\parbox{5in}{
\begin{center}
\large \textbf{Declaration}
\end{center}
\small{ 
\par\noindent The work presented in this dissertation has been carried out by me under the guidance of 
Dr. Anosh Joseph at the Indian Institute of Science Education and Research Mohali.\newline

\noindent This work has not been submitted in part or in full for a degree, a diploma,
or a fellowship to any other university or institute. Whenever contributions
of others are involved, every effort is made to indicate this clearly, with due
acknowledgement of collaborative research and discussions. This thesis is a
bonafide record of original work done by me and all sources listed within
have been detailed in the bibliography.
}
\vspace{1cm}
\begin{flushright}
 Nikhil Tanwar
 
 (Candidate)
\end{flushright}

\begin{flushright}
 Dated: \today
\end{flushright}
\small{ \par\noindent In my capacity as the supervisor of the candidate’s project work, I certify
that the above statements by the candidate are true to the best of my
knowledge.}

\vspace{1cm}
\begin{flushright}
 Dr. Anosh Joseph
 
 (Supervisor)
\end{flushright}
}

\clearpage
\parbox{5in}{
\begin{center}
\large \textbf{Acknowledgment}
\end{center}
\small{ 
\par\noindent I express my sincere gratitude to my supervisor, Dr. Anosh Joseph for his patience, motivation, prudent comments, valuable suggestions, and beneficial information,
which have helped me remarkably in my research and in writing this thesis. His enormous knowledge and thorough experience in lattice field theory have enabled me to complete this research successfully. I am incredibly thankful to him for
sparing his precious time to guide me, clarifying my queries, correcting me during my research and sharing his workstation without which simulations would have never been completed. \\[0.20 cm]

Besides my advisor, I wish to thank the rest of my thesis committee members, Dr. Kinjalk Lochan and Dr. Sanjib Dey, for their insightful comments and encouragement. \\[0.20 cm]

I extend my sincere thanks to Mr. Arpith Kumar for giving his valuable time for the discussion sessions and for helping me during the simulation of the BFSS Matrix model. I also extend my sincere thanks to Mr. Adeeb Mev for the collaboration we made for the simulation of the BFSS matrix model. \\[0.20 cm]

I acknowledge IISER Mohali for providing me with the best infrastructure and environment for carrying out this project. I am also thankful to the Department of Science and Technology (DST), Government of India, for supporting me with the institute fellowship during the past five years. \\[0.20 cm]

I want to thank my batchmates and friends for interesting peer discussions throughout the five years of the course. I am thankful to Nilangshu Bhattacharyya, Amit Suthar, Adarsh Bhardwaj, Anubhav Jindal, Apoorv Gaurav, Debanjan Chowdhury, Debjit Ghosh, Gaurav Singh, Ishan Sarkar, Paresh Nath Das, Satyam Prakash, Swastik P G, Vidur Sury and Vivek Jadhav. I would also like to thank all my friends listed above for their continuous moral support and motivation throughout this journey without which it would have been a very tough journey. \\[0.20 cm]

I would also like to thank and acknowledge my parents and my brother Dhruv Tanwar for their love and support throughout the years of my study. This would not have been possible without them.

}}
\clearpage

\pagenumbering{roman}
\listoffigures \addcontentsline{toc}{chapter}{List of Figures} 
\clearpage
\listoftables \addcontentsline{toc}{chapter}{List of Tables}
\clearpage
\begin{center}
 \Large \bf Abstract
\end{center}
In this thesis, we studied the bosonic BFSS and IKKT matrix models using Monte Carlo simulations. First, we explored some toy models to check the validity of the numerical simulations. Then we simulated the BFSS matrix model using Hamiltonian Monte Carlo (HMC) algorithm. In the BFSS matrix model, we used the Polyakov loop as an order parameter to investigate the large-$N$ behaviour of this model at different temperatures. Our simulations confirmed that the model exhibits a confinement-deconfinement phase transition as the temperature of the system is varied. Besides the Polyakov loop, other observables such as internal energy and extent of space were also computed. In the bosonic IKKT model, we studied the spontaneous symmetry breaking (SSB) of $SO(10)$ symmetry using the moment of inertia tensor and found that there is no SSB of $SO(10)$ symmetry in this model. Besides the eigenvalues of the moment of inertia tensor, other observable such as extent of spacetime was also computed. We also studied the simulation theory of the phase-quenched IKKT model. \addcontentsline{toc}{chapter}{Abstract}
\cleardoublepage

\tableofcontents
\mainmatter

\newpage
% newtheorem defined ------------------------------------------------------------------------------------------------------
\newtheorem{defn}{Definition}[chapter]
\newtheorem{lemma}{Lemma}[chapter]
\newtheorem{theo}{Theorem}[chapter]
\newtheorem{meth}{Method}[chapter]
\newtheorem{protocol}{Protocol}[chapter]
\newtheorem{cond}{Condition}[section]
\newtheorem{remark}{Remark}[chapter]
\newtheorem{coro}{Corollary}[chapter]
\newtheorem{prop1}{Proposition}[chapter]
\newtheorem{theorem}{Theorem}
\newtheorem{const}{Construction}[chapter]
% Definition --------------------------------------------------------------------------------------------------------------
\def\proof{\paragraph{Proof}}
\fontfamily{cmss}
\def\ket{\rangle}
\def\bra{\langle}
\def\exam{\paragraph{Example}}
\def\defin{\paragraph{Definition}}
%--------------------------------------------------------------------------------------------------------------------------

\newpage
\pagenumbering{arabic}
% Chapter 1 ----------------------------------------------------------------------------------------------------------------------------
 % How to give reference to the Appendix
% Footnote hyperlink goes to page 1??

%%%%%%
\chapter{Introduction}\addcontentsline{}{×}{×}
%%%%%%

String theory is believed to be a very promising candidate for the theory of everything. It attempts to unify the theory of gravity with other three fundamental forces of nature: Electromagnetism, Strong and Weak nuclear forces. It replaces the point like particles, which is the fundamental assumption of the Standard Model of particle physics, by one-dimensional objects called {\it strings}. All the matter particles as well as the force carrying particles are made up of these strings and the theory explains how these strings propagate in spacetime and interact with each other. In string theory, each of the fundamental particle is represented by the unique vibrational frequency of the string.

Till today there are five consistent superstring theories: type I, type IIA, type IIB, heterotic $SO(32)$ and heterotic $E_8 \times E_8$. These theories differ whether the strings are open or closed, whether the strings are oriented or not and on how they treat electrical charges. All of these theories are in 10-dimensional spacetime and all the calculations are done perturbatively in terms of string length $l_s$ and string coupling $g_s$.

It has been found that all of these different string theories are connected to each other through some transformation (T-duality, S-duality etc.) and this led to the conjecture in 1995 by Edward Witten \cite{Witten_1995} that they are part of some bigger theory called {\it M-theory}.  M-theory is still undiscovered but it is known that it lives in 11 dimensions and it is a non-perturbative theory. The M in M-theory is undefined and sometimes it is referred to as ``membrane'' or ``matrix.''

String theory was first studied in the late 1960s as a model of strong interaction and from 1974 onwards proper study of string theory started and since then it has been evolving. Even today, there is a debate over its validity as no part of this theory has been experimentally verified.

\newpage

%%%%%%%%%%%%%%%
\section{BFSS Matrix Model}
%%%%%%%%%%%%%%%

BFSS Model was conjectured in 1996 by T. Banks, W. Fischler, S. H. Shenker, and L. Susskind \cite{banks1997m}. This model is one dimensional supersymmetric Yang-Mills theory and this theory acts as the low energy effective description of $D0$-branes of type IIA superstring theory \cite{Filev_2016} and in the limit $N \to \infty$ it is believed that this theory will represent M-theory. One of the methods to obtain the action of this theory is by dimensionally reducing the $\mathcal{N} =1$ super Yang-Mills theory in 9+1 dimensions with gauge group $SU(N)$ to 0+1 dimensions (See Appendix \ref{appen1}). The resulting action is given by
\begin{equation}
S = \frac{1}{2 g^2} \int dt \,  \Tr \Bigg( (D_t X_i)^2 + \frac{1}{2} [X_i, X_j]^2 -  i \Psi^{T}_\alpha C_{10} \gamma^0_{\alpha \beta} D_t \Psi_\beta + \Psi^T_\alpha C_{10} \gamma^i_{\alpha \beta} [X_i, \Psi_\beta] \Bigg).
\end{equation}
where $D_t = \partial_t - i [A(t), \cdot]$; $i, j = 1, 2, \cdots, 9$; $\alpha, \, \beta = 1,\, 2, \, \cdots, \, 32$; $X^i$ are $N \times N$ traceless Hermitian matrices; $g$ is the one-dimensional Yang-Mills coupling; $A(t)$ is the one-dimensional gauge field; $\Psi$ is a 32-component Majorana-Weyl fermion with each component of the fermion being an $N \times N$ traceless Hermitian matrix; $C_{10}$ is the charge conjugation matrix in the ten dimensions; and $\gamma^i$ are the gamma matrices in ten dimensions. The action is invariant under the following set of gauge transformations
\begin{eqnarray}
X^i(t) &\longrightarrow& V(t) X^i(t) V^{\dagger}(t), \\
A(t) &\longrightarrow& V(t) \left( A(t) + i \partial_t \right) V^{\dagger}(t), \\
\Psi_\alpha(t) &\longrightarrow& V(t) \, \Psi_\alpha(t) \, V^{\dagger}(t),
\end{eqnarray}
where $V(t) \in SU(N)$.

%%%%%%%%%%%%%%%
\section{BFSS Model and Black Hole thermodynamics }
%%%%%%%%%%%%%%%

One of the motivations to study the BFSS model is the connection of this model with the black hole states. The finite temperature states of BFSS model is related to the black hole states of type IIA supergravity \cite{klebanov1998schwarzschild}  \cite{banks1998schwarzschild}. The connection between BFSS matrix model and the black hole states has also been discussed in the Ref. \cite{kabat2001black} \cite{kabat2001black2}. In these two references, the authors calculated the entropy of a black hole in strongly coupled quantum mechanics beginning with the Bekenstein-Hawking entropy of a ten-dimensional non-extremal back hole. The free energy of the black hole in terms of the parameters of the gauge theory (here the BFSS model) can be written as
\begin{align}
\frac{F}{T} = -4.115 N^2 \left( \frac{T^3}{g^2_{ym} N} \right)^{3/5}.
\end{align}
The above formula is of interest for a couple of reasons. First, its dependence on the 't Hooft coupling. The limit $N \to \infty$ corresponds to the region where supergravity is valid and therefore it is the appropriate limit to  look for black holes in the BFSS model. Second reason being that the free energy depends on $N^2$, which is the expected behaviour of a gauge theory in the deconfined phase.

It would be important to reproduce this formula from the matrix model. To get this behaviour in the matrix model, one must look in the strong coupling, which is not accessible in the perturbation theory. That is, in a regime where it is highly impossible to carry out an analytical calculation to derive this expression. Thus we are left with numerical simulations as these technique does not have any such restrictions.  

%%%%%%%%%%%%%%%
\section{Phase Transition}
%%%%%%%%%%%%%%%

For a system with a density of states that grows exponentially,
\begin{align}
\rho(E) \sim e^{\beta_H E}
\end{align}
there exist an upper limit to the temperature known as the {\it Hagedorn} temperature. Above this temperature the partition function diverges \cite{furuuchi2003fivebrane} 
\begin{align}
\lim \limits_{T \to T_H^-} \Tr( e^{- \beta H} ) \to \infty.
\end{align}

Above this cutoff temperature $T_H$, partition function does not exist. However, it can be made to exist if we keep $N$ large but finite. Performing this breaks the exponential growth in the asymptotic density of states at some large but finite energy value. For temperature greater than $T_H$, the entropy and the energy are dominated by the states at and above the cutoff scale and the free energy jumps from $\mathcal{O}(1)$ to $\mathcal{O}(N^2)$ \cite{hadizadeh2005free}. Thus 
\begin{eqnarray}
\lim \limits_{N \to \infty} \frac{F}{N^2} &=& 0 \qquad \textrm{(Confined phase)}, \\
\lim \limits_{N \to \infty} \frac{F}{N^2} &\neq& 0 \qquad \textrm{(Deconfined phase)}.
\end{eqnarray}

The transition from the confined phase to deconfined phase is known as confinement-deconfinement phase transition or deconfinement phase transition. In the confined phase, the quantum states of the Hamiltonian should be singlets under the gauge symmetry \cite{SEMENOFF_2004}. This condition fails as soon as the system reaches the Hagedorn temperature. This type of transition is found in large-$N$ gauge theories, such as weakly coupled Yang-Mills theory and it is also expected to be found in the BFSS model \cite{SEMENOFF_2004}.

This confinement-deconfinement phase transition in the matrix models is associated with the breakdown of the center symmetry, i.e., $ A(t) \to A(t) + k \mathbb{1}$. The order parameter for this symmetry breaking is the {\it Polyakov loop} \cite{polyakov1978thermal}. Polyakov loop is defined as the trace of the holonomy of the gauge field around the finite temperature Euclidean time circle
\begin{align}
P = \frac{1}{N} \Tr( P( e^{i \oint dt \, A(t)} ) ).
\end{align}
This operator is gauge invariant as it is a special case of the Wilson line operator. The expectation value of this operator is zero in confined phase and it jumps to a non-zero value as we cross the phase transition point and enters the deconfined phase. This is simply because Polyakov loop is a unitary matrix and its eigenvalues are uniformly distributed on the unit circle in the confined phase and in the deconfined phase the eigenvalues clump towards a single point. We have
\begin{eqnarray}
\left\langle P \right\rangle &=& 0 \qquad \textrm{(Confined phase)}, \\ 
\left\langle P \right\rangle &\neq& 0 \qquad \textrm{(Deconfined phase)}.
\end{eqnarray}
We use this as an order parameter for our simulations of the BFSS model.

%%%%%%%%%%%%%%%
\section{IKKT Matrix Model}
%%%%%%%%%%%%%%%

IKKT model was conjectured in 1997 by N. Ishibashi, H. Kawai, Y. Kitazawa, A. Tsuchiya \cite{Ishibashi_1997}. This model was proposed as a non-perturbative formulation of superstring theory. This model is also known as type IIB matrix model. The action of this model can be obtained by dimensionally reducing the $\mathcal{N} =1$ super Yang-Mills theory in 9+1 dimensions with gauge group $SU(N)$ to $0+0$ dimensions (See Appendix \ref{appen1}). The resulting Euclidean action is given by
\begin{align}
S_{\rm E} = -\frac{1}{4 g^2} \Tr([X_i, X_j]^2) - \frac{1}{2 g^2} \Tr( \Psi^T_\alpha C_{10} \gamma^i_{\alpha \beta} [X_i, \Psi_\beta] ),
\end{align}
where $i, j = 1, 2, \cdots, 10$; $\alpha, \, \beta = 1, \, 2, \, \cdots, \, 32$; $X^i$ are $N \times N$ traceless Hermitian matrices; $g$ is the zero-dimensional Yang-Mills coupling; $\Psi$ is a 32-component Majorana-Weyl fermion with each component of the fermion being an $N \times N$ traceless Hermitian matrix; $C_{10}$ is the Euclidean charge conjugation matrix in ten dimensions; and $\gamma^i$ are the Euclidean gamma matrices in ten dimensions. The action is invariant under the following set of gauge transformations
\begin{eqnarray}
X_i &\longrightarrow& V \, X_i \, V^{\dagger}, \\
\Psi_\alpha &\longrightarrow& V \, \Psi_\alpha(x) \, V^{\dagger}.
\end{eqnarray}

\newpage

%%%%%%%%%%%%%%%
\section{Motivation for IKKT Model}
%%%%%%%%%%%%%%%

One of the main motivations to study the IKKT model is that in the large-$N$ limit, spacetime emerges dynamically in this model from the eigenvalues of the ten bosonic matrices and this tells us that dynamical compactification can be viewed as a purely non-perturbative effect \cite{Aoki_1998}. The evidence for the same came from the simulations of the Lorentzian version of the IKKT model \cite{Nishimura_2019, nishimura2020new}. In these simulations it has been found that continuous time emerges dynamically and three-dimensional space undergoes expansion after a critical time with the other six spatial dimensions remaining small. The expansion is exponential at early times \cite{Ito_2014}, which becomes a power law at later times \cite{Ito_2015}. These two results provide evidence that a realistic cosmological scenario may also arise dynamically from this model.

On the other hand, we can also look for dynamical compactification of extra dimensions in the Euclidean model by spontaneous symmetry breaking (SSB) of the $SO(10)$ due to the fluctuations of the phase of the Pffafian for $SO(d)$  symmetric configurations with larger $d$ \cite{Nishimura_2000}. It has been showed that SSB does not take place in phase-quenched model, so SSB is attributed to the effect of the phase of the Pffafian \cite{Ambj_rn_2000}. It was also shown that in the IKKT model that the $SO(3)$ symmetric vacuum has the lowest free energy, which implies SSB to $SO(3)$ \cite{Nishimura_2011}. Recent work on this model using complex Langevin algorithm showed that there is SSB of $SO(10)$ to $SO(3) \times SO(7)$ \cite{anagnostopoulos2020complex}. They further showed that the $SO(7)$ symmetry is also broken into smaller groups.

%%%%%%%%%%%%%%%
\section{Path Integral and Statistical Mechanics}
%%%%%%%%%%%%%%%

In this section will introduce how the path integrals in quantum field theory and statistical mechanics are connected. We will explain this using the example of a real scalar field.

Let us consider a theory containing a scalar field defined by the following action
\begin{align}
S[\phi(x)] = \int \limits_{0}^{t} dt{'} \int d^3 x \left( \frac{1}{2} \partial_{\mu} \phi \partial^{\mu} \phi - V(\phi) \right),
\end{align}
where $V(\phi)$ is the potential functional. Then the probability amplitude to go from the field configurations $\phi_1(\vec{x})$ to $\phi_2(\vec{x})$ in time $t$ is given by the path integral $G(\phi_2(\vec{x}), \phi_1(\vec{x}); t)$. We have
\begin{eqnarray}
G(\phi_2(\vec{x}),\phi_1(\vec{x}); t) &=& \bra \phi_2(\vec{x}) | e^{- i H t} | \phi_1(\vec{x}) \ket, \\
G(\phi_2(\vec{x}),\phi_1(\vec{x}); t) &=& \int \limits_{ \phi(\vec{x}, 0) = \phi_1 (\vec{x}) }^{ \phi(\vec{x},t) = \phi_2(\vec{x}) } \, \mathcal{D} \phi  \, \,  e^{i S[\phi(x)]},
\end{eqnarray}  
where $H$ is the Hamiltonian of this system. The partition function for the theory of scalar field is given by
\begin{align}
Z = \Tr( e^{- \beta H} ) = \int \mathcal{D} \phi(\vec{x}) \,\, \bra  \phi(\vec{x}) | e^{- \beta H} | \phi(\vec{x}) \ket. \label{eq1.4}
\end{align}

If we perform a Wick rotation (see Appendix \ref{appen1}), i.e., $t = - i t_E$, then we go to the Euclidean signature from the Minkowski signature and we can use that to write
\begin{align}
\bra \phi(\vec{x}) | e^{- \beta H} | \phi(\vec{x}) \ket = \int \limits_{ \phi(\vec{x}, 0) = \phi(\vec{x}) }^{ \phi(\vec{x}, \beta) = \phi(\vec{x}) } \, \mathcal{D} \phi  \, \,  e^{- S_E[\phi(x)]} = G(\phi(\vec{x}),\phi(\vec{x}); -i \beta),
\end{align}
where 
\begin{equation}
S_E[\phi(x)] = \displaystyle \int \limits_{0}^{\beta} dt_E \int d^3 x \left( \frac{1}{2} \partial_{\mu} \phi \partial_{\mu} \phi + V(\phi) \right)
\end{equation}
is the Euclidean action. If we substitute this in Eq. \eqref{eq1.4}, the partition function becomes
\begin{eqnarray}
Z &=& \int \mathcal{D} \phi(\vec{x}) \,\, \bra  \phi(\vec{x}) | e^{- \beta H} | \phi(\vec{x}) \ket \nonumber \\
&=& \int \mathcal{D} \phi(\vec{x}) \int \limits_{ \phi(\vec{x}, 0) = \phi(\vec{x}) }^{ \phi(\vec{x},\beta) = \phi(\vec{x}) } \, \mathcal{D} \phi  \, \,  e^{- S_E[\phi(x)]} \nonumber \\
&=& \oint \limits_{PBC} \mathcal{D} \phi \,\, e^{- S_E [\phi]},
\end{eqnarray}
where PBC stands for periodic boundary conditions, i.e., $\phi(\vec{x}, 0) = \phi(\vec{x}, \beta)$. If we perform the same calculation for fermionic fields we will get the same result but in place of periodic boundary conditions we will have anti-periodic boundary conditions, i.e., $\psi(\vec{x}, 0) = -\psi(\vec{x}, \beta)$. Using this formalism we can study the physics of finite temperature quantum field theory. This formalism is also useful in exploring strongly coupled quantum field theories using Monte Carlo simulations of the Euclidean model.

So a general result for fields in flat Minkowski spacetime follows from here: {\it Euclidean quantum field theory in $(D+1)$-dimensional spacetime with $0 \leq t_E < \beta$ is the same as the quantum statistical mechanics in $D$-dimensional space} \cite{zee2010quantum}.

% Chapter 2 -----------------------------------------------------------------------------------------------------------------------------
 % How to give reference to the Appendix
% Footnote hyperlink goes to page 1??

%%%%%%
\chapter{Markov Chain Monte Carlo Algorithms}
\addcontentsline{}{×}{×}
%%%%%%

This chapter introduces the computational techniques required to sample data from a probability distribution (PD) that is known up to some multiplicative constants. We will see how to apply these algorithms to the Euclidean path integrals in this and later chapters. To begin, consider a general partition function of a particle in one dimension in some potential $V(x)$, given by
\begin{align}
Z = \int \mathcal{D}x(t) \,\, e^{- S_E [x(t)]},
\end{align}
where the Euclidean action for a general potential $V(x)$ is given by
\begin{align}
S_E [x(t)] = \int\limits_{0}^{\beta} dt \left( \dfrac{1}{2} \dot{x}^2 + V(x) \right),
\end{align}
with periodic boundary conditions $x(t + \beta) = x(t)$ (for bosonic fields). Since we are using the canonical ensemble, the system is at temperature given by $T = \beta^{-1}$. Now if we were able to calculate $Z$, then we can calculate thermodynamic quantities like free energy, energy, and other quantities using $Z$ and derivatives of it.

But for a general potential $V(x)$ it cannot be solved exactly. So we need to resort to other methods to calculate the expectation values. One method is to use perturbative calculations using Feynman diagrams. But this method breaks down for strong-coupling terms present in the potential function as the contribution of the higher order diagrams are comparable to the previous diagrams. So we need to resort to non-perturbative methods for those cases.

One of the methods to evaluate these expectation values is through numerical simulations. But our system is continuous, and so the partition function is to be integrated over dynamical variables at all times. The standard way to do this is to divide the time variable into slices and evaluate each spatial integral at a fixed time \cite{peskin1995introduction}. First we discretize the time variable by dividing the time variable into a lattice of $T$ points with uniform lattice spacing $a$ with $\beta = a T$. Now the dynamical variable $x(t)$ is no longer continuous and exists only on the discrete lattice points $x_t$. So our partition function on the lattice becomes
\begin{align}
Z = \int \left( \prod\limits_{t=1}^T dx_t \right) e^{- S_{\rm lat} (x_t)},
\end{align}
where $S_{\rm lat}$ is the lattice action. We can construct this action by discretizing $S_E$. For this, we need to replace integrals with finite sums and derivatives with finite differences. The prescription to do this is as follows
\begin{align}
x(t) \,\, &\longrightarrow \,\, x_t,\label{eq2.4} \\
\frac{\partial x}{\partial t} \,\, &\longrightarrow \,\, \frac{x_t - x_{t-1}}{a},\label{eq2.5} \\
\int \limits_{0}^{\beta} \,\, &\longrightarrow \,\, a \sum\limits_{t=1}^{T},\label{eq2.6} \\
V(x(t)) \,\, &\longrightarrow \,\, V(x_t).\label{eq2.7}
\end{align}

This prescription works well only for bosonic fields. The description to put gauge fields and fermionic fields on the lattice will be discussed in the later chapters. Since bosonic fields follow periodic boundary conditions we also need to maintain that condition on the lattice. On the lattice, periodic boundary condition becomes $x_t = x_{t + T}$. So with this prescription lattice action $S_{\rm lat}$ is given by
\begin{align}
S_{\rm lat} = a \sum \limits_{t=1}^T \left[ \frac{1}{2} \left(  \frac{x_t - x_{t-1}}{a}\right)^2 + V(x_t) \right],
\end{align}
with $x_0 = x_T$. The above method describes the behaviour of the partition function up to some multiplicative constant provided that enough number of lattice sites are used so that system behaves as if it is a continuous system.

Now we can use the above form of the partition function to calculate the expectation values of observables. For some observable $\mathcal{O}$, the expectation value $\left \langle \mathcal{O} \right \rangle$ is given by
\begin{align}
\left \langle \mathcal{O} \right \rangle = \dfrac{\displaystyle\int \left( \prod\limits_{t=1}^T dx_t \right) \,\mathcal{O}(x_t) \, e^{- S_{\rm lat} (x_t)}}{\displaystyle\int \left( \prod\limits_{t=1}^T dx_t \right) e^{- S_{\rm lat} (x_t)}}. 
\label{eq2.9}
\end{align}

Now we can evaluate this expression numerically. But if we use simple numerical integration techniques (e.g., Gauss' quadrature, Simpson's rule), evaluating the above integral becomes a computationally heavy task. We can understand this using a simple argument. For one-dimensional integral, we need to divide the integration region into grids, let us say, $N$ grids. If we now consider a $d$-dimensional integral then we would require $N^d$ grids to evaluate the integral numerically. For better accuracy we need $N$ to be large, so the complexity of the algorithm is $\mathcal{O}(N^d)$. For example, if we take 20 lattice points and the number of grids to be 100, then we need to sum $100^{20} = 10^{40}$ terms, which is a huge number. Apart from this, there will be regions in the integration domain where the contribution to integral will be negligible as the weight factor here is $e^{- S_{\rm lat} (x_t)}$. This way, we are going to waste computational resources on these integration regions. So we require more clever methods to evaluate the above integral. Such methods may give more importance to the volume of the integration region that contributes the most to the integral.

%%%%%%
\section{Importance Sampling}
%%%%%%

In any Monte Carlo integration, the error is proportional to the standard deviation of the integrand and inversely proportional to the sample size. One way to reduce the error is to reduce the standard deviation as it does not cost more computational power. This is done by sampling the important region of the integral.
 
Let us say we have an integral
\begin{align}
I = \int \limits_a^b f(x) dx,
\end{align}
and we have a probability distribution $p(x)$ over this domain. That is, 
\begin{align}
\int \limits_a^b p(x) dx = 1.
\end{align}

Then we can write the integral in this form
\begin{align}
I = \int \limits_{a}^{b} \frac{f(x)}{p(x)}\, p(x) \, dx.
\end{align}

Now if we sample $N$ data points from probability distribution $p(x)$, then we can approximate the above integral as
\begin{align}
I \approx \frac{1}{N} \sum \limits_{i=1}^{N} \frac{f(x_i)}{p(x_i)}, \qquad \qquad I = \lim_{N \rightarrow \infty} \frac{1}{N} \sum \limits_{i=1}^{N} \frac{f(x_i)}{p(x_i)}.
\end{align}

Now we need to choose the probability distribution $p(x)$ such that the above term varies slowly as this will decrease the standard deviation. Right choice of $p(x)$ will give a good result even for small number of data points. For example, if we have $f(x)$ sharply peaked at some point then choosing the probability distribution $p(x)$ which is also sharply peaked near that point will give a nice estimate of the integral as compared to the probability distribution which has vanishing tail in that region.

%%%%%%
\section{Metropolis Algorithm}
%%%%%%

Our main aim is to evaluate the expectation values of observables like Eq. \eqref{eq2.9}. Since these are all infinite-dimensional integrals for the continuous case, so to evaluate these, we put the system on the lattice and get a finite-dimensional integral to solve. As already discussed in the introduction of this Chapter, the dimension of the integral is large, and we need numerical techniques to evaluate it. So we replace the integral with sum and try to get the estimate of it. Then the quantity in Eq.\eqref{eq2.9} becomes
\begin{align}
\left \langle \mathcal{O} \right\rangle \simeq \left\langle \mathcal{O} \right\rangle_n = \frac{\displaystyle\sum_{s=1}^{n} \mathcal{O}(\left\lbrace x_t \right\rbrace_s) e^{- S_{\rm lat}(\left\lbrace x_t \right\rbrace_s)}}{\displaystyle\sum_{s=1}^{n} e^{- S_{\rm lat}(\left\lbrace x_t \right\rbrace_s)}},
\end{align}
where $\left \lbrace x_t \right \rbrace_s$ is one of the microstates allowed to the system and $n$ is the total number of microstates. For our case, the number of microstates are infinite and we are estimating the integral with only finite number of terms. So we require some method to sample the important region of the integral.

We will use importance sampling to get the field configurations which contribute largely to the integral. So we  need to introduce a probability distribution $p_s$, which denotes the probability of getting microstate $\left \lbrace x_t \right \rbrace_s$ and the above average value can be written as
\begin{align}
\left \langle \mathcal{O} \right \rangle = \frac{\displaystyle\sum_{s} \frac{\mathcal{O}\left(\left\lbrace x_t \right\rbrace_s\right)}{p_s}\,\, p_s \,\, e^{- S_{lat}\left(\left\lbrace x_t \right\rbrace_s\right)}}{\displaystyle\sum_{s} \frac{e^{- S_{lat}\left(\left\lbrace x_t \right\rbrace_s\right)}}{p_s}\,\, p_s}.
\label{eq2.15}
\end{align}

Now we generate the microstate $s$ with probability $p_s$ and approximate $\left \langle \mathcal{O} \right \rangle$ with $\left \langle \mathcal{O} \right \rangle_n$ as
\begin{align}
\left \langle \mathcal{O} \right \rangle_n = \frac{\displaystyle\sum_{s=1}^{n} \frac{\mathcal{O} \left(\left\lbrace x_t \right\rbrace_s\right)}{p_s}\,\, e^{- S_{lat}\left(\left\lbrace x_t \right\rbrace_s\right)}}{\displaystyle\sum_{s=1}^{n} \frac{e^{- S_{lat}\left(\left\lbrace x_t \right\rbrace_s\right)}}{p_s}}.
\label{eq2.16}	
\end{align}

If we look back at Eq.\eqref{eq2.15} then we can clearly see that the sum will be dominated by those terms whose exponential term is large as the function $e^{-x}$ decays very fast. So we can choose the probability distribution $p_s \propto e^{- S_{lat}\left(\left\lbrace x_t \right\rbrace_s\right)}$ as the required probability distribution. The exact probability distribution $p_s$ is given by
\begin{align}
p_s = \frac{e^{- S_{lat}\left(\left\lbrace x_t \right\rbrace_s\right)}}{\displaystyle\sum_{s=1}^{n} e^{- S_{lat}\left(\left\lbrace x_t \right\rbrace_s\right)}}.
\end{align}

Now if we substitute this in Eq. \eqref{eq2.16}, the average value $\left\langle \mathcal{O} \right\rangle_n$ just becomes the arithmetic average
\begin{align}
\left \langle \mathcal{O} \right \rangle_n = \frac{1}{n} \sum \limits_{s=1}^{n} \mathcal{O} \left(\left\lbrace x_t \right\rbrace_s\right).
\label{eq2.18}
\end{align}

Now we require some method to sample data according to probability distribution $p_s$. Once we get the means of generating field configurations according to probability distribution $p_s$, we can calculate expectation value of any observable using Eq. \eqref{eq2.18}. 

{\it Metropolis algorithm} \cite{metropolis1953equation} is one of the algorithms to sample data from probability distributions which are known up to some multiplicative constant. Metropolis algorithm uses Markov chain to generate these configurations and in the limit of Markov chain length going to infinity, it converges to the exact probability distribution. During this process the system passes through the region of configuration space which contains most of the values contributing to the expectation value.

Now let us describe the algorithm. Suppose the probability distribution we want to sample data from is $P(\vec{x} = \left\lbrace x_1,\,x_2,\,\dotsc,\,x_n\right\rbrace)$, where $P(\vec{x})$ depends on $n$ variables.
\begin{itemize}
\item \textbf{Step 1}: Choose an initial state say $\vec{x}_0 = \left\lbrace x_1^{0},\,x_2^{0},\,\dotsc,x_n^{0} \right\rbrace$.
\item \textbf{Step 2}: Then we choose a distribution $Q(\vec{y}_2|\vec{y}_1)$ to generate a small random change in the previously accepted state. The distribution $Q(\vec{y}_2 | \vec{y}_1)$ is symmetric, i.e., $Q(\vec{y}_2|\vec{y}_1) = Q(\vec{y}_1|\vec{y}_2)$. The new proposed state is $\vec{x}_{\rm pro} = \vec{x}_{n} + \Delta\vec{x}\,\,$ where $\Delta\vec{x}$ is distributed according to $Q(\vec{x}_{\rm pro}|\vec{x}_n)$.
\item \textbf{Step 3}: Let $A(\vec{x}_{\rm pro},\vec{x}_n) = {\rm min} \left\lbrace 1,\dfrac{P(\vec{x}_{\rm pro})}{P(\vec{x}_n)} \right\rbrace$. $A(\vec{x}_{\rm pro},\vec{x}_n)$ is the probability of accepting the state $\vec{x}_{pro}$ starting from $\vec{x}_{n}$. We accept the state $\vec{x}_{\rm pro}$ with probability $A(\vec{x}_{\rm pro},\vec{x}_n)$. If the state is accepted then $\vec{x}_{n+1} = \vec{x}_{\rm pro}$ otherwise $\vec{x}_{n+1} = \vec{x}_n$.
\item \textbf{Step 4}: Repeat Steps 2 and 3.
\end{itemize}

This algorithm makes a random walk in the configuration space sometimes accepting the state and sometimes remaining in the same place. Intuitively this algorithm accepts the more probable states by making use of $A(\vec{x}_{\rm pro},\vec{x}_n)$. So states with relatively larger probability will always be accepted but states with relatively smaller probability will be accepted with probability $A(\vec{x}_{\rm pro},\vec{x}_n)$. So more points are sampled from high-density regions of $P(\vec{x})$, while visiting low-density region for very less number of times.

Computationally \textbf{Step 3} is implemented by choosing a random number $\alpha\in[0,1]$ from a uniform random number generator. We used ran3 function from the text \emph{Numerical Recipes} \cite{2002numerical} for this purpose. If $\alpha \le A(\vec{x}_{pro},\vec{x}_n)$ then we accept the proposed state and otherwise reject the proposed state and keep the original state. 

Necessary and sufficient conditions for this algorithm to converge are the {\it detailed balanced condition} and {\it ergodicity} \cite{robert2004monte}. Ergodicity simply means that the system should be able to reach from one state to another state in finite number of steps. Detailed balance condition is that the probability to reach from $\vec{x}_1$ to $\vec{x}_2$ is same as the probability to reach from $\vec{x}_2$ to $\vec{x}_1$, i.e., $P(\vec{x}_1) \, Q(\vec{x}_2|\vec{x}_1) \, A(\vec{x}_2, \vec{x}_1) = P(\vec{x}_2) \, Q(\vec{x}_1|\vec{x}_2) \, A(\vec{x}_1, \vec{x}_2)$. To prove this, first note that for Metropolis algorithm $Q(\vec{x}_2|\vec{x}_1) = Q(\vec{x}_1|\vec{x}_2)$ and $A(\vec{x}_2, \vec{x}_1) = {\rm min} \left\lbrace  1, \dfrac{P(\vec{x}_2)}{P(\vec{x}_1)}\right\rbrace$.
\begin{align}
\text{If } P(\vec{x}_2) > P(\vec{x}_1)\,\,\Rightarrow A(\vec{x}_2, \vec{x}_1) & = 1 \,\, {\rm~and~} \,\, A(\vec{x}_1, \vec{x}_2) = \dfrac{P(\vec{x}_1)}{P(\vec{x}_2)}\\
P(\vec{x}_1) \, A(\vec{x}_2, \vec{x}_1) & = P(\vec{x}_1)\\
P(\vec{x}_2) \, A(\vec{x}_1, \vec{x}_2) & = P(\vec{x}_2) \, \dfrac{P(\vec{x}_1)}{P(\vec{x}_2)} = P(\vec{x}_1)\\
\text{So, } P(\vec{x}_1) \, Q(\vec{x}_2|\vec{x}_1) \, A(\vec{x}_2, \vec{x}_1) & = P(\vec{x}_2) \, Q(\vec{x}_1|\vec{x}_2) \, A(\vec{x}_1, \vec{x}_2)
\end{align}
The proof of ergodicity condition is given in Ref. \cite{robert2004monte}.

%%%%%%
\section{Hamiltonian Monte Carlo (HMC) Algorithm}
%%%%%%

Metropolis algorithm is very old and it has some drawbacks. First, the convergence rate is slow and second, it has high autocorrelation in the generated states. To overcome these problems, many algorithms had been developed - e.g., heat bath algorithm. But these algorithms also have some problems; they cannot be applied to all the models. So to rescue us from this HMC comes into play.

Metropolis algorithm uses random walk to move through the configuration space but HMC uses directed paths to move through the configuration space while sampling. It makes use of Hamilton's equations for this purpose.

For this algorithm, first we need to introduce conjugate momentum variable for each degree of freedom i.e., field variables and a fictitious time $\tau$ to evolve the field and conjugate momentum variables using Hamilton's equations. Let us say the probability distribution we want to sample data from is $P(\vec{x} = \left\lbrace x_1,\,x_2,\,\dotsc,\,x_n\right\rbrace)$. We can also write this probability distribution as $P(\vec{x}) = e^{-\left\lbrace-\log(P(\vec{x}))\right\rbrace}$. Now for each variable $x_i$ we introduce a conjugate momentum $p_i$ and defines the Hamiltonian of the system as
\begin{align}
H(\vec{x},\vec{p}) = \frac{1}{2} \sum \limits_{i=1}^{n} p_{i}^2 - \log(P(\vec{x})).
\label{eq2.23}
\end{align} 
 
The joint probability distribution is given by $\pi (\vec{x},\vec{p}) = e^{-H(\vec{x}, \vec{p})}$. If we integrate out the momentum variables $p_i$ from this joint probability distribution $\pi (\vec{x}, \vec{p})$ we get the original probability distribution $P(\vec{x})$. Hamilton's equations are as follows
\begin{align}
\frac{d \vec{x}}{d \tau} = \frac{\partial H}{\partial \vec{p}}, \qquad \qquad \frac{d \vec{p}}{d \tau} = - \frac{\partial H}{\partial \vec{x}}.
\end{align}

Hamilton's equations are used to evolve these field variables and their conjugate momenta in the fictitious time $\tau$ numerically. This is done using symplectic integrators i.e., those numerical algorithms for solving Hamilton's equation which preserves the time reversibility and volume of phase space property. The simplest symplectic integrator is the {\it leapfrog integrator/algorithm}.

The steps involved in the leapfrog algorithm are
\begin{align}
p_i\left(\tau + \frac{\epsilon}{2}\right) &= p_i(\tau) - \frac{\epsilon}{2} \frac{\partial H}{\partial q_i} (\vec{q}(\tau)),\label{eq2.25} \\
q_i(\tau + \epsilon) &= q_i(\tau) + \epsilon p_i\left(\tau +  \frac{\epsilon}{2}\right),\label{eq2.26} \\
p_i(\tau + \epsilon) &= p_i\left(\tau + \frac{\epsilon}{2}\right) - \frac{\epsilon}{2} \frac{\partial H}{\partial q_i}(\vec{q}(\tau + \epsilon))\label{eq2.27}.
\end{align}

Now let us describe the algorithm. Suppose the probability distribution we want to sample data from is $P(\vec{x} = \left\lbrace x_1,\,x_2,\,\dotsc,\,x_n\right\rbrace)$, where $P(\vec{x})$ depends on $n$ variables.
\begin{itemize}
\item \textbf{Step 1}: First we form the Hamiltonian as given in Eq. \eqref{eq2.23}. Then we choose some initial state say $\vec{x}_0 = \left\lbrace x_1^{0},\,x_2^{0},\,\dotsc,x_n^{0}  \right\rbrace = \vec{x}_0(0)$.
\item \textbf{Step 2}: Then we choose each $p_i = p_i(0)$ randomly from $\mathcal{N}(0, 1)$ i.e., normal distribution with $\mu = 0$ and $\sigma =1$. Then we calculate $H_i = H(\vec{x}_0(0), \vec{p}(0))$.
\item \textbf{Step 3}: Now we evolve the $\vec{x}_i \,\, \text{and} \,\, \vec{p}$ using leapfrog algorithm for $n$ time steps with $\epsilon$ being the time resolution parameter. The $\vec{x}_0(n \epsilon)$ is the new proposed state. Then we calculate $H_f = H(\vec{x}_i(n \epsilon),\vec{p}(n \epsilon))$.
\item \textbf{Step 4}: Now we apply the Metropolis test. $A(\vec{x}_i(n \epsilon),\vec{x}_i(0)) = {\rm min} \left\lbrace 1,\dfrac{\pi(\vec{x}_i(n \epsilon), \vec{p}(n \epsilon))}{\pi(\vec{x}_i(0),\vec{p}(0))} \right\rbrace = e^{-(H_f - H_i)}$. We accept the state $\vec{x}_i(n \epsilon)$ with the probability $A(\vec{x}_i(n \epsilon), \vec{x}_i(0))$. If the state is accepted then $\vec{x}_{i+1} = \vec{x}_i(n \epsilon)$, otherwise $\vec{x}_{i + 1} = \vec{x}_i(0)$.
\item \textbf{Step 5}: Repeat Steps 2 to 4.
\end{itemize}

We know that Hamilton's equations are time-reversible if $H(p, q) = H(-p, q)$ so we build the Hamiltonian Eq. \eqref{eq2.23}, which follows this property. This property is required for the detailed balance condition \cite{neal2011mcmc}. The volume of phase space also remains conserved in Hamiltonian dynamics. We need this because the joint probability distribution $\pi(\vec{q}, \vec{p})$ remains covariant under the Hamiltonian dynamics as the Jacobian of the transformation is 1. Since HMC uses Metropolis test for the Markov chain formation so both the sufficient and necessary conditions are satisfied. The proofs are given in Refs. \cite{neal2011mcmc} and \cite{betancourt2017conceptual}.

One of the tests to check whether the HMC is implemented properly is to look for the expectation value of $e^{-\Delta H}$\cite{Ydri:2015zba}. The value of this is given by
\begin{align}
\left\langle e^{-\Delta H} \right\rangle = \left\langle e^{-(H(\vec{x}',\vec{p}') - H(\vec{x},\vec{p}))} \right\rangle &= \frac{1}{Z} \int\, d\vec{p}\, d\vec{x}\,\,\, e^{-H(\vec{x},\vec{p})} \,\, e^{-(H(\vec{x}',\vec{p}') - H(\vec{x},\vec{p}))}\\
&= \frac{1}{Z} \int \, d\vec{p}\, d\vec{x}\,\,\, e^{-H(\vec{x}',\vec{p}')}\\
&= \frac{1}{Z} \int \, d\vec{p}'\, d\vec{x}'\,\,\, e^{-H(\vec{x}',\vec{p}')} \qquad {\rm as~} \frac{\partial(\vec{p}',\vec{x}')}{\partial(\vec{p},\vec{x})} = 1\\
\left\langle e^{-\Delta H} \right\rangle &= \frac{1}{Z}\,\, Z = 1. \label{eq2.31}
\end{align}
Eq. \eqref{eq2.31} acts as one of the tests for the HMC algorithm. If the value of $e^{- \Delta H}$ fluctuates away from 1 then it is clear indication that the algorithm is not implemented correctly.

%%%%%%
\section{Statistical Error in MCMC Algorithms}
%%%%%%

After thermalization, we can use the field configurations to calculate the expectation values of observables using Eq.\eqref{eq2.18}. But the error in the expectation value depends on whether the states used to calculate the expectation are correlated or not. If the states $\vec{x}_1, \vec{x}_2, \dotsc, \vec{x}_N$ are not correlated, then error in the expectation of $f$ is given by \cite{gattringer2009quantum}
\begin{align}
\delta f = \frac{\sigma}{\sqrt{N}}, 
\end{align}
where $\sigma$ is the standard deviation in the expectation value of $f$. We can get the uncorrelated states by finding the autocorrelation length and taking the states for expectation values after the autocorrelation length. Autocorrelation length is where autocorrelation becomes nearly zero.

%%%%%%
\section{Simulation Details}
%%%%%%

For all Monte Carlo simulations certain features remain common. We will discuss some of these basic features here and additional features will be discussed in the later chapters.

%%%%%%
\subsection{Thermalization and Initial Condition}
%%%%%%

We can initialize our simulation with any field configuration for both the algorithms, and we will get the same result from any starting point as the algorithm respects the ergodic property. But sometimes the algorithm may take longer time to converge to the desired probability distribution for some starting points. So, if we have some {\it a priori} information regrading the probability distribution then we can use that for selecting the initial condition. But mostly we do not have {\it a priori} information so we mostly make a {\it cold start} for the simulation. A cold start is the initial configuration in which all the field configurations are set to zero (for bosonic fields only). If we start with a random field configuration, then we call it a {\it hot start}. Sometimes people use cold start for some fields and hot start for some fields, this is known as mixed-field configuration. Mostly in our simulations, we used cold start but for some simulations we used mixed-field configurations as well.

After choosing the initial field configurations, we run the simulation to allow the distribution to reach the equilibrium distribution and after that we use it to generate/sample data. This process is known as {\it thermalization}. Thermalization is achieved when the value of the observable starts to oscillate about some value. Then we can use the data after thermalization to calculate the expectation values. Thermalization occurs faster in HMC as compared to Metropolis algorithm. We will see this in the later chapters.

%%%%%%
\subsection{Sweeps and Updating}
%%%%%%

On the lattice we choose to update only one field variable at one lattice site at a time and repeat this for other lattice sites. For Metropolis algorithm it is better to update the field variable at one lattice site multiple times before moving to the next field variable. This is done so that the system converges to the required distribution faster. For HMC it does not give any better result by making multiple changes at one lattice site. So we update the field variables only once at each lattice site for HMC.

When we have updated the whole lattice with the new configuration it is known as a {\it sweep}. We collect all the information of the observables after each sweep and store this data in a file. Every thermalization plots are ploted against the sweep number. To prevent any data loss we store the field configurations after every 1000 sweeps. This way we did not need to thermalize the system again and again and we can restart the simulation with this previous stored configuration.

%%%%%%
\subsection{Autocorrelation}
%%%%%%

It is important to note that field configurations generated by the Markov process are not independent, each configuration is dependent on the previous configuration. So we cannot directly evaluate our observable from these correlated field configurations as this will increase error bars and it would require more field configurations to reduce this error bars. So what we do is that we run the simulation for enough number of times and store the values of observables. Then we calculate the {\it normalized autocorrelation} to get to know how many sweeps are required to generate statistically independent field configuration. Autocorrelation and normalized autocorrelation are defined as
\begin{align}
\Gamma_{\tau} &= \frac{1}{N - \tau} \sum \limits_{i=1}^{N-\tau} (f_i - \left\langle f \right\rangle) (f_{i + \tau} - \left\langle f \right\rangle) \,\,\,
&{\rm Autocorrelation}, \\
\rho_{\tau} &= \frac{\Gamma_{\tau}}{\Gamma_{0}} & {\rm Normalized \,\,\, Autocorrelation},
\end{align}
where $N$ is the number of data points after the thermalization, $\tau$ is the number of sweeps for which we are checking the correlation. $\Gamma_0$ is simply the variance. Autocorrelation length is defined as the minimum number of sweeps required to get the uncorrelated field configuration. So we find the autocorrelation length by looking at the values of the normalized autocorrelation where it crosses 0 or just near 0. After getting the autocorrelation length we calculate the expectation value of the observable and error bars from these uncorrelated field configurations. From now on we will use the term autocorrelation for the normalized autocorrelation.

% Chapter 3 -----------------------------------------------------------------------------------------------------------------------------
 % How to give reference to the Appendix
% Footnote hyperlink goes to page 1??

%%%%%%
\chapter{Harmonic Oscillator and Harmonic Oscillator with Commutator Potential}
\addcontentsline{}{×}{×}
%%%%%%

In this chapter, we are going to simulate two models, one using Metropolis algorithm and the other one using HMC algorithm. From these simulations we will see that the HMC algorithm converges to the required distribution faster compared to the Metropolis algorithm.

We start this chapter with a simple model: a collection of $N^2$ uncoupled-harmonic oscillators. The Euclidean action for this model is
\begin{align}
S_E = \frac{1}{2} \int \limits_{0}^{\beta} dt \,\, \Tr\left( \dot{X}^2 + m^2 X^2 \right),
\end{align}
where $X$ is a $N \times N$ Hermitian matrix with periodic boundary condition $X(t) = X(t + \beta)$, with $\beta$ being the period of the system. We can easily see, by expanding the trace, that it is indeed the Euclidean action of $N^2$ uncoupled-harmonic oscillators. This model is exactly solvable, but our aim here is to simulate this using Metropolis algorithm and verify the simulation results with the analytical results. Now, using the discretization procedure from the previous chapter, Eqs. \eqref{eq2.4} - \ref{eq2.7}, the Euclidean action on the lattice takes the form
\begin{align}
\label{eq3.2}
S_{\rm lat} = \frac{a}{2} \sum \limits_{t=1}^{T} \,\,  \Tr \left[ \left( \frac{X_t - X_{t-1}}{a} \right)^2 + m^2 {X_t}^2 \right], 
\end{align}
where $X_0 = X_T$. Our aim is to generate the field configurations whose probability distribution is proportional to $e^{-S_{\rm lat}}$. We will use Metropolis algorithm to sample data from this probability distribution.

%%%%%%
\section{Metropolis Algorithm for Harmonic Oscillator}
%%%%%%

This section introduces us on how to apply Metropolis algorithm to generate field configurations according to the given probability distribution. The idea is to propose a random change in the field configurations and then compare it with the previous field configurations. We can either propose a change in the field configuration at one lattice site or we can propose the change in all the field configurations. It is easier to make a change in a single field configuration at a time as each time we need to calculate the difference in the action. That is, 
\begin{equation}
\Delta S = S(X_t^{\rm pro}) - S(X_t^{\rm old}).
\end{equation}
To calculate this, we can see from Eq. \eqref{eq3.2} that the terms which contribute to this difference only contains terms with $X_t$ and we do not need to calculate the whole action to calculate this change.

We can separate the terms containing $X_t$ from the action and we get the local part of the action $S_{\rm loc}$ containing $X_t$
\begin{align}
S_{\rm loc} = \frac{a}{2} \,\, \Tr \left[ \left( \frac{2}{a^2} + m^2 \right)X_t^2  - \frac{2 X_t}{a^2} \left( X_{t-1} + X_{t+1} \right) \right].
\end{align}
Now we need to propose a random change from a probability distribution which is symmetric i.e., $Q(x|y) = Q(y|x)$. We can use any distribution with this property but the easiest one is the uniform distribution. Thus the random change proposed to the field configuration has the form
\begin{align}
X_t^{\rm pro} = X_t^{\rm old} + \delta Y_t,
\end{align}
where $Y_t$ is a random Hermitian matrix with each element, both real and complex, taken from a uniform probability distribution between $(-1,1)$, and $\delta$ is a configurable parameter which will allow us to set the acceptance rate for this algorithm. Now we need to compare the proposed state with the old state. For that we need to calculate the acceptance probability, which is just $e^{-\Delta S}$ here. Thus the acceptance probability is
\begin{align}
e^{- \Delta S} = e^{-(S_{\rm loc}(X_t^{\rm pro}) - S_{\rm loc}(X_t^{\rm old}))}.
\end{align}
We accept the proposed state with the probability $e^{- \Delta S}$. For that we take a random number $\alpha$ from a uniform random number generator between $(0,1)$. If $e^{- \Delta S} \geq \alpha$, then we accept the proposed state otherwise we reject the proposed state and propose a new state. We continue this procedure until we have enough data to calculate the expectation values of observables.

In the above the parameter $\delta$ controls the efficiency of the algorithm. It is known as the jump parameter. We look at the effectiveness of the algorithm by looking at the \emph{acceptance rate}. Acceptance rate is defined as the ratio of the total number of accepted states to the total number of proposed states. If $\delta$ is small, then the acceptance rate will be high, and it will take small steps in the configuration space. This way we need to run the simulation for longer time to get enough uncorrelated data for the expectation value of the observable. If $\delta$ is large, then it will take large steps in the configuration space, and the acceptance rate will be low. Again we need to simulate for a longer time to get the desired result. So we need some value in between these large and small values to get the acceptance between 65-85\%. Acceptance rate between this range is good enough to get the data sampled according to the probability distribution.

%%%%%%
\section{Simulation Details and Results}
%%%%%%

In this section, we will discuss the important quantities which are needed to monitor the reliability of simulations and to compare the simulation data with the analytical results. 

In simulation everything is measured in unit of a length $a_0$. Since action needs to be dimensionless in natural units, so $X$ has a dimension of $[L^{3/2}]$, mass $m$ has a dimension of $[L^{-1}]$. So $a$ is measured in units of $a_0$, $X$ is measured in units of $a_0^{3/2}$ and mass $m$ is measured in units of $a_0^{-1}$. We use the following parameters for the simulations: $N = 16$, $a = 1$, $m = 1$, $T = 50$ and $\beta = 50$.

%%%%%%
\subsection{Thermalization}
%%%%%%

\begin{figure}[h!]
\centering
\includegraphics[width = 0.95\textwidth]{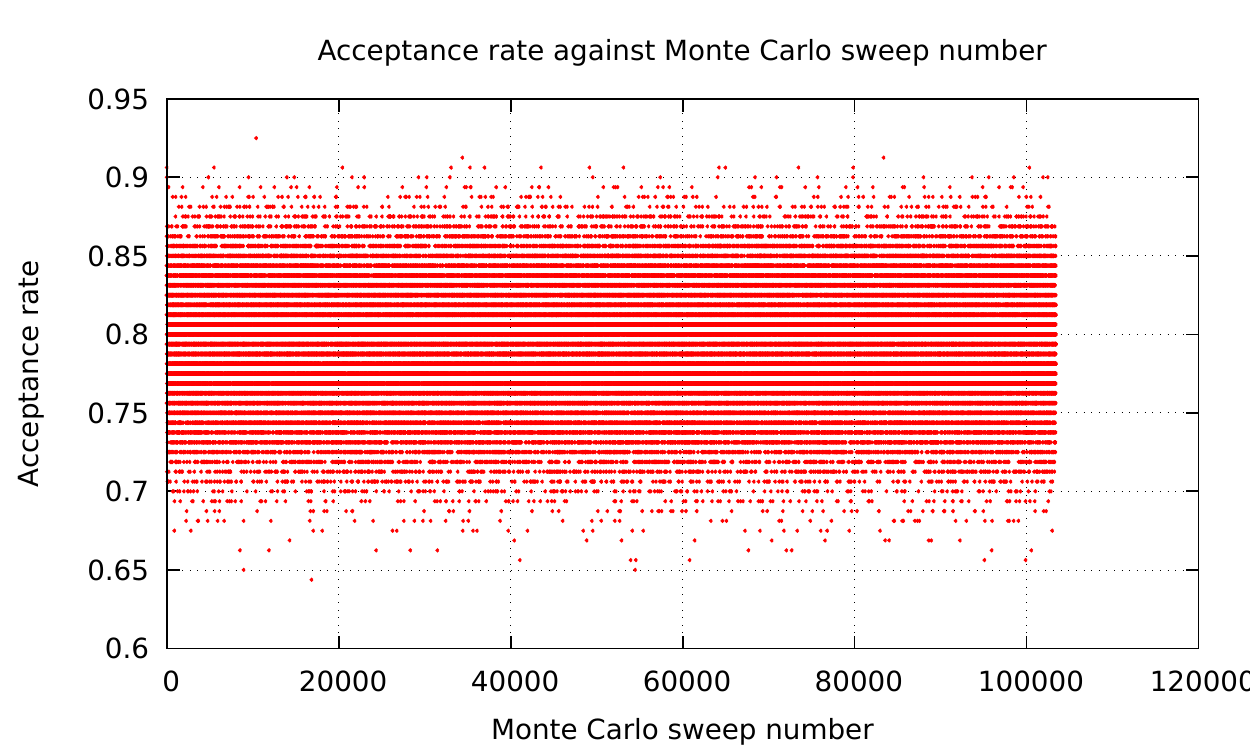}
\caption[Acceptance rate against Monte Carlo sweep number.]{Acceptance rate against Monte Carlo sweep number. We can see that most of the values of the acceptance rate lies between 70\% and 85\%, which is good enough to get the desired results.}
\label{fig3.1}
\end{figure}

Let us discuss the initial conditions for the simulations first. For this model, we started our simulation with a cold start, i.e., all the field configurations are set to zero matrices. We also need to choose the value of the parameter $\delta$ so that we get the acceptance rate between 65\% and 85\%. In Fig. \ref{fig3.1} we show the acceptance rate for this simulation. 
  
\begin{figure}[h!]
\centering
\includegraphics[width = 0.95\textwidth]{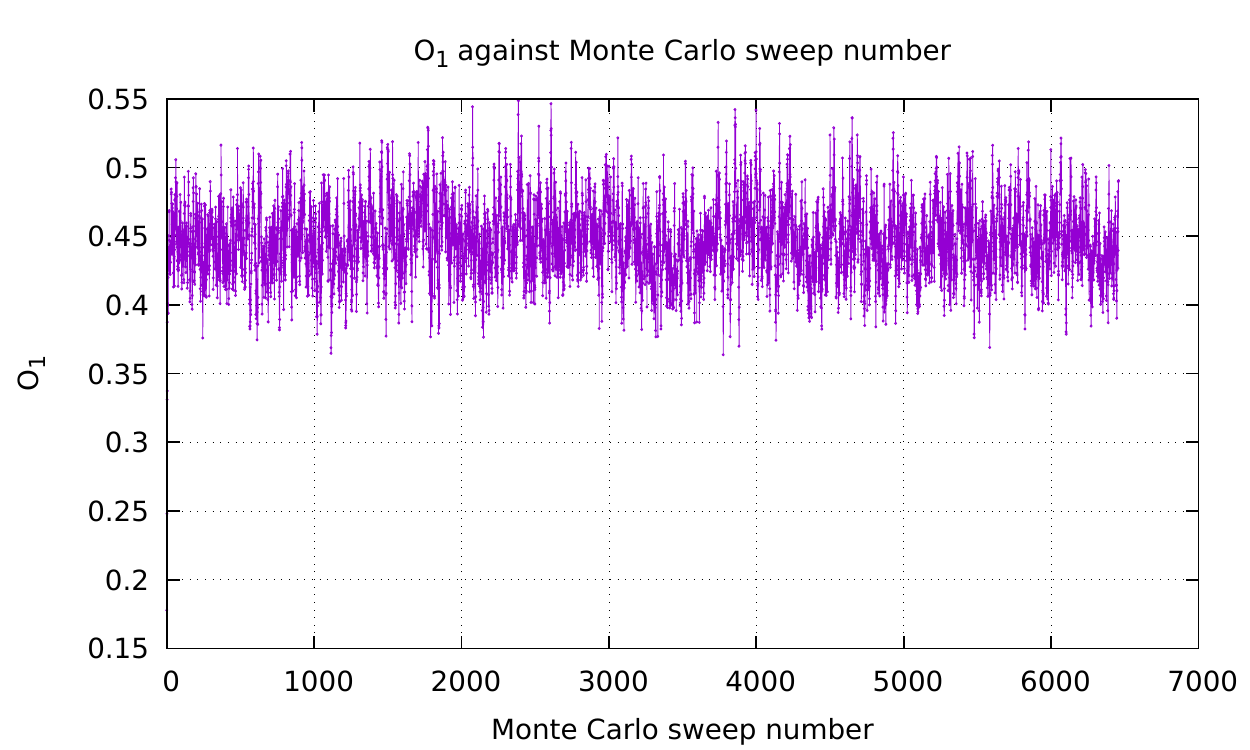}
\caption[Run-time history of observable $O_1$ against Monte Carlo sweep number.]{Run-time history of observable $O_1$ against Monte Carlo sweep number. Here we only plotted the run-time history after the system has been thermalized and discarded all the previous values. We can clearly see that the value of the observable $O_1$ is oscillating around 0.45.}
\label{fig3.2}
\end{figure}

After that, we update the field configurations at a single lattice site for multiple times (in our case we updated field at each site 50 times) to make the system thermalize faster. We stored the value of the observables after every sweep. Now we need to decide whether the system has been thermalized or not. For that what we do is to simulate the system for large number of sweeps, say 10000. Then we plot one of the observables against the number of sweeps to see whether the system has been thermalized or not. When the value of the observable starts to oscillate about some value, then we are sure that the system has been thermalized. Let us look at the run time history of the observable $O_1$ defined as
\begin{align}
O_1 \equiv \frac{1}{N^2 \beta} \int \limits_0^{\beta} dt \,\, \Tr(X^2) = \frac{1}{N^2 T} \sum \limits_{t=1}^T \Tr(X_t^2) \qquad \textrm{(Discretized version)}.
\end{align}
Now we collected the data for this observable after the system has been thermalized. We can see in Fig. \ref{fig3.2} that the value of $O_1$ is oscillating about 0.45. This observation leads us to the conclusion that the system has been thermalized and now we can collect the data for our calculation purpose.

%%%%%%
\subsection{Autocorrelation}
%%%%%%

After thermalization, we collected data for the calculation purpose. It is clear from Fig. \ref{fig3.2} that the value of $O_1$ after every sweep is not statistically independent and it takes multiple sweeps to get a statistically independent value of the observable. We use the autocorrelation to get to know after how many sweeps the value of the observable are statistically independent.
 
\begin{figure}[!htb]
\centering
\includegraphics[width = \textwidth]{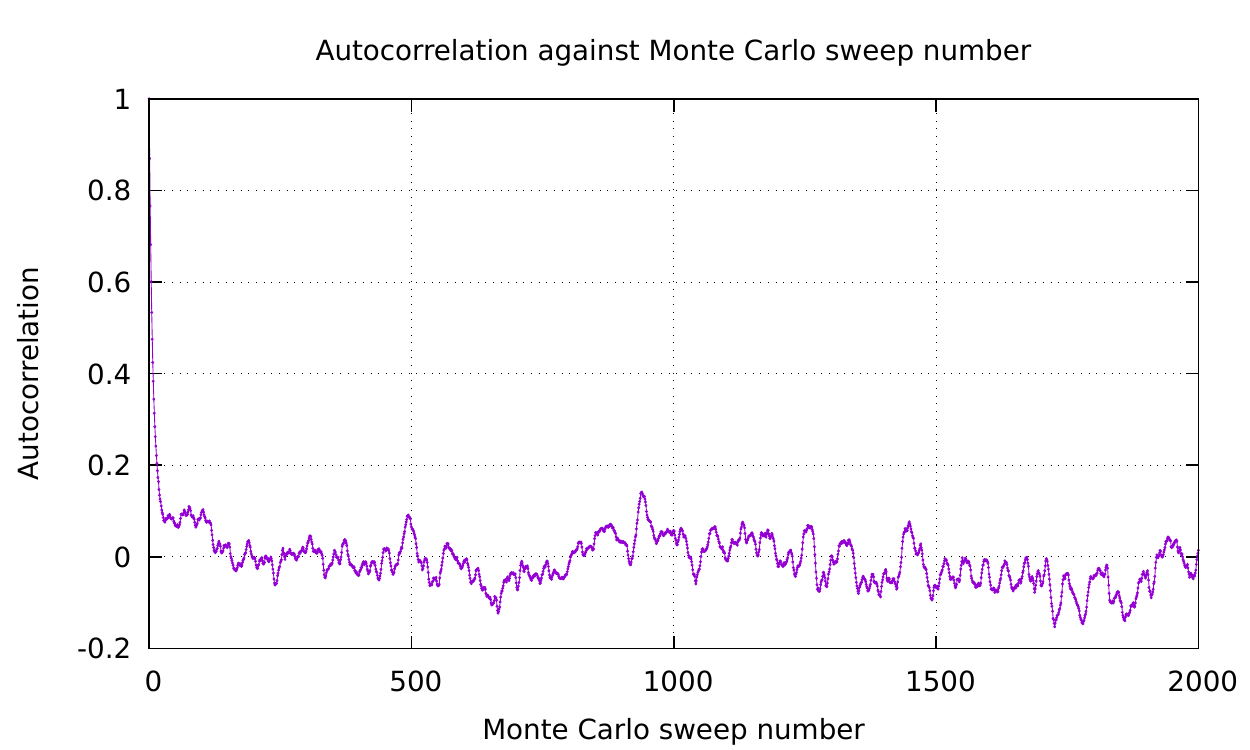}
\caption[Normalized autocorrelation against Monte Carlo sweep number for observable $O_1$.]{Normalized autocorrelation against Monte Carlo sweep number for observable $O_1$. It starts from 1 and decreases exponentially towards 0 and then oscillates about 0. It crosses 0 value after 158 sweeps, so the autocorrelation length is 158 for $O_1$.}
\label{fig3.3}
\end{figure}

Fig. \ref{fig3.3} shows the plot of normalized autocorrelation of the observable $O_1$. We can see that the plot decreases exponentially and after that, it just oscillates about zero. The point where the plot crosses the zero in the graph, we defined that thing as the autocorrelation length. So from this autocorrelation plot, we can see that the autocorrelation length for observable $O_1$ is 158. Autocorrelation is an observable dependent quantity. It will be different for other observables. Hence the autocorrelation length will also be different for other observables.
 
For calculating any expectation value using Eq. \ref{eq2.18}, we need to take a gap of at least the autocorrelation length so that we only take the average of the uncorrelated data.

%%%%%%
\subsection{Two-point Correlation Function}
%%%%%%

\begin{figure}[h!]
\centering
\includegraphics[width = \textwidth]{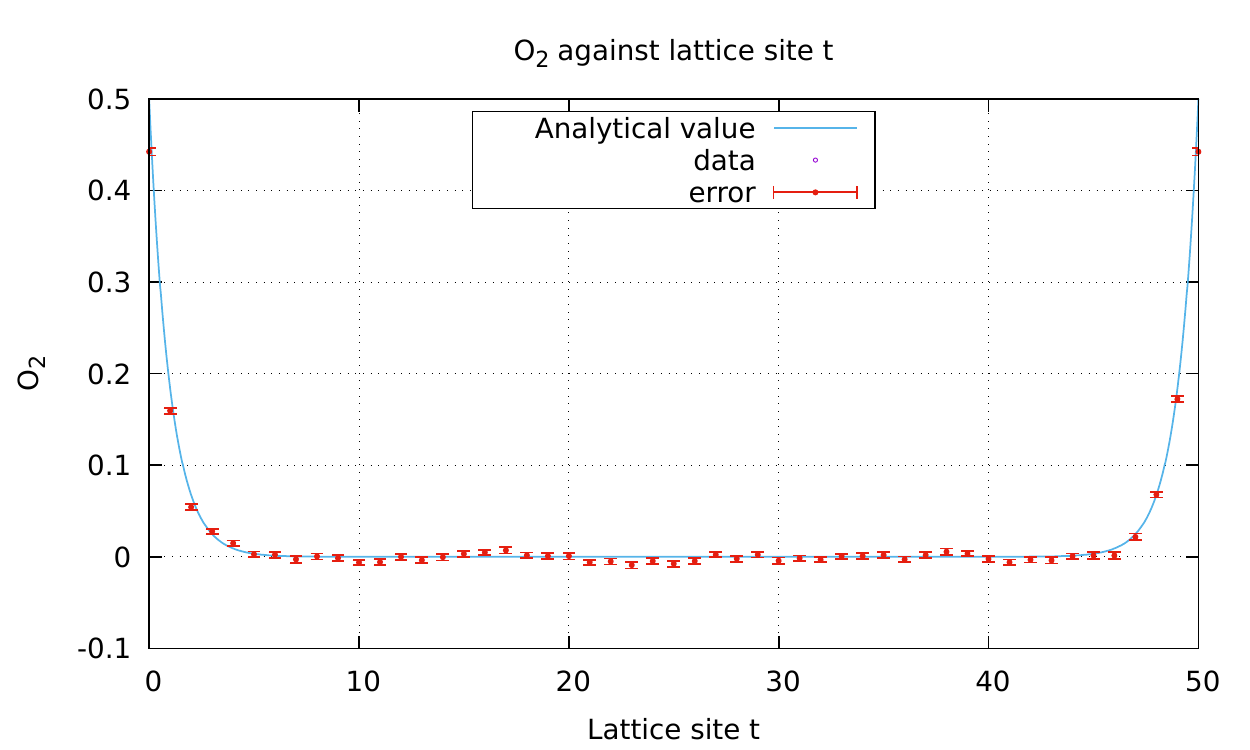}
\caption[$\left \langle O_2(t) \right\rangle$ against lattice site $t$.]{$\left \langle O_2(t) \right\rangle$ against lattice site $t$. We simulated this on a lattice with 50 sites with lattice spacing $a = 1$. We can see that the simulated result agrees with the theoretical result.}
\label{fig3.4}
\end{figure}

One of the observables which we want to look at is the two-point correlation function. It is defined as
\begin{align}
O_2(t) \equiv \frac{1}{N^2} \Tr(X(0)X(t)) = \frac{1}{N^2} \Tr(X_0 X_t) \qquad \textrm{(Discretized form)}.
\end{align}
The theoretical result for the expectation value of $O_2$ is
\begin{align}
\left\langle O_2(t) \right\rangle = \frac{e^{-mt} + e^{-m(\beta-t)}}{2m (1-e^{-\beta m})}.
\end{align}
Fig. \ref{fig3.4} shows the simulated result along with the theoretical result. We can see that it matches the theoretical result at most of the points but at some points it lies inside the error range. This result is convincing enough to believe that the algorithm is in working order. 

%%%%%%
\section{HMC for Harmonic Oscillator with Commutator Potential}
%%%%%%

Now we introduce more complexity into the harmonic oscillator potential by adding a commutator square term into the action. The Euclidean action is given by
\begin{align}
\label{eq3.9}
S_E = \frac{N}{2 \lambda} \int \limits_{0}^\beta \, dt \, \Tr \left[ {\dot{X^i}}^2 - \sum \limits_{\substack{i, j = 1\\ i < j}}^{d} [X^i, X^j]^2 + m^2 {X^i}^2 \right], 
\end{align}
where $i, j$ varies from 1 to $d$, $X^i$ are $N \times N$ Hermitian matrices, $\lambda$ is the 't Hooft coupling defined as $\lambda = N g_{\rm YM}^2$ where $g_{\rm YM}$ is the one-dimensional Yang-Mills coupling, $m$ is the mass. We are also using Einstein's summation convention that the repeated indices are summed over. In the above $d$ is the number of spatial dimensions we want in our system. Right now we are just studying how to implement the algorithm properly so we stick to three spatial dimensions for this simulation. That is, $d = 3$ in our simulations. (Note that this theory can be considered as the one obtained by dimensional reduction of a four-dimensional mother theory down to one dimension.)

Ideally, we need to simulate this model without mass term, but the model possesses flat directions. If we have a commutator potential square term $\Tr([X^i, X^j]^2)$ then the system has flat directions associated with it \cite{Joseph_2015}. Flat directions are those field configurations which mutually commute with each other, i.e., $[X^i, X^j] = 0$ but the value of $\Tr(X_i^2)$ keeps on increasing without an upper bound. This makes the simulation unstable, and we need to remove this to get the appropriate results. We can remove the flat directions in two ways. One way is to set the zero mode of the system. This is done by setting the global trace of the fields to zero.
\begin{align}
\label{eq3.10}
\sum \limits_{t=1}^T \Tr(X_t^{i}) = 0 \qquad \textrm{for each $i$}.
\end{align}
But for simulations it is much easier to set the local trace to zero i.e., $\Tr(X_t^i) = 0$ for each $t$ and $i$. This will automatically satisfy the global trace condition Eq. \eqref{eq3.10}.

Another way is to add a mass term in the Euclidean action itself as we have added in Eq. \eqref{eq3.9}. Then we simulate the model for different mass values and extrapolate it towards zero mass limit to get the value of the observable at zero-mass value. We will simulate this model by applying both the methods to remove the flat directions and compare the extrapolated results with the exact zero-mass result.

Using the discretization procedure mentioned in Eqs. \eqref{eq2.4} - \eqref{eq2.7}, the Euclidean action becomes
\begin{align}
S_{\rm lat} = \frac{N a}{2 \lambda} \sum \limits_{t=1}^T \,\, \Tr \left[ \left( \frac{X_{t}^{i} - X_{t-1}^{i}}{a} \right)^2 - \sum \limits_{\substack{i, j = 1 \\ i < j}}^d [X_t^i, X_t^j]^2 + m^2 {X_t^i}^2 \right]
\end{align}
with $X_0^i = X_T^i$.

Now we need to setup the HMC algorithm for this model. First we need to introduce the momentum variable for each field variable $X_t^i$. We introduce the momentum matrix $P_t^i$ for each $X_t^i$ matrix. Note that both $X_t^i$ and $P_t^i$ are Hermitian matrices. Now we introduce the Hamiltonian for this system. The Hamiltonian for this system is given by
\begin{align}
H =& \sum \limits_{t=1}^T \frac{1}{2} \Tr({P_t^i}^2) + S_{\rm lat}, \\
H =& \sum \limits_{t=1}^T  \Tr \left[ \frac{1}{2}{P_t^i}^2 + \frac{Na}{2 \lambda} \left \lbrace \left( \frac{X_{t}^{i} - X_{t-1}^{i}}{a} \right)^2 - \sum \limits_{\substack{i, j = 1 \\ i < j}}^d [X_{t}^{i}, X_{t}^{j}]^2 + m^2 {X_{t}^{i}}^2 \right \rbrace \right].
\end{align}

Our next task is to set the initial conditions for the field variables and take the momentum value randomly from the $\mathcal{N}(0,1)$ distribution. Details about how to generate random numbers distributed as Gaussian distribution can be found in Ref. \cite{Ydri:2015zba}. After that we need to evolve the field and momentum variables according to the Hamilton's equations. 

We need to find the forces for each momentum variable. We have
\begin{equation}
(\dot{X_t^i})_{rs} = \frac{\partial H}{\partial (P_t^i)_{sr}} \hspace{0.5cm}  {\rm and} \hspace{0.5cm}  (\dot{P_t^i})_{rs} = - \frac{\partial H}{\partial (X_t^i)_{sr}},
\end{equation}
where the dot on the field and momentum variables are time derivatives with respect to the fictitious time $\tau$. We then have the equation of motion
\begin{eqnarray}
(\dot{X_t^i})_{rs} &=& (P_t^i)_{rs}, \\
(\dot{P_t^i})_{rs} &=& - \frac{N a}{\lambda} \Bigg[ \left( \frac{2}{a^2} + m^2 \right) {(X_t^i)}_{rs} - \frac{1}{a^2} \left[ {(X_{t-1}^i)}_{rs} + {(X_{t+1}^i)}_{rs} \right] \nonumber \\
&& \hspace{2cm} -  \sum \limits_{\substack{j=1\\j\neq i}}^{d} [X_t^j,[X_t^i,X_t^j]]_{rs} \Bigg].
\end{eqnarray}

In matrix form these equations simply become
\begin{align}
\dot{X_t^i} =& P_t^i, \\
\dot{P_t^i} =& - \frac{N a}{\lambda} \Bigg[ \left( \frac{2}{a^2} + m^2 \right) X_t^i -\frac{1}{a^2}\left(X_{t-1}^i + X_{t+1}^i\right) -  \sum \limits_{\substack{j=1\\j\neq i}}^{d} [X_t^j,[X_t^i,X_t^j]] \Bigg].
\end{align}

We will evolve the above equations using the leapfrog algorithm described in the previous chapter, Eq. \eqref{eq2.25} - \eqref{eq2.27}. The equations are simply
\begin{align}
P_t^i\left(\tau + \frac{\epsilon}{2}\right) &= P_t^i(\tau) - \frac{\epsilon}{2} \frac{\partial H}{\partial X_t^i} (X_t(\tau)), \\
X_t^i(\tau + \epsilon) &= X_t^i(\tau) + \epsilon P_t^i \left(\tau +  \frac{\epsilon}{2}\right), \\
P_t^i(\tau + \epsilon) &= P_t^i \left(\tau + \frac{\epsilon}{2}\right) - \frac{\epsilon}{2} \frac{\partial H}{\partial X_t^i}(X_t(\tau + \epsilon)),
\end{align}
where $\epsilon$ is a tunable parameter. We also have one more tunable parameter here which is the number of times we evolve the fields using the leapfrog algorithm. We call this parameter $n$. So we evolve the fields for $n$ times and for $n \epsilon$ amount of fictitious time. Both $\epsilon$ and $n$ must be tuned to get the desirable acceptance rate in the simulations.

After evolving the fields and their conjugate momenta, we need to compare it with the initial field and their momenta. For that, we need to calculate the acceptance probability for this algorithm. Acceptance probability for this algorithm is simply $e^{-\Delta H} = \exp{-(H(X^i(n\epsilon), P^i(n\epsilon)) - H(X^i(0), P^i(0)))}$. We accept the proposed state with the probability $e^{-\Delta H}$. For that, we take a random number $\alpha$ from a uniform random number distribution between $(0,1)$. If $\alpha \leq e^{- \Delta H}$, then we accept the proposed state, otherwise we reject the proposed state and propose a new state and continue this procedure until we have enough data to calculate the expectation value of the observable.

For this algorithm also we require the acceptance rate between 65-85\% to move through the configuration space optimally. This acceptance rate can be achieved by tuning $\epsilon$ and $n$ parameters appropriately.

%%%%%%
\section{Simulation Details and Results}
%%%%%%

Similar to the Metropolis algorithm, we need to monitor some of the important things for the HMC algorithm also. We will discuss these in this section and later will show some results for this model.

In this simulation each variable and parameter has been measured in the unit of 't Hooft coupling $\lambda_0$. Since the Yang-Mills coupling $g_{\rm YM}^2$ has a dimension of [$L^{-3}$] in one dimension, so $X$ can be measured in the unit of $\lambda_0^{1/3}$, $a$ can be measured in units of $\lambda_0^{-1/3}$ and mass $m$ can be measured in units of $\lambda_0^{1/3}$. Thus, every variable and parameter appearing in Eq. \eqref{eq3.9} are dimensionless. Values of the parameter used for this simulations are: $N = 8$, $a = 0.5$, $T = 10$, $\beta = 5$, $d = 3$ and $\lambda = 1$. We have taken $m = 0.01$ for the results of the next two sub-sections.

%%%%%% 
\subsection{Initial Conditions, Acceptance Rate and Thermalization}
%%%%%%

\begin{figure}[!h]
\centering
\includegraphics[width = 0.85\textwidth]{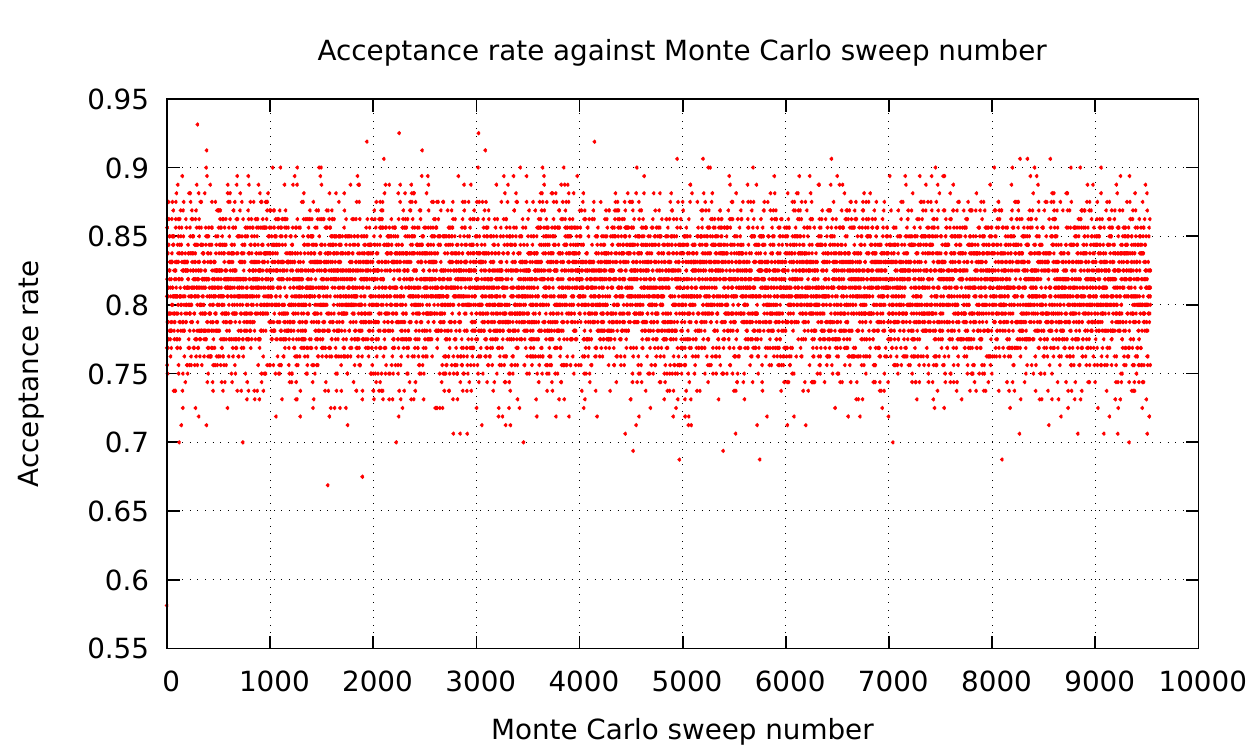}
\caption[Acceptance rate against Monte Carlo sweep number.]{Acceptance rate against Monte Carlo sweep number. We can see that most of the data points are concentrated between 75 to 85\%, which is what we want for good simulation results.}
\label{fig3.5}
\end{figure}

We started the simulation with a cold start, i.e., all the field configurations are set to zero matrices initially. To explore the configuration space more efficiently, we choose our tunable parameter $\epsilon$ and $n$ so that we get the acceptance rate between 65-85\%. Fig. \ref{fig3.5} shows the acceptance rate against Monte Carlo sweeps for this simulation. Tuning these parameters is a very crucial step for the HMC algorithm. Sometimes a slight change in these parameters can make the simulation unstable.

\begin{figure}[!h]
\centering
\includegraphics[width = 0.95\textwidth]{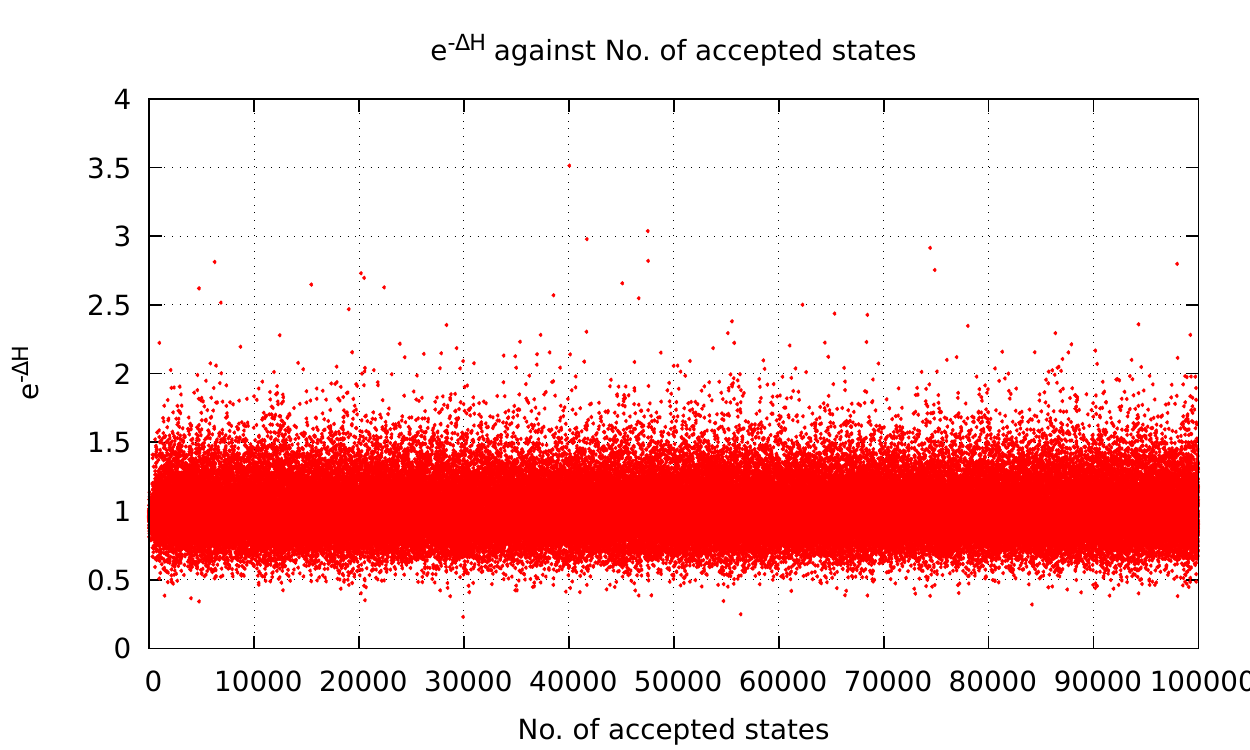}
\caption[The plot of $e^{-\Delta H}$ against the number of accepted states.]{The plot of $e^{-\Delta H}$ against the number of accepted states. In this plot we can see that most of the values are concentrated near 1 which is a clear indication that HMC algorithm has been implemented correctly.}
\label{fig3.6}
\end{figure}

After tuning these parameters, we monitor the quantity $e^{-\Delta H}$. As mentioned in the previous chapter, this is one of the tests to check whether the HMC has been implemented properly or not. Fig. \ref{fig3.6} shows the plot of $e^{-\Delta H}$ against number of accepted states. We can clearly see that the data points oscillate about 1 in this plot. The average value of the $e^{-\Delta H}$ comes out to be $1.10195 \pm 0.29467$, which clearly shows that the expected value, 1, lies within the error range. This is enough to show that the algorithm is working fine.

\begin{figure}[!h]
\centering
\includegraphics[width = 0.95\textwidth]{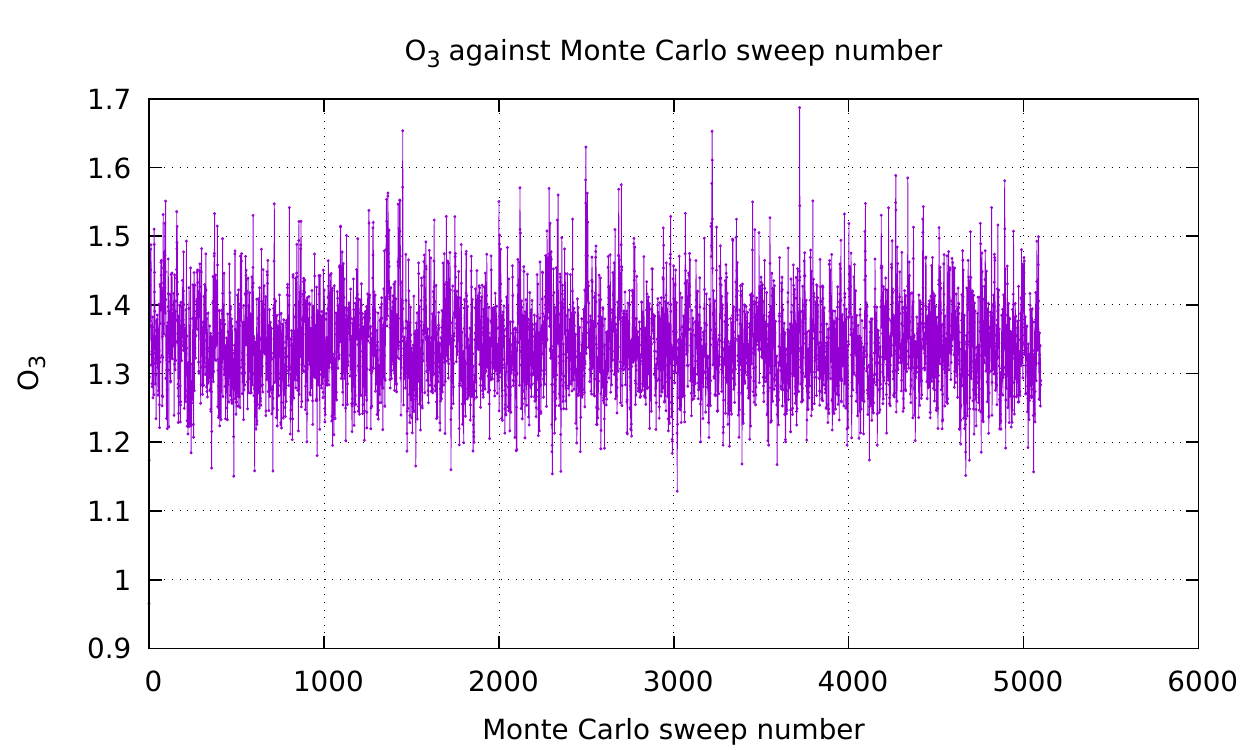}
\caption[Run-time history of observable $O_3$ against Monte Carlo sweep number.]{Run-time history of observable $O_3$ against Monte Carlo sweep number. The data tell us that it took only about 10-12 sweeps for the system to thermalize.}
\label{fig3.7}
\end{figure}

Since this is also a Monte Carlo simulation and uses Markov chain to converge to the desired distribution so we will expect that the value of the observable first start from some value and after some sweeps it will thermalize and then oscillate about some value. The feature appears here also but the system thermalizes much faster for this algorithm as compared to the Metropolis algorithm. In Fig. \ref{fig3.7} we show Monte Carlo time history of the observable $O_3$ defined as
\begin{align}
O_3 \equiv \frac{1}{N \beta} \int \limits_0^\beta dt \,\, \Tr({X^i}^2) = \frac{1}{N T} \sum \limits_{t=1}^T \Tr({X_t^i}^2) \qquad \textrm{(Discretized form)}.
\end{align}

%%%%%%
\subsection{Autocorrelation}
%%%%%%

\begin{figure}[!h]
\centering
\includegraphics[width = \textwidth]{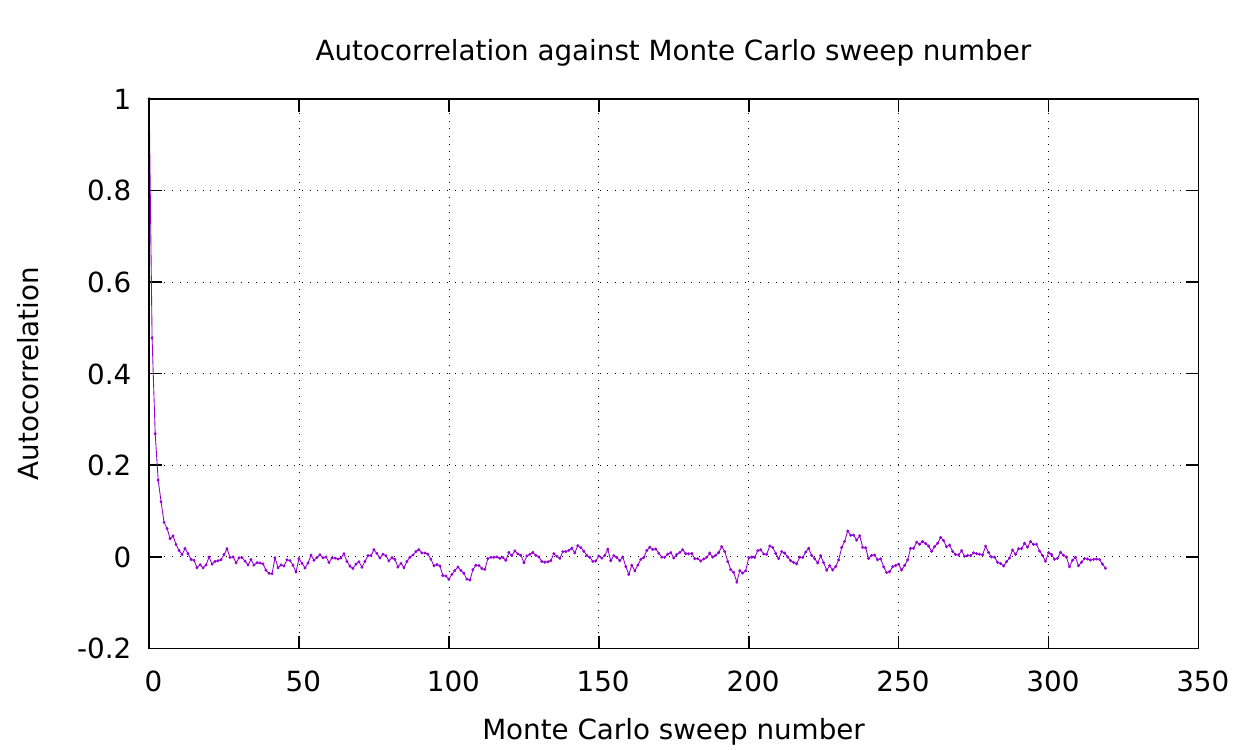}
\caption[Normalized autocorrelation against Monte Carlo sweep number.]{Normalized autocorrelation against Monte Carlo sweep number. If we compare it with Fig. \ref{fig3.3} we see that the data obtained through HMC algorithm are not that correlated compared to the data obtained through the Metropolis algorithm.}
\label{fig3.8}
\end{figure}

Looking at the plot of the run-time history of the observable $O_3$, it is clear that the data points are correlated with each other, and we need to skip a certain number of consecutive steps to get uncorrelated data. Fig. \ref{fig3.8} shows the plot of normalized autocorrelation against the number of sweeps for the observable $O_3$. Autocorrelation quickly decays and oscillate about zero after about 15-16 sweeps, which shows that the autocorrelation between the data points is quite less. Autocorrelation length of this observable comes out to be 15. 

%%%%%%
\subsection{Result}
%%%%%%

\begin{figure}[!h]
\centering
\includegraphics[width=\textwidth]{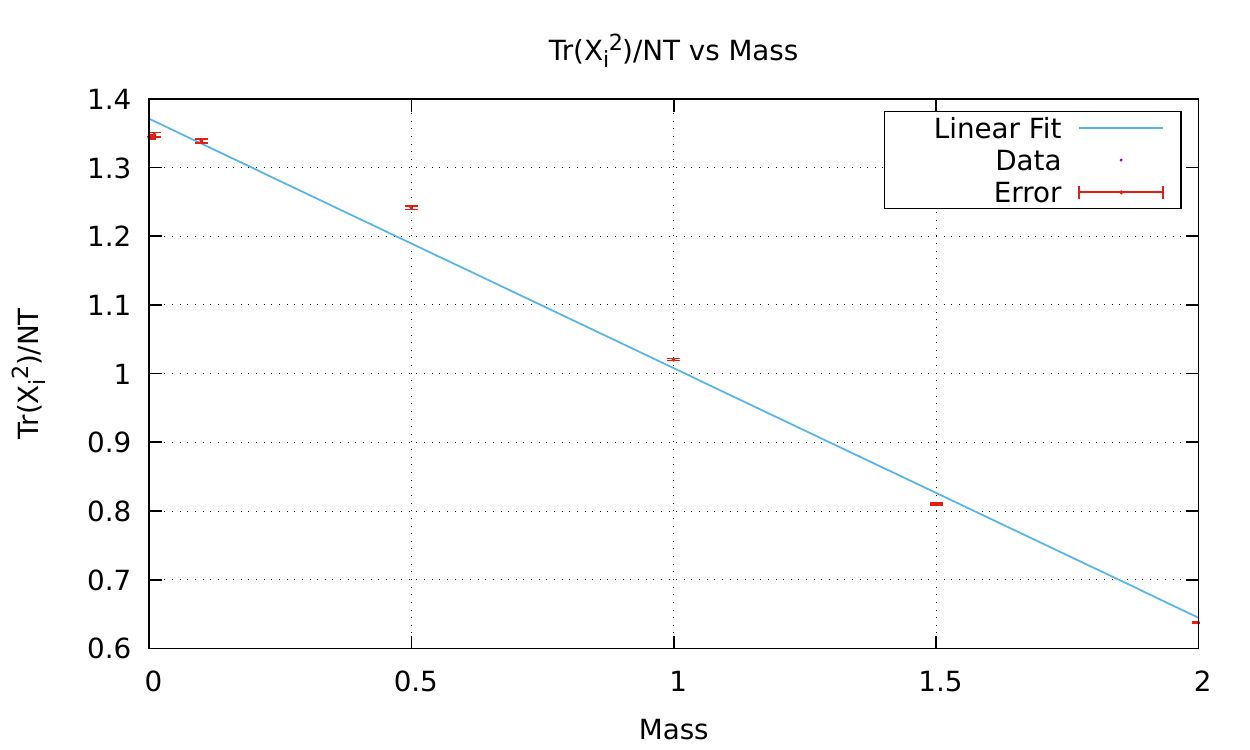}
\caption[The observable $O_3$ against $m$.]{The observable $O_3$ against $m$. The plot also contains the linear fit to the data.}
\label{fig3.9}
\end{figure}

We simulated this model by removing the flat directions in two ways. First, we simulated the model at $m = 0$ by applying the local trace condition. We calculated the value of the observable $O_3$ from this simulation. Then we simulated this model by adding different mass terms and then linearly extrapolated it towards zero mass limit. Table \ref{tab3.1} shows the value of $O_3$ for different mass $m$. Fig. \ref{fig3.9} shows the plot of $O_3$ against mass $m$. The straight line is the linear fit to the data.

\begin{table}[!t]
\centering
\begin{tabular}{ |c|c| }
  \hline
  $m$ & $O_3$ \\
  \hline \hline
  0.0 &  1.34458(303) \\
  0.01 & 1.34794(315) \\
  0.1 &  1.33892(291) \\
  0.5 &  1.24151(258) \\
  1.0 &  1.02051(145) \\
  1.5 &  0.81008(106)\\
  2.0 &  0.63752(81)\\
  \hline
\end{tabular}
\caption{The value of observable $O_3$ for different $m$ values.}
\label{tab3.1}
\end{table}

Equation of the linear fit is given by
\begin{align}
O_3 = -0.3630(151)~m + 1.3709(156).
\end{align}

From the linear fit we get the value of $O_3$ at $m = 0$ to be $1.3709(156)$, which clearly differs from the simulation at $m = 0$ case given in Table \ref{tab3.1}. This may be because the linear extrapolation is only the first-order correction and we need to know how the observable behaves for different mass values.  If we have some idea on how the observable behaves with different mass, then we can use that information for extrapolation.

%%%%%%
\section{Remarks}
%%%%%%

In this chapter, we simulated two different models, harmonic oscillator and harmonic oscillator with commutator potential, using two different algorithms. The second model is more complicated as compared to the first one. We have observed that using HMC algorithm on the second model not only made the system to thermalize much faster but also the autocorrelation between the data is quite less as compared to the simulation done using the Metropolis algorithm. Although we have not simulated the same model with both the algorithms but the second model is more complicated as compared to the first one. So from here, we conclude that HMC algorithm is more efficient as compared to the Metropolis algorithm. From the next chapter onwards, we will use the HMC algorithm for the simulations.

% Chapter4 -----------------------------------------------------------------------------------------------------------------------------
 % How to give reference to the Appendix
% Footnote hyperlink goes to page 1??

%%%%%%
\chapter{Bosonic $D = 4$ and BFSS Matrix Models}
\addcontentsline{}{×}{×}
%%%%%%

In this chapter, we are first going to discuss how to implement gauge fields on the lattice and then apply these techniques to simulate the bosonic $D = 4$ and bosonic BFSS models. Simulations including fermions will be discussed in the next chapter.

%%%%%%
\section{Gauge Theory on a Lattice}
%%%%%%

Most of the contents of this section can be found in Ref. \cite{gattringer2009quantum}. They have performed these calculations for lattice QCD but we will do it for more general gauge group $SU(N)$.

In this section we try to describe how to implement gauge fields on a lattice by using an example of Yang-Mills theory coupled to a complex scalar field in four spacetime dimensions. We take the gauge group of the gauge field to be $SU(N)$. The scalar field transforms in the fundamental representation of this gauge group. The metric of the four-dimensional spacetime is $\eta_{\mu \nu} = {\rm diag} (1,-1,-1,-1)$. 

The action of this theory is given by
\begin{equation}
S = \int d^4x \left \lbrace -\frac{1}{2 g^2} \Tr (F_{\mu \nu} F^{\mu \nu}) + {(D_{\mu} \Phi)}^{\dagger} D^\mu \Phi - m^2 \Phi^{\dagger} \Phi \right \rbrace, 
\end{equation}
where
\begin{eqnarray}
D_\mu &=& \partial_\mu + i A_\mu, \\
F_{\mu \nu} &=& \partial_{[\mu} A_{\nu]} + i [A_{\mu}, A_{\nu}].
\end{eqnarray}

The path integral for this model is given by
\begin{equation}
Z = \int \mathcal{D} A_\mu \,\, \mathcal{D} \Phi \,\, \exp{i S}.
\end{equation}

After performing the Wick rotation we will get the Euclidean action for this model. The details of this is given in Appendix \ref{appen1}. The metric after Wick rotation becomes $\eta_{\mu \nu} = {\rm diag} (1,1,1,1)$. The Euclidean action is given by
\begin{align}
S_E = \int d^4x_E \,\, \left \lbrace \frac{1}{2 g^2} \Tr(F_{\mu \nu}^{2})_E + {(D_{\mu} \Phi)}^{\dagger} D_{\mu} \Phi + m^2 {\Phi}^{\dagger} \Phi \right \rbrace, 
\label{eq4.5}
\end{align}
where
\begin{eqnarray}
{F_{\mu \nu}}_E &=& \partial_{[\mu}A_{\nu]} + i [A_{\mu}, A_{\nu}], \\
D_\mu &=& \partial_\mu + i A_\mu.
\end{eqnarray}

While working in Euclidean spacetime we note that we do not need to change the sign when we go from lower index to upper index and vice-versa. Let us represent all tensor quantities using only the lower indices. Another item that has changed is that now both $\mu$ and $\nu$ will run from 1 to 4 instead of 0 to 3. As both the fields are bosonic type they will satisfy periodic boundary conditions in Euclidean time $t_E$, i.e., 
\begin{eqnarray}
A(\vec{x}, t_E + \beta) &=& A(\vec{x}, t_E), \\
\Phi(\vec{x}, t_E + \beta) &=& \Phi(\vec{x}, t_E).
\end{eqnarray}
From now on we will suppress the subscript $E$ on the expressions and we will get to know the distinction between the Euclidean and Minkowski terms from the context. 

Now we can discretize the above action to put the theory on the lattice. We cannot discretize this theory using a naive lattice discretization as this theory is locally gauge invariant under the gauge transformation
\begin{eqnarray}
\Phi(x) &\longrightarrow& V(x) \Phi(x), \\
A_\mu(x) &\longrightarrow& V(x) A_\mu (x) V^\dagger(x) + i \partial_\mu V(x) V^\dagger(x).
\end{eqnarray}
In the above equations we have $V(x) = \exp(-i \Lambda(x))$ with $\Lambda(x)$ belonging to the Lie algebra $\mathfrak{su}(N)$, $x \in \mathbb{R}^4$ and $V^{\dagger}(x) = V^{-1}(x)$.

Our aim is to make the lattice action locally gauge invariant such that in the limit of lattice spacing $a$ going to zero we get back the action of the continuum theory. To do this let us first forget about the gauge field for now and focus on the free scalar field whose Euclidean action is given by
\begin{equation}
S = \int d^4 x \left \lbrace \partial_\mu \Phi^\dagger \partial_\mu \Phi + m^2 \Phi^\dagger \Phi \right \rbrace.
\end{equation}

As no gauge field is present we can simply discretize the action to get the lattice action
\begin{align}
S_{\rm lat} = a^4 \sum \limits_{n \in \Lambda} \left[  \sum \limits_{\mu = 1}^4 \frac{ \left( \Phi_{n + \hat{\mu}}^\dagger - \Phi_n^\dagger \right) \left( \Phi_{n + \hat{\mu}} - \Phi_n \right)}{a^2} + m^2 \Phi_n^\dagger \Phi_n \right], 
\label{eq4.11}
\end{align}
where $\Lambda = \left \lbrace n = (n_1, n_2, n_3, n_4) | \,\, n_1, n_2, n_3 = 1, 2, \cdots, N_s; \, \, n_4 = 1, 2, \cdots , T \,\, {\rm and} \,\, x = n a \right \rbrace$, $\hat{\mu}$ is the basis vector on the lattice; $\hat{1} = (1, 0, 0, 0)$ and similarly for $\hat{2},\,\hat{3}$ and $\hat{4}$. We also apply periodic boundary conditions in all the four directions. It is easy to see that the term $ \Phi_n^\dagger \Phi_n$ is gauge invariant as both of these belong to the same lattice point and the gauge transformation cancels each other leaving this term invariant. Now we need to check whether the term $\Phi_{n + \hat{\mu}}^{\dagger} \Phi_n$ remains invariant or not. We do the required calculation by transforming the field at both of these lattice points by appropriate transformations
\begin{eqnarray}
\Phi_n &\longrightarrow& V(n) \Phi_n, \\
\Phi_{n + \hat{\mu}}^{\dagger} &\longrightarrow& \Phi_{n + \hat{\mu}}^{\dagger} V^{\dagger}(n + \hat{\mu}), \\
\Phi_{n + \hat{\mu}}^{\dagger} \Phi_{n} &\longrightarrow& \Phi_{n + \hat{\mu}}^{\dagger} V^{\dagger}(n + \hat{\mu}) V(n) \Phi_{n},
\end{eqnarray}
where both $V^{\dagger}(n + \hat{\mu})$ and $V(n)$ belong to $SU(N)$.  We note that $V(n + \hat{\mu})$ and $V(n)$ can be different in general. Thus the above term is not locally gauge invariant and this is the only term (and a complex conjugate term also) in the action Eq. \eqref{eq4.11} which is not locally gauge invariant. We need to add something in between $\Phi_{n + \hat{\mu}}^{\dagger}$ and $\Phi_{n}$ which cancels these gauge transformations. We know one such object which transforms in the opposite direction to that of the scalar field. It is the path ordering of the exponential of the line integral of gauge field along some curve $\gamma$ connecting points from $x$ to $y$. It is called the {\it Wilson line} integral or the {\it gauge transporter}. The expression of the operator and the transformation properties are given by
\begin{eqnarray}
U(x, y) &=& P \left\lbrace \exp \left[ i \int \limits_{\gamma} dz_{\mu}  \, A_{\mu}(z) \right] \right \rbrace, \\
U(x, y) &\longrightarrow& V(x) U(x, y) V^{\dagger}(y),
\end{eqnarray}
where $P$ stands for path ordering. If we have an analog of this for the lattice then we can use that to make $\Phi_{n + \hat{\mu}}^{\dagger} \Phi_{n}$ locally gauge invariant. If we replace $x$ with $n$ and $y$ with $n + \hat{\mu}$ and approximate the line integral with $i a A_{\mu}(n)$ then to the lowest order of approximation we do not need any path ordering and only higher order terms will have path ordering terms. Thus we get
\begin{eqnarray}
U(n, n + \hat{\mu}) &=& \exp{i a A_{\mu}(n)} = U_{\mu}(n) \label{eq4.16}, \\
U_{\mu}(n) &\longrightarrow& V(n) U_\mu(n) V^{\dagger}(n+ \hat{\mu}).
\end{eqnarray}

This operator $U_{\mu}(n)$ is known as the {\it link field}. Also note that $U_{-\mu}(n) = U(n, n - \hat{\mu}) = \exp{-i a A_{\mu}(n)} = U_{\mu}^{\dagger}(n - \hat{\mu})$. We note that $U_{\mu}(n)$ has a directional nature, i.e., $U_{\mu}(n)$ connects $n$ to $n + \hat{\mu}$ and $U_{-\mu}(n)$ connects $n$ to $n - \hat{\mu}$. So if we add $U(n + \hat{\mu}, n) = U_{-\mu}(n + \hat{\mu}) = U_{\mu}^{\dagger}(n)$ in between $\Phi_{n + \hat{\mu}}^{\dagger}$ and $\Phi_{n}$ we get $\Phi_{n + \hat{\mu}}^{\dagger} U_{\mu}^{\dagger}(n) \Phi_n$. We can easily check that it is locally gauge invariant. The calculations are provided below
\begin{equation}
\Phi_{n + \hat{\mu}}^{\dagger} U_{\mu}^{\dagger}(n) \Phi_n \longrightarrow \Phi_{n + \hat{\mu}}^{\dagger} V^{\dagger}(n + \hat{\mu}) V(n + \hat{\mu}) U_{\mu}^{\dagger}(n) V^{\dagger}(n) V(n) \Phi_{n} = \Phi_{n + \hat{\mu}}^{\dagger} U_{\mu}^{\dagger}(n) \Phi_n.
\end{equation}

Now we are left with our last task to check whether in the limit $a \longrightarrow 0$ we get the correct continuum action or not. So let us insert the required link fields in the action Eq. \eqref{eq4.11} and take the limit $a \longrightarrow 0$. This is done by first replacing $\Phi_{n + \hat{\mu}}^{\dagger}$ with $\Phi_{n + \hat{\mu}}^{\dagger} U_{\mu}^{\dagger}(n)$. Then the action becomes
\begin{equation}
S_{\rm sc} = a^4 \sum \limits_{n \in \Lambda} \left[ \sum \limits_{\mu = 1}^{4} \frac{\Big( \Phi_{n + \hat{\mu}}^{\dagger} U_{\mu}^{\dagger}(n) - \Phi_{n}^{\dagger} \Big) \Big( U_{\mu}(n) \Phi_{n + \hat{\mu}} - \Phi_{n} \Big)}{a^2} + m^2 \Phi_{n}^{\dagger} \Phi_{n} \right]. \label{eq4.19}
\end{equation}

From Eq. \eqref{eq4.16} it is clear that $U_{\mu}(n) = \mathbb{1} + i a A_{\mu}(n) + \mathcal{O}(a^2)$ and $U_{\mu}^{\dagger}(n) = \mathbb{1} - i a A_{\mu}(n) + \mathcal{O}(a^2)$ as $A_{\mu}(n) \in \mathfrak{su}(N)$. Substituting this in Eq. \eqref{eq4.19} and collecting the terms we get
\begin{eqnarray}
S_{\rm sc} &=& a^4 \sum \limits_{n \in \Lambda} \Bigg[  \sum \limits_{\mu = 1}^{4} \left( \frac{ \Phi_{n + \hat{\mu}}^{\dagger} \left\lbrace \mathbb{1} - i a A_{\mu}(n) + \mathcal{O}(a^2) \right\rbrace - \Phi_{n}^{\dagger}}{a} \right) \nonumber \\
&& ~~~~~~~~ \times \left( \frac{ \left\lbrace \mathbb{1} + i a A_{\mu}(n) + \mathcal{O}(a^2) \right\rbrace \Phi_{n + \hat{\mu}} - \Phi_{n}}{a} \right) + m^2 \Phi_{n}^{\dagger} \Phi_{n} \Bigg] \nonumber \\
&=& a^4 \sum \limits_{n \in \Lambda} \Bigg[ \sum \limits_{\mu = 1}^{4} \left( \frac{ \Phi_{n + \hat{\mu}}^{\dagger} - \Phi_{n}^{\dagger} }{a} - i \Phi_{n + \hat{\mu}}^{\dagger} A_{\mu}(n) \right) \left( \frac{\Phi_{n + \hat{\mu}} - \Phi_{n}}{a} + i A_{\mu}(n) \Phi_{n + \hat{\mu}} \right) \nonumber \\
&& ~~~~~~~~ + m^2 \Phi_{n}^{\dagger} \Phi_{n} + \mathcal{O}(a) \Bigg].
\end{eqnarray}

Simplifying further we get
\begin{eqnarray}
S &=& \lim_{a \rightarrow \, 0} a^4 \sum \limits_{n \in \Lambda} \Bigg[ \sum \limits_{\mu = 1}^{4} \left( \frac{ \Phi_{n + \hat{\mu}}^{\dagger} - \Phi_{n}^{\dagger} }{a} - i \Phi_{n + \hat{\mu}}^{\dagger} A_{\mu}(n) \right) \nonumber \\
&& ~~~~~~~~ \times \left( \frac{\Phi_{n + \hat{\mu}} - \Phi_{n}}{a} + i A_{\mu}(n) \Phi_{n + \hat{\mu}} \right) + m^2 \Phi_{n}^{\dagger} \Phi_{n} + \mathcal{O}(a) \Bigg] \nonumber \\
&=& \int d^4 x \left\lbrace \left( \partial_{\mu} \Phi^{\dagger}(x) - i \Phi^{\dagger}(x) A_{\mu}(x) \right) \left( \partial_{\mu} \Phi(x) + i A_{\mu}(x) \Phi(x) \right) + m^2 \Phi^{\dagger}(x) \Phi(x) \right\rbrace \nonumber \\
&=& \int d^4 x \left\lbrace \left( D_{\mu} \Phi(x) \right)^{\dagger} D_{\mu} \Phi(x) + m^2 \Phi^{\dagger}(x) \Phi(x) \right\rbrace. \label{eq4.20}
\end{eqnarray}

If we compare Eq. \eqref{eq4.20} with Eq. \eqref{eq4.5} we can see that we have got all the parts of the action apart from the kinetic term for the gauge field. For the kinetic term we need to introduce one more operator, which is a combination of four link fields, known as the {\it plaquette}. It has the form
\begin{eqnarray}
U_{\mu \nu}(n) &=& U_{\mu}(n) U_{\nu}(n + \hat{\mu}) U_{- \mu}(n + \hat{\mu} + \hat{\nu}) U_{- \nu}(n + \nu) \nonumber \\
&=& U_{\mu}(n) U_{\nu}(n + \hat{\mu}) U_{\mu}^{\dagger}(n + \hat{\nu}) U_{\nu}^{\dagger}(n).
\end{eqnarray}

\begin{figure}[!h]
\centering
\includegraphics[scale=0.3]{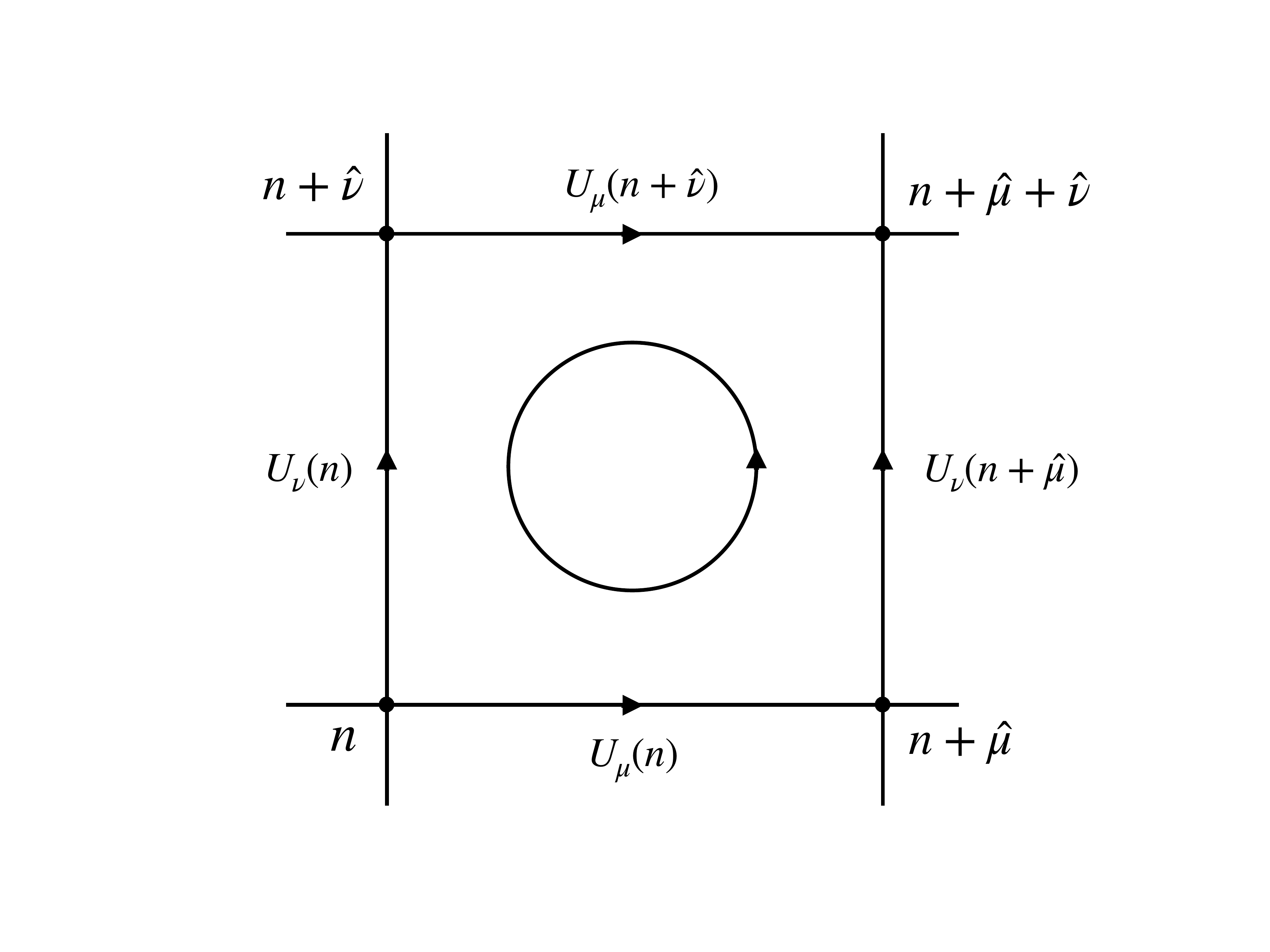}
\caption[The link variables form the plaquette $U_{\mu \nu}$.]{The link variables form the plaquette $U_{\mu \nu}$. The circle indicates the direction we need to follow to form the plaquette. This figure is taken from Ref. \cite{Joseph:2019zer}.}
\label{fig4.1}
\end{figure}

Pictorially we can understand $U_{\mu \nu}(n)$ with the help of Fig. \ref{fig4.1}. The trace of $U_{\mu \nu}(n)$ is a gauge invariant object. We can see this easily
\begin{eqnarray}
\Tr(U_{\mu \nu}(n)) &\longrightarrow& \Tr(V(n) U_{\mu}(n) U_{\nu}(n + \hat{\mu}) U_{\mu}^{\dagger}(n + \hat{\nu}) U_{\nu}^{\dagger}(n) V^{\dagger}(n) ) \nonumber \\
&\longrightarrow& \Tr(V^{\dagger}(n) V(n) U_{\mu \nu}(n)) = \Tr( U_{\mu \nu}(n) ).
\end{eqnarray}

Using the plaquette we can form the kinetic term for the gauge field. This lattice action was first introduced by K. G. Wilson \cite{wilson1974confinement} in 1974. Wilson gauge action is given by
\begin{align}
S_G = \frac{2}{g^2} \sum \limits_{n \in \Lambda} \sum \limits_{\mu < \nu} \mathfrak{Re} \left[ \Tr(\mathbb{1} - U_{\mu \nu}(n)) \right], \label{eq4.23}
\end{align}
where we need to consider all the plaquettes with only one orientation. Now our task is to show that this converges to the continuum action in the limit $a \rightarrow 0$. For that we need to use Baker-Campbell-Hausdorff formula to evaluate the value of the plaquette. Baker-Campbell-Hausdorff formula is given by
\begin{align}
\exp(X) \exp(Y) =  \exp(X + Y + \frac{1}{2}[X,Y] + \cdots).
\end{align}

Then using this 
\begin{eqnarray}
U_{\mu \nu}(n) &=& \exp(i a A_{\mu}(n)) \exp(i a A_{\nu}(n + \hat{\mu})) \exp(-i a A_{\mu}(n + \hat{\nu})) \exp(-i a A_{\nu}(n)) \nonumber \\
&=& \exp \Big\lbrace i a(A_{\mu}(n) + A_{\nu}(n + \hat{\mu})) - \frac{a^2}{2}[ A_{\mu}(n), A_{\nu}(n + \hat{\mu})] \Big\rbrace \nonumber \\
&& \times \exp \Big\lbrace -i a(A_{\mu}(n + \hat{\nu}) + A_{\nu}(n)) - \frac{a^2}{2}[A_{\mu}(n + \hat{\nu}), A_{\nu}(n)] \Big\rbrace.
\end{eqnarray}

Expanding the terms
\begin{eqnarray}
U_{\mu \nu}(n) &=& \exp \Bigg\lbrace i a ( A_{\nu}(n + \hat{\mu}) - A_{\nu}(n) + (A_{\mu}(n + \hat{\nu}) - A_{\mu}(n) ) ) \nonumber \\
&& + \frac{a^2}{2} ( [A_{\mu}(n), A_{\mu}(n + \hat{\nu})] \Bigg. \nonumber \\
&& \Bigg. + [A_{\nu}(n + \hat{\mu}), A_{\nu}(n)] + [A_{\mu}(n), A_{\nu}(n)] + [A_{\nu}(n + \hat{\mu}), A_{\mu}(n + \hat{\nu})] \nonumber \\
&& - [A_{\mu}(n), A_{\nu}(n + \hat{\mu})] - [A_{\mu}(n + \hat{\nu}), A_{\nu}(n)] ) \Bigg\rbrace.
\end{eqnarray}

In the limit $a \rightarrow 0$, $U_{\mu \nu}(n)$ takes the form
\begin{eqnarray}
U_{\mu \nu}(n) &=& \exp \Big\lbrace i a^2 (\partial_{\mu} A_{\nu}(n) - \partial_{\nu} A_{\mu}(n) + i [A_{\mu}(n), A_{\nu}(n)]) \Big\rbrace \nonumber \\
&=& \exp \{ i a^2 F_{\mu \nu}(n) \}.
\end{eqnarray}

Expanding $U_{\mu \nu}(n)$ up to order $a^4$ we get
\begin{align}
U_{\mu \nu}(n) = \mathbb{1} + i a^2 F_{\mu \nu}(n) - \frac{a^4}{2} F_{\mu \nu}^2(n).
\end{align}

Substituting the value of $U_{\mu \nu}(n)$ in Eq \ref{eq4.23} we get
\begin{eqnarray}
S_G &=& \frac{2}{g^2} \sum \limits_{n \in \Lambda} \sum \limits_{\mu < \nu} \mathfrak{Re} \left[ \Tr( i a^2 F_{\mu \nu}(n) - \frac{a^4}{2} F_{\mu \nu}^2(n) ) \right] \nonumber  \\
&=& \frac{1}{g^2} \sum \limits_{n \in \Lambda} \sum \limits_{\mu < \nu} \mathfrak{Re} \left[ \Tr( i a^2 F_{\mu \nu}(n) - \frac{a^4}{2} F_{\mu \nu}^2(n) - i a^2 F_{\mu \nu}(n) - \frac{a^4}{2} F_{\mu \nu}^2(n) ) \right].
\end{eqnarray}

That is,
\begin{eqnarray}
S_G &=& \frac{a^4}{g^2} \sum \limits_{n \in \Lambda} \sum \limits_{\mu < \nu} \Tr( F_{\mu \nu}^2(n) ), \nonumber \\
&=& \frac{a^4}{2 g^2} \sum \limits_{n \in \Lambda} \sum \limits_{\mu, \nu} \Tr( F_{\mu \nu}^2(n) ). \label{eq4.28} \\
\end{eqnarray}

Taking $a \to 0$ limit
\begin{eqnarray}
S &=& \lim_{a \rightarrow 0} S_G = \lim_{a \rightarrow 0} \frac{a^4}{2 g^2} \sum \limits_{n \in \Lambda} \sum \limits_{\mu, \nu} \Tr( F_{\mu \nu}^2(n) ) \nonumber \\
&=& \frac{1}{2 g^2} \int d^4 x \Tr(F_{\mu \nu}^2(x)).
\end{eqnarray}

Thus, in the limit $a \rightarrow 0$, Wilson gauge action converges to the correct continuum gauge action. The full action of the theory on the lattice is the sum of the Eq. \eqref{eq4.19} and Eq. \eqref{eq4.28}. That is,
\begin{align}
S_{\rm Full} = S_{\rm sc} + S_G.
\end{align}

Apart from the lattice action we also need to take care of the measure for the integral in the partition function. In the continuum theory we have gauge fields in the measure of the partition function but on the lattice we have link fields, which belong to the Lie group $SU(N)$, as compared to the gauge fields, which belong to the Lie algebra $\mathfrak{su}(N)$. So, for the invariance of the measure on the lattice we need to choose that measure which remains invariant under both the left and right multiplication by the group element. One such measure for evaluating integral on the group manifold $SU(N)$ is the {\it Haar measure}. More details of this measure can be found in Ref. \cite{gattringer2009quantum}. One important property of the Haar measure is 
\begin{align}
\mathcal{D}U_{\mu}(n) = \mathcal{D}(V(n) U_{\mu}(n)) = \mathcal{D}(U_{\mu}(n) V^{\dagger}(n + \hat{\mu})) = \mathcal{D} (V(n) U_{\mu}(n) V^{\dagger}(n + \hat{\mu})). \label{eq4.31}
\end{align}

%%%%%%
\section{$D=4$ Model}
%%%%%%

In this section we will discuss the bosonic $D = 4$ model. It is obtained by dimensionally reducing the $3+1$-dimensional Yang-Mills model with gauge group $SU(N)$ to $0+1$ dimensions. The details on these are provided in Appendix \ref{appen1}. The Euclidean action of the model is given by
\begin{equation}
S_E = \frac{N}{\lambda} \int \limits_{0}^{\beta} dt \Tr \left\lbrace \frac{1}{2} (D_t X^i)^2 - \frac{1}{4} [X^i, X^j]^2 \right\rbrace, \label{eq4.32} \\
\end{equation}
where $D_t X^i = \partial_t X^i - i [A(t), X^i]$ is the covariant derivative. As always repeated indices are summed over, $i, j = 1, 2, \cdots, d$, $\lambda = N g^2$ where $\lambda$ is the 't Hooft coupling and $g$ is the one-dimensional Yang-Mills coupling, $X^i$ are $N \times N$ traceless Hermitian matrices, $A(t)$ is the gauge field. We also have periodic boundary conditions for both the scalar and gauge field, i.e., $X^i(t + \beta) = X^i(t)$ and $A(t + \beta) = A(t)$. For this model, $d = 3$ since we have three scalar fields for the $D = 4$ model but we retained $d$ because BFSS model action is also given by Eq. \eqref{eq4.32} but in BFSS model we have $d = 9$, i.e., nine scalar fields. The action Eq. \eqref{eq4.32} remains invariant under the following gauge transformation
\begin{eqnarray}
X^i(t) &\longrightarrow& V(t) X^i(t) V^{\dagger}(t), \\
A(t) &\longrightarrow& V(t) \left( A(t) + i \partial_t \right) V^{\dagger}(t),
\end{eqnarray}
where $V(t) \in SU(N)$.

The partition function is given by
\begin{align}
Z = \int \mathcal{D} A \, \mathcal{D} X ~ \exp{- S_E}.
\end{align}

Now we need to put this theory on the lattice. The whole procedure is described in Ref. \cite{Filev_2016}. We will use the same arguments here also. First, note that here we only have temporal direction and so we will have a one-dimensional lattice and that lattice is given by $\Lambda = \{n | n = 0, 1, \cdots, T-1; t = na \}$ and imposing boundary conditions implies that we need to identify $0$ with $T$ and $\beta = a T$. To discretize the covariant derivate we need to introduce the link fields given by
\begin{align}
U_{n, n+1} = P \left\lbrace \exp \left[ i \int \limits_{na}^{(n + 1)a} dt A(t) \right] \right\rbrace = \exp{i a A(t)}.
\end{align}

If we discretize the ordinary derivate we get
\begin{align}
\partial_t X^i_n \longrightarrow \frac{X^i_{n+1} - X^i_{n}}{a}.
\end{align}

We need to bring the field at $t_{n+1}$ back to $t_n$ and for that we use the link field $U_{n, n+1}$. Then the discretized ordinary derivative becomes the discretized covariant derivative. It is given by
\begin{align}
\mathcal{D}_t X^i_{n} \longrightarrow \frac{U_{n, n+1}X^i_{n+1}U^{\dagger}_{n, n+1} - X^i_{n}}{a}.
\end{align}

Substituting this back in the discretized action we get
\begin{align}
S = \frac{Na}{2 \lambda} \sum_{n=0}^{T-1} \Tr \left\lbrace \left( \frac{U_{n, n+1}X^i_{n+1}U^{\dagger}_{n, n+1} - X^i_{n}}{a}\right)^2 - \frac{1}{2} [X^i_n, X^j_n] \right\rbrace.
\end{align}

Simplifying the above equation we get
\begin{align}
S = \frac{N}{\lambda} \sum \limits_{n=0}^{T-1} \Tr \left\lbrace -\frac{1}{a} U_{n, n+1}X^i_{n+1} U^{\dagger}_{n, n+1} X^i_{n} + \frac{1}{a} (X^i_n)^2 - \frac{a}{4} [X^i_n, X^j_n] \right\rbrace.
\end{align}

Now we use the $SU(N)$ symmetry to simply the action further. If we change the fields by the following set of transformations
\begin{eqnarray}
X_0^i &\longrightarrow& X_0^i, \\
X_1^i &\longrightarrow& U_{0, 1}^{\dagger} X_1^i U_{0, 1}, \nonumber \\
X_2^i &\longrightarrow& (U_{0, 1} U_{1, 2})^{\dagger} X_2^i (U_{0, 1} U_{1, 2}), \nonumber \\
\vdots \nonumber \\
X_{T-1}^i &\longrightarrow& (U_{0, 1} U_{1, 2} \cdots U_{T-2, T-1} )^{\dagger} X_{T-1}^i (U_{0, 1} U_{1, 2} \cdots U_{T-2, T-1}), \nonumber
\end{eqnarray}
and also if we define 
\begin{equation}
W = (U_{0, 1} U_{1, 2} \cdots U_{T-2, T-1} U_{T-1, 0}),
\end{equation}
we can express the action in the following way
\begin{eqnarray}
S &=& - \frac{N}{a \lambda} \Tr \left\lbrace \sum \limits_{n=0}^{T-2} X^i_{n+1} X^i_{n} + W X^i_0 W^{\dagger} X^i_{T-1}  \right\rbrace \nonumber \\
&& \hspace{2cm} + \frac{N}{\lambda} \sum \limits_{n=0}^{T-1} \Tr \left\lbrace \frac{1}{a} (X^i_n)^2 - \frac{a}{4} [X^i_n,X^j_n] \right\rbrace.
\end{eqnarray}

Upon using the result that every (finite-dimensional) Hermitian matrix can be diagonalized, we can diagonalize the $W$ matrix by using the gauge symmetry of the field $X_n^i$. Let $W$ diagonlized by $V$ matrix then 
\begin{equation}
W =  V D V^{\dagger}
\end{equation}
where 
\begin{equation}
D = {\rm diag} \{ e^{i \theta_1}, e^{i \theta_2}, \cdots, e^{\theta_N}\}.
\end{equation} 

If we transform $X_n^i \longrightarrow V X_n^i V^{\dagger}$ then the action becomes
\begin{align}
S = - \frac{N}{a \lambda} \Tr \left\lbrace \sum \limits_{n=0}^{T-2} X^i_{n+1} X^i_{n} + D X^i_0 D^{\dagger} X^i_{T-1}  \right\rbrace + \frac{N}{\lambda} \sum \limits_{n=0}^{T-1} \Tr \left\lbrace \frac{1}{a} (X^i_n)^2 - \frac{a}{4} [X^i_n,X^j_n] \right\rbrace. \label{eq4.43}
\end{align}

We will use the action Eq. \eqref{eq4.43} for the simulation purpose. Now let us take a look at the integration measure for this system. The integration measure for the scalar fields will remain the same but we will see that the integration measure for the link fields will get much simplified. The integration measure for the link fields is given by
\begin{align}
\prod \limits_{n=0}^{T-1} \mathcal{D} U_{n,n+1}  = \prod \limits_{n=1}^{T-1} \mathcal{D}U_{n, n+1} \mathcal{D} U_{0,1}.
\end{align}

Now using the property of the Haar measure, Eq. \ref{eq4.31}, and 
\begin{equation}
U_{0, 1} = W ( U_{1, 2} \cdots U_{T-2, T-1} U_{T-1, 0} )^{\dagger},
\end{equation} 
we get $\mathcal{D} U_{0, 1} = \mathcal{D}W$. If we look at Eq. \eqref{eq4.43}, the action only depends on the diagonal matrix $D$ which is the diagonalized form of $W$. So we can integrate the extra variables out and the integration measure simplifies to
\begin{eqnarray}
Z &=& \int \prod \limits_{n=0}^{T-1} \mathcal{D} U_{n,n+1} \, \mathcal{D} X \,\, e^{-S[X,D]} \nonumber \\
&=& \int \prod \limits_{n=1}^{T-1} \mathcal{D} U_{n, n+1} \mathcal{D} W \mathcal{D} X \,\, e^{-S[X, D]} \nonumber \\
&=& \left( {\rm Vol} (SU(N)) \right)^{T-1} \int \mathcal{D} W \mathcal{D} X \,\, e^{-S[X, D]} \nonumber \\
&\propto& \int \prod \limits_{i=1}^{N} d \theta_i \prod \limits_{j > k}^{N} \left| e^{i \theta_j} - e^{i \theta_k} \right|^2 \mathcal{D} X e^{-S[X, D]} \nonumber \\
&\propto& \int \prod \limits_{i=1}^{N} d \theta_i \prod \limits_{j > k}^{N} \sin^2 \left(\frac{\theta_j - \theta_k}{2}\right) \mathcal{D}X e^{-S[X, D]} \nonumber \\
&\propto& \int \prod \limits_{i=1}^{N} d \theta_i \, \mathcal{D}X e^{-S[X, D(\theta)] - S_{\rm FP}[\theta]}, \label{eq4.45}
\end{eqnarray}
where $S_{\rm FP}[\theta]$ is known as Faddeev-Popov term and it is given by
\begin{align}
S_{\rm FP}[\theta] = - 2 \sum \limits_{j < k}^{N} \ln \left| \sin \left( \frac{\theta_j - \theta_k}{2} \right) \right| = - \sum \limits_{j \ne k}^{N} \ln \left| \sin \left( \frac{\theta_j - \theta_k}{2} \right) \right|.
\end{align}
The proportionality constant in Eq. \eqref{eq4.45} does not bother us because the proportionality constant cancels out from the numerator and denominator when we calculate the values of the observables.

%%%%%%
\subsection{HMC for $D = 4$ Model}
%%%%%%

Now we will introduce the conjugate momenta for every field variables. Momentum conjugate to $X^i_n$ is $P^i_n$, where $P^i_n$ is an $N \times N$ traceless Hermitian matrix and the momentum conjugate to $\theta_k$ is $P_k$. The Hamiltonian for the molecular dynamics part of the HMC is
\begin{align}
H = \sum \limits_{n=0}^{T-1} \frac{1}{2} \Tr{{P_i^n}^2} + \sum \limits_{j=1}^{N} \frac{1}{2} P_j^2 + S[X, D(\theta)] + S_{\rm FP}[\theta]. \label{eq4.47}
\end{align}

The Hamilton's equations are given by
\begin{eqnarray}
\dot{X}_{n, rs}^i &=& \frac{\partial H}{\partial P_{n, sr}^i} = P_{n, rs}^i, \qquad {\dot{P}_{n, rs}^i} = - \frac{\partial H}{\partial X_{n, sr}^i}, \\
\dot{\theta}_{j} &=& \frac{\partial H}{\partial P_j} = P_j, \qquad \dot{P}_{j} = - \frac{\partial H}{\partial \theta_j},
\end{eqnarray}
where the dot represents time derivative with respect to the fictitious time $\tau$. These forces can be calculated using the Hamiltonian, Eq. \eqref{eq4.47}, and are given by
\begin{eqnarray}
- \frac{\partial H}{\partial X_{0, sr}^i} &=& \frac{N}{a \lambda} \left( X_1^i - 2 X_0^i + D^{\dagger} X_{T-1}^i D \right)_{rs} + \frac{N a}{\lambda} [X_0^j, [X_0^i, X_0^j]]_{rs}, \nonumber \\
- \frac{\partial H}{\partial X_{n, sr}^i} &=& \frac{N}{a \lambda} \left( X_{n+1}^i - 2 X_n^i + X_{n-1}^i \right)_{rs} + \frac{N a}{\lambda} [X_n^j, [X_n^i, X_n^j]]_{rs} ~~{\rm for}~~ n = 1, 2, \cdots, T-2 , \nonumber \\
- \frac{\partial H}{\partial X_{T-1, sr}^i} &=& \frac{N}{a \lambda} \left( D X_0^i D^{\dagger} - 2 X_{T-1}^i + X_{T-2}^i \right)_{rs} + \frac{N a}{\lambda} [X_{T-1}^j, [X_{T-1}^i, X_{T-1}^j]]_{rs}, \nonumber \\
- \frac{\partial H}{\partial \theta_j} &=& \frac{2 N}{a \lambda} \sum \limits_{k=1}^{N} \mathfrak{Re} \left( i X^i_{T-1, kj} X^i_{0, jk} e^{i(\theta_{j} - \theta_{k})} \right) + \sum \limits_{\substack{k=1\\ k \neq j}}^{N} \cot \left( \frac{\theta_{j} - \theta_k}{2} \right).
\end{eqnarray}
Since $D \in SU(N)$, all the $\theta$ variables will not be independent and is constrained by the equation 
\begin{equation}
\sum \limits_{i=1}^{N} \theta_i =0
\end{equation} 
as $\det(D)=1$.
 
%%%%%% 
\subsection{Observables}
%%%%%%
 
The observables we are interested in the  $D = 4$ model are the Polyakov loop $|P|$, the extent of space $R^2$, and the internal energy $E$. The definition and the discretized forms of these observables are given below
\begin{itemize}
\item Polyakov loop $|P|$
\begin{align}
\left\langle |P| \right\rangle \equiv \frac{1}{N} \left\langle \left| P( e^{ i \oint dt A(t)} ) \right| \right\rangle = \left\langle \left| \frac{\Tr(D)}{N} \right| \right\rangle \quad \textrm{(On lattice)}
\end{align}
\item Extent of space $R^2$
\begin{eqnarray}
\left\langle R^2 \right\rangle &\equiv& \left\langle \frac{\lambda}{N \beta} \int \limits_{0}^{\beta} dt \, \Tr({X^i}^2) \right\rangle \nonumber \\
&=& \left\langle \frac{\lambda}{N T} \sum \limits_{n=0}^{T-1} \Tr({X^i_n}^2) \right\rangle \quad \textrm{(Discretized form)}
\end{eqnarray}
\item Internal energy $E$
\begin{eqnarray}
\left\langle \frac{E}{N^2} \right\rangle &\equiv& \left\langle - \frac{3 \lambda}{4 N \beta} \int \limits_{0}^{\beta} dt \, \Tr([X^i, X^j]^2) \right\rangle \nonumber \\
&=& \left\langle - \frac{3 \lambda}{4 N T} \sum \limits_{n=0}^{T-1} \Tr( [X^i_n, X^j_n]^2 ) \right\rangle \quad \textrm{(Discretized form)}
\end{eqnarray}
\end{itemize}

%%%%%%
\subsection{Simulation Details} 
\label{subsec4.2.3}
%%%%%%

In the simulations each variable and parameter has been measured in the unit of 't Hooft coupling $\lambda_0$. Since the Yang-Mills coupling $g_{\rm YM}^2$ has a dimension of [$L^{-3}$] in one dimension, so $X^i_n$ and $A(n)$ can be measured in the unit of $\lambda_0^{1/3}$ and $a$ can be measured in units of $\lambda_0^{-1/3}$. Thus, every variable and parameter appearing in Eq. \eqref{eq4.43} are dimensionless. Value of the parameters used in the simulations are: $N = 4$, $T = 20$, $d = 3$ and $\lambda = 1$.

The simulations can be performed in two ways. On way is to fix the lattice spacing $a$ and then perform the simulation for different temperatures. However, for large temperature values lattice size becomes large and thus it takes quite a lot of time to simulate the model. Another way is to fix the lattice size $T$ and vary the lattice spacing $a$. In this method, the lattice size remains constant which ensures that it will take the same amount of simulation time for different temperatures but the lattice spacing will increases with increasing temperature so we need to choose $T$ such that $a$ remains small enough for the simulation. We performed the simulation by fixing $T = 20$ and got pretty good results.

The simulations which we are performing can be quite heavy. It takes a lot of time so we cannot risk to run it in a single go without saving the data and field configurations in between. Incidents such as power loss or system crash may happen resulting in a loss of the final field configuration forcing us to rerun the whole simulation again. In order to overcome this we have written a script in {\tt bash} to stop the code after every 100 sweeps and then save the data and field configurations and then rerun again. In this way we will not face any loss of data.

As we have already mentioned, the simulations of this model is quite heavy; so we run the simulations for each temperature on a single core of the computer. We run the simulations for all temperatures simultaneously using multiple cores at the same time. This procedure saves us quite a lot of time. (Another method for saving simulation time is to write a code suitable for parallel computing.)

We run the simulations for 17 different temperature values and for each temperature we run it for 1600K sweeps. The results are provided below. Our results are in excellent agreement with the results given in Ref. \cite{Hanada_2007}.

%%%%%%
\subsubsection{Polyakov Loop $|P|$}
%%%%%%

\begin{figure}[!h]
\centering
\includegraphics[width=12cm]{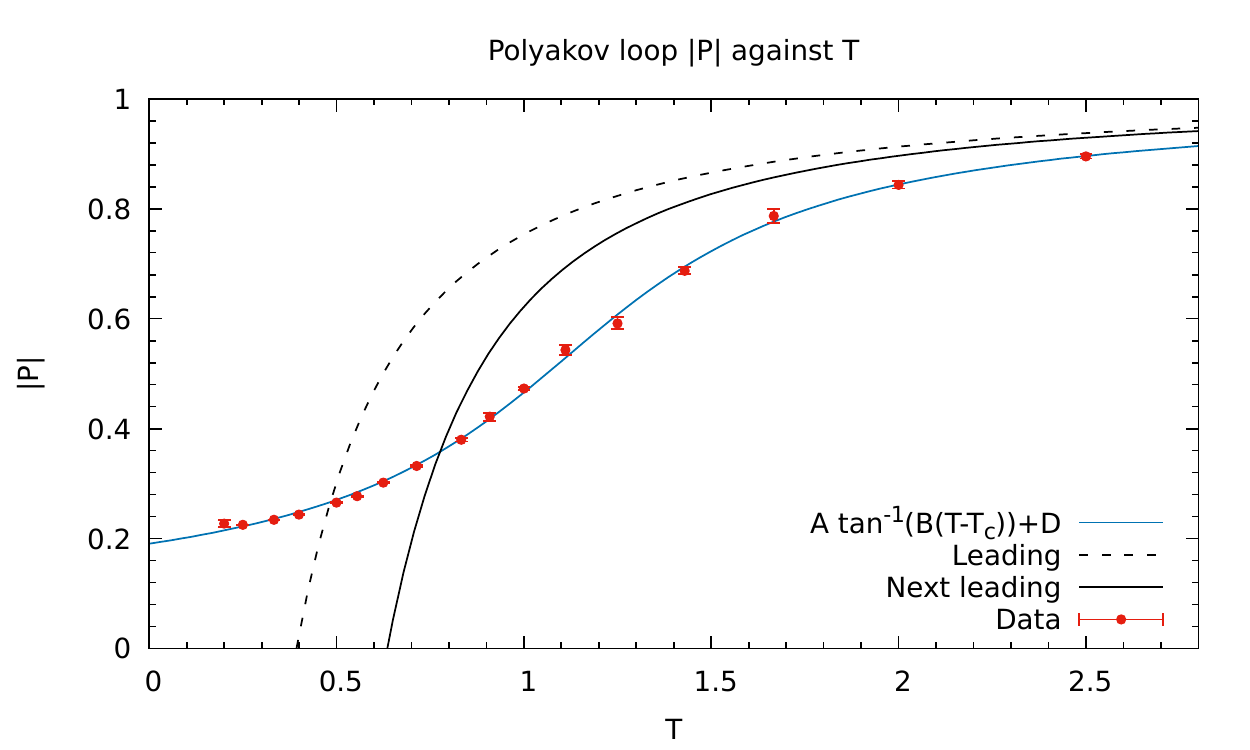}
\caption[Polyakov loop against temperature $T$.]{Polyakov loop against temperature $T$. The dotted and bold black curves show the leading order and the next-to-leading order terms of the high temperature expansion (HTE). The blue curve is the fitted curve to the plot. The phase transition occurs at $T_c \approx 1.11$.}
\label{fig4.2}
\end{figure}

We have already mentioned in the Chapter 1 that Polyakov loop $|P|$ acts as an order parameter for the confinement/deconfinement phase transition. Fig. \ref{fig4.2} shows the plot of the Polyakov loop against temperature $T$. The dotted and the bold black curves show the leading and next-to-leading terms of the high temperature expansion (HTE) given in Ref. \cite{Kawahara_2007}. The system remains in the confined phase for small temperature and after crossing the critical temperature $T_c$, the system moves to the deconfined phase. In the limit $N \rightarrow \infty $ the plot will be given by the Heaviside step function with the discontinuity at the critical temperature $T_c$. So for a finite system we fit our plot of the Polyakov loop with a suitable function that in the limit should converge to the step function. Our fitting function is given by
\begin{align}
f(T) = A \tan^{-1}(B(T- T_c)) + D,
\end{align}
where $A, B, T_c$ and $D$ are the fit parameters. If we choose $A = \dfrac{1}{\pi}$, $D = 0.5$ and take the limit $B \rightarrow \infty$ then
\begin{align}
\lim_{B \rightarrow \infty} \frac{1}{\pi} \tan^{-1}( B (T - T_c)) + 0.5 = \Theta (T - T_c)
\end{align}
We see that it does converge to the required limit; so the function $f(T)$ is the correct choice. The values of fit parameters obtained after the fit are provided in Table \ref{tab4.1}. From this we conclude that the critical temperature $T_c = 1.116(13)$.

\begin{table}[!h]
\centering
\begin{tabular}{ |c|c| }
  \hline
  Parameter & Fit Value \\
  \hline \hline
  $A$ &  0.3000712(9052) \\
  $B$ & 1.92222(11280) \\
  $T_c$ &  1.11612(1320) \\
  $D$ &  0.53172(485) \\
  \hline
\end{tabular}
\caption{Values of the fitting parameters $A$, $B$, $T_c$ and $D$.}
\label{tab4.1}
\end{table}

\begin{figure}[!h]
\begin{subfigure}{0.5\textwidth}
	\centering
	\includegraphics[width = \textwidth , height = 1.1\textwidth ]{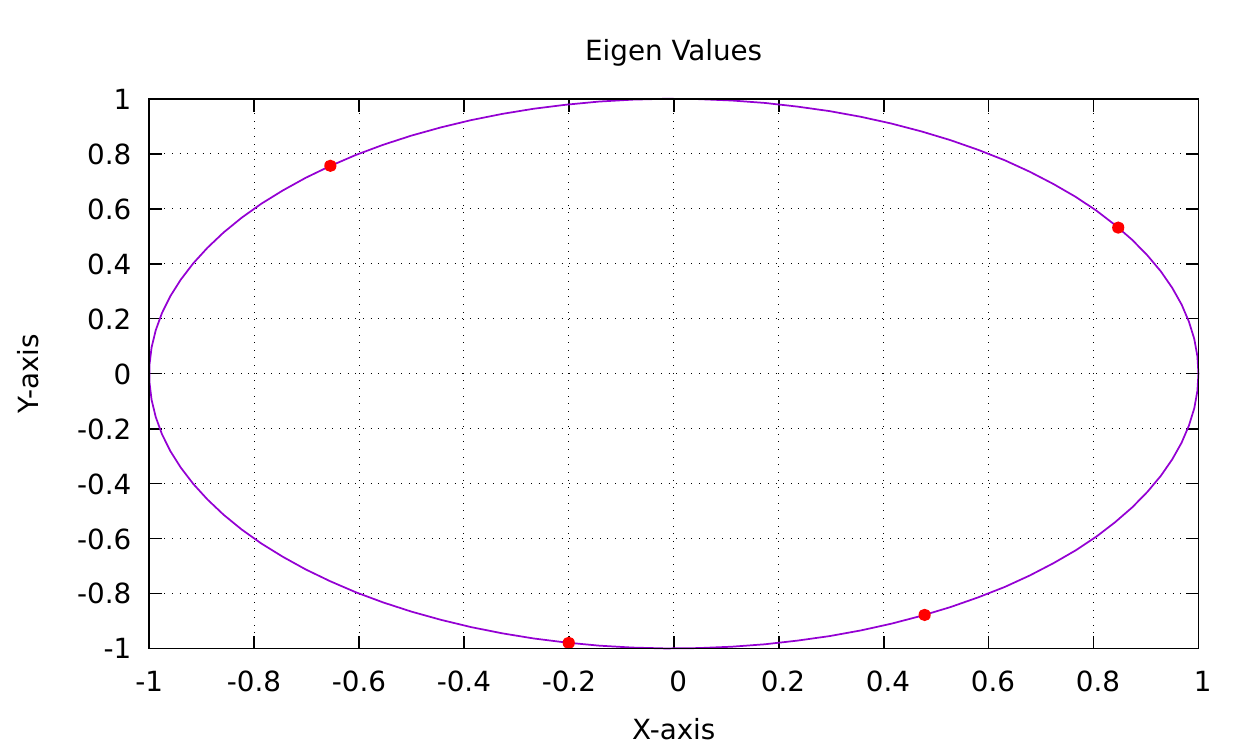}
	\caption{Confined phase $T = 0.25$.}
	\end{subfigure}%
	\hfill
	\begin{subfigure}{0.5\textwidth}
	\centering
	\includegraphics[width = \textwidth , height = 1.1\textwidth]{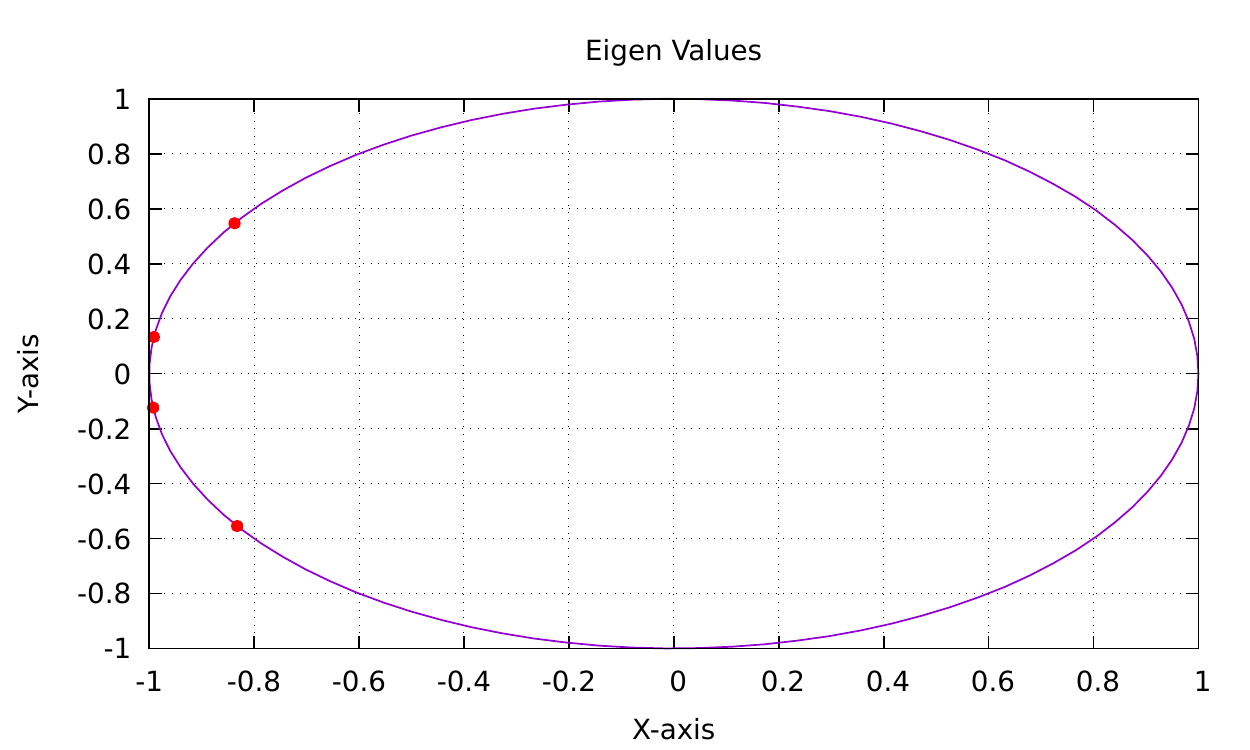}
	\caption{Deconfined phase $T = 2.5$.}
	\end{subfigure}
	
	\caption[Distribution of the eigenvalues of the Polyakov loop operator.]{Distribution of the eigenvalues of the Polyakov loop operator in the confined and deconfined phases.}
	\label{fig4.3}
\end{figure}

Fig. \ref{fig4.3} shows the distribution of the eigenvalues of the Polyakov loop operator in the confined and deconfined phase. In the confined phase, the eigenvalues spread uniformly on the unit circle, whereas they tend to cluster near one point in the deconfined phase. 

%%%%%%
\subsubsection{Internal Energy $E$ and Extend of Space $R^2$}
%%%%%%

\begin{figure}
\centering
\includegraphics[width=12cm]{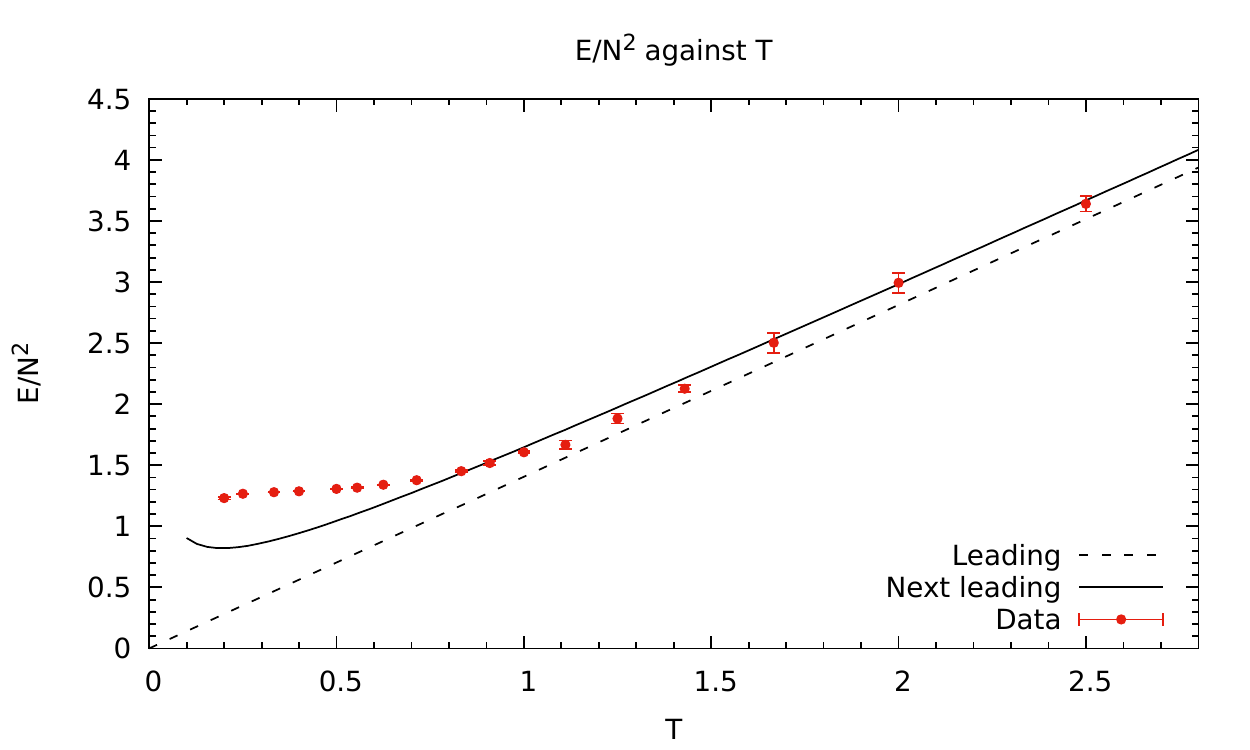}
\caption[$\frac{E}{N^2}$ against temperature $T$.]{$\frac{E}{N^2}$ against temperature $T$. The dotted and bold black curves show the leading order and the next-to-leading order terms of the high temperature expansion (HTE).}
\label{fig4.4}
\end{figure}

\begin{figure}
\centering
\includegraphics[width=12cm]{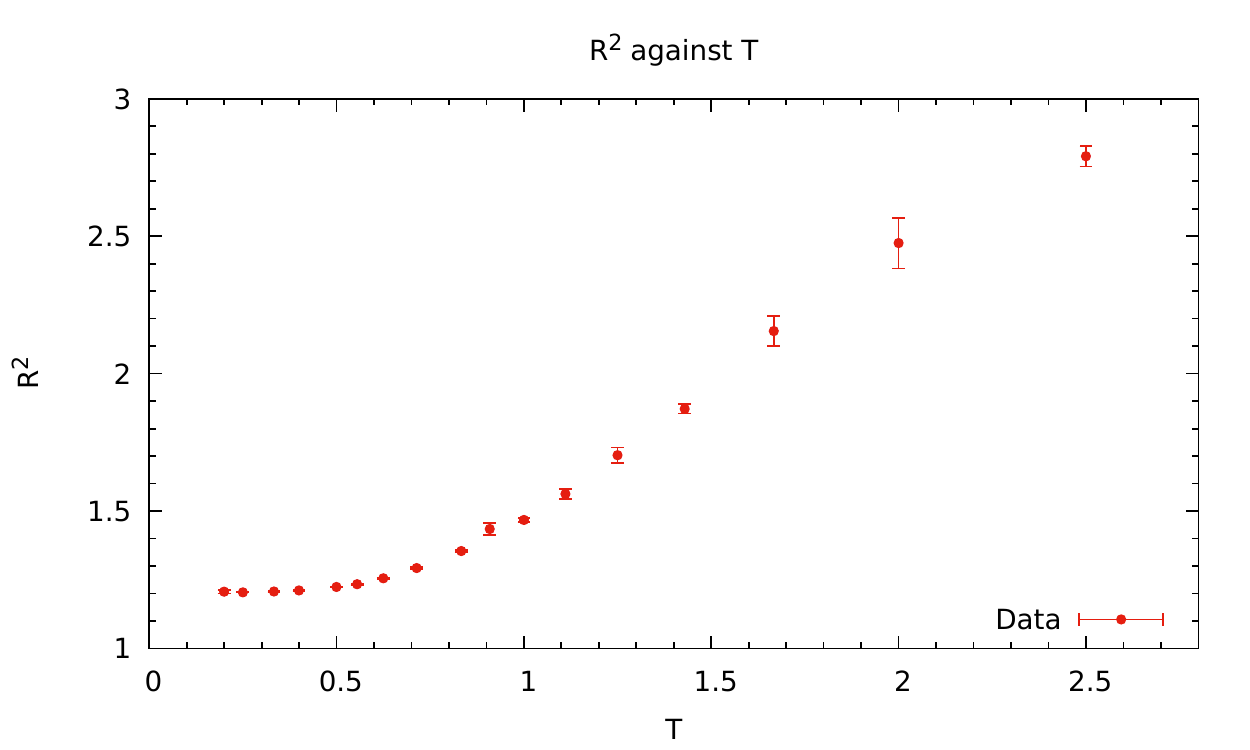}
\caption[Plot of $R^2$ against temperature $T$.]{Plot of $R^2$ against temperature $T$.}
\label{fig4.5}
\end{figure}

Fig. \ref{fig4.4} shows the plot of the scaled internal energy $\frac{E}{N^2}$ against temperature $T$. If we had simulated this for larger $N$ then we would clearly see a kink near the critical temperature, which would directly show that the system undergoes a phase transition. But for smaller $N$ this feature is not visible in the plot. The plot agrees well with the high temperature expansion (HTE). In Fig. \ref{fig4.5} we show the plot of the extent of space $R^2$ against temp $T$.

%%%%%%
\section{BFSS Matrix Model}
%%%%%%

The action of this model can be obtained by dimensionally reducing the $9+1$-dimensional $\mathcal{N} = 1$ Super Yang-Mills theory to $0+1$ dimensions. The details of this model are given in Appendix \ref{appen1}. The Euclidean action of the BFSS model is given by
\begin{align}
S_E = \frac{N}{2 \lambda} \int \limits_{0}^{\beta} dt \, \Tr \left\lbrace (D_t X^i)^2 - \frac{1}{2} [X^i, X^j]^2 + \psi^{T} C_9 D_t \psi - \psi^T C_9 \gamma^i [X^i, \psi] \right \rbrace,
\end{align}
where $D_t = \partial_t - i [A(t), \cdot]$; $i, j = 1, 2, \cdots, 9$; $X^i$ are $N \times N$ traceless Hermitian matrices; $\lambda$ is the 't Hooft coupling; $\psi$ is a sixteen component Majorana fermion and each component of the fermion is an $N \times N$ traceless Hermitian matrix; $C_9$ is the Euclidean charge conjugation matrix in the nine dimensions; and $\gamma^i$ are the Euclidean gamma matrices in nine dimensions. Fermionic part of the action is obtained by taking a particular choice for the Majorana-Weyl fermion and the gamma matrices. The details are given in Ref. \cite{Filev_2016}.

In this section we will simulate the quenched BFSS model, i.e., we will remove the fermions from the theory and we just simulate the bosonic part of this model. The Euclidean action of the bosonic part is given by
\begin{align}
S_E = \frac{N}{2 \lambda} \int \limits_{0}^{\beta} dt \, \Tr \left\lbrace (D_t X^i)^2 - \frac{1}{2} [X^i, X^j]^2 \right\rbrace.
\end{align}

We see that the bosonic action is same as of the $D = 4$ model but here we have nine scalar fields instead of three scalar fields. So the set of procedures from discretizing the action to applying the HMC remains the same. However, we discuss the important points below.

%%%%%%
\subsection{HMC for Bosonic BFSS Matrix Model}
%%%%%%

We introduce the momentum conjugate to $X^i_n$ to be $P^i_n$, where $P^i_n$ is an $N \times N$ traceless Hermitian matrix and momentum conjugate to $\theta_k$ is $P_k$. The Hamiltonian for the molecular dynamics part of HMC is
\begin{align}
H = \sum \limits_{n=0}^{T-1} \frac{1}{2} \Tr{{P_i^n}^2} + \sum \limits_{j=1}^{N} \frac{1}{2} P_j^2 + S[X, D(\theta)] + S_{FP}[\theta]. \label{eq4.58}
\end{align}

The Hamilton's equations are given by
\begin{eqnarray}
\dot{X}_{n, rs}^i &=& \frac{\partial H}{\partial P_{n, sr}^i} = P_{n, rs}^i, \qquad {\dot{P}_{n, rs}^i} = - \frac{\partial H}{\partial X_{n, sr}^i}, \\
\dot{\theta}_{j} &=& \frac{\partial H}{\partial P_j} = P_j, \qquad \dot{P}_{j} = - \frac{\partial H}{\partial \theta_j},
\end{eqnarray}
where the dot represents time derivative with respect to the fictitious time $\tau$. 

The forces can be calculated using the Hamiltonian, Eq. \eqref{eq4.58}, and are given by
\begin{eqnarray}
- \frac{\partial H}{\partial X_{0, sr}^i} &=& \frac{N}{a \lambda} \left( X_1^i - 2 X_0^i + D^{\dagger} X_{T-1}^i D \right)_{rs} + \frac{N a}{\lambda} [X_0^j, [X_0^i, X_0^j]]_{rs}, \nonumber \\
- \frac{\partial H}{\partial X_{n, sr}^i} &=& \frac{N}{a \lambda} \left( X_{n+1}^i - 2 X_n^i + X_{n-1}^i \right)_{rs} + \frac{N a}{\lambda} [X_n^j, [X_n^i, X_n^j]]_{rs}~~\textrm{for}~~n = 1, 2, \cdots, T-2 , \nonumber \\
- \frac{\partial H}{\partial X_{T-1, sr}^i} &=& \frac{N}{a \lambda} \left( D X_0^i D^{\dagger} - 2 X_{T-1}^i + X_{T-2}^i \right)_{rs} + \frac{N a}{\lambda} [X_{T-1}^j, [X_{T-1}^i, X_{T-1}^j]]_{rs}, \nonumber \\
- \frac{\partial H}{\partial \theta_j} &=& \frac{2 N}{a \lambda} \sum \limits_{k=1}^{N} \mathfrak{Re} \left( i X^i_{T-1, kj} X^i_{0, jk} e^{i(\theta_{j} - \theta_{k})} \right) + \sum \limits_{\substack{k = 1\\ k \neq j}}^{N} \cot \left( \frac{\theta_{j} - \theta_k}{2} \right).
\end{eqnarray}
Since $D \in SU(N)$, all the $\theta$ variables will not be independent and is constrained by the equation 
\begin{equation}
\sum \limits_{i=1}^{N} \theta_i =0 \nonumber
\end{equation} 
as $\det(D) = 1$.

%%%%%%
\subsection{Observables}
%%%%%%
 
The observables we are interested in the bosonic BFSS model are the Polyakov loop $|P|$, the extent of space $R^2$ and the internal energy $E$. The definitions and the discretized forms of the observables are given below
\begin{itemize}
\item Polyakov loop $|P|$
\begin{align}
\left\langle |P| \right\rangle \equiv \frac{1}{N}  \left\langle \left| P( e^{ i \oint dt A(t)} ) \right| \right\rangle = \left\langle \left| \frac{\Tr(D)}{N} \right| \right\rangle ~~~~ \textrm{(On lattice)}
\end{align}
\item Extent of space $R^2$
\begin{align}
\left\langle R^2 \right\rangle \equiv \left\langle \frac{\lambda}{N \beta} \int \limits_{0}^{\beta} dt \, \Tr({X^i}^2) \right\rangle = \left\langle \frac{\lambda}{N T} \sum \limits_{n=0}^{T-1} \Tr({X^i_n}^2) \right\rangle ~~~~ \textrm{(Discretized form)}
\end{align}
\item Internal energy $E$
\begin{eqnarray}
\left\langle \frac{E}{N^2} \right\rangle &\equiv& \left\langle - \frac{3 \lambda}{4 N \beta} \int \limits_{0}^{\beta} dt \, \Tr([X^i,X^j]^2) \right\rangle \nonumber \\
&=& \left\langle - \frac{3 \lambda}{4 N T} \sum \limits_{n=0}^{T-1} \Tr( [X^i_n, X^j_n]^2 ) \right\rangle ~~~~ \textrm{(Discretized form)}
\end{eqnarray}
\end{itemize}

%%%%%%
\subsection{Simulation Details}
%%%%%%

In simulations each variable and parameter has been measured in the unit of the 't Hooft coupling $\lambda_0$. Since the Yang-Mills coupling $g_{\rm YM}^2$ has a dimension of [$L^{-3}$] in one dimension, we can measure $X^i_n$ and $A(n)$ in the unit of $\lambda_0^{1/3}$, and $a$ can be measured in units of $\lambda_0^{-1/3}$. Thus, each variable and parameter appearing in Eq. \eqref{eq4.43} is dimensionless. The values of the parameters used for this simulations are: $N = 4$, $T = 10$, $d = 9$ and $\lambda = 1$. All the other details of the simulations are the same as the $D = 4$ model. (See Sec. \ref{subsec4.2.3} for those details.)

We performed the simulations for 31 different temperature values each running for 800K Monte Carlo sweeps. The critical temperature $T_c$ comes out to be $0.9054(20)$. Our results are in excellent agreement with those given in Refs. \cite{Kawahara_2007_2} and \cite{Filev_2016}.

%%%%%%
\subsubsection{Polyakov Loop $|P|$}
%%%%%%

\begin{figure}[!h]
\centering
\includegraphics[width = 12cm]{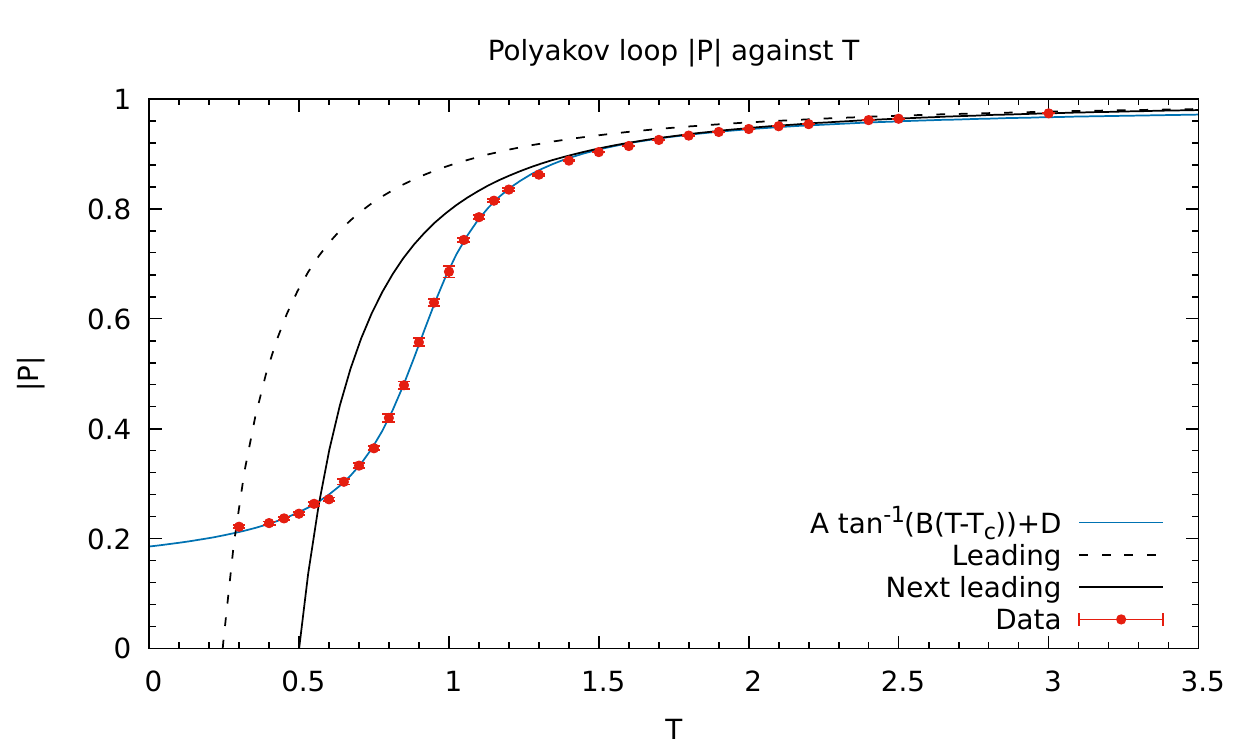}
\caption[The Polyakov loop against temperature $T$.]{The Polyakov loop against temperature $T$. The dotted and the bold black curves show the leading and the next-to-leading terms of the high temperature expansion (HTE) given in Ref.  \cite{Kawahara_2007}. The blue curve is the fitted curve to the data. The phase transition occurs at $T_c \approx 0.905 $.}
\label{fig4.6}
\end{figure}

Fig. \ref{fig4.6} shows the plot of the Polyakov loop $|P|$ against temperature $T$. The values of the parameters for the fitted curve are given in Table \ref{tab4.2}. From this we conclude that the critical temperature $T_c = 0.9054(20)$.

\begin{table}[!h]
\centering
\begin{tabular}{ |c|c| }
  \hline
  Parameter & Fit Value \\
  \hline \hline
  $A$ &  0.273982(1569) \\
  $B$ & 5.37305(8625) \\
  $T_c$ &  0.905404(2010) \\
  $D$ &  0.560378(1439) \\
  \hline
\end{tabular}
\caption{Values of the fitting parameters $A$, $B$, $T_c$ and $D$.}
\label{tab4.2}
\end{table}

%%%%%%
\subsubsection{Internal Energy $E$ and Extent of Space $R^2$}
%%%%%%

\begin{figure}
\centering
\includegraphics[width=12cm]{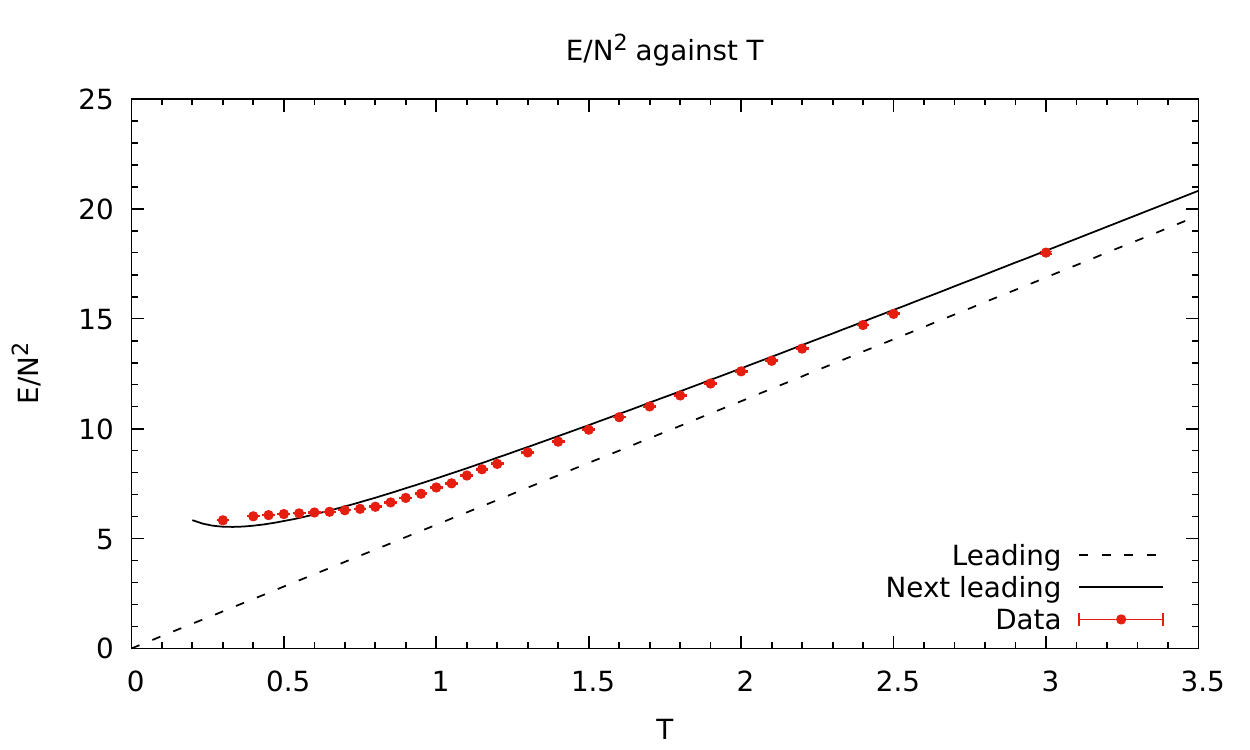}
\caption[The plot of $\frac{E}{N^2}$ against temperature $T$.]{The plot of $\frac{E}{N^2}$ against temperature $T$. The dotted and bold black curves show the leading and the next-to-leading order terms of the high temperature expansion (HTE).}
\label{fig4.7}
\end{figure}

\begin{figure}[!t]
\centering
\includegraphics[width=12cm]{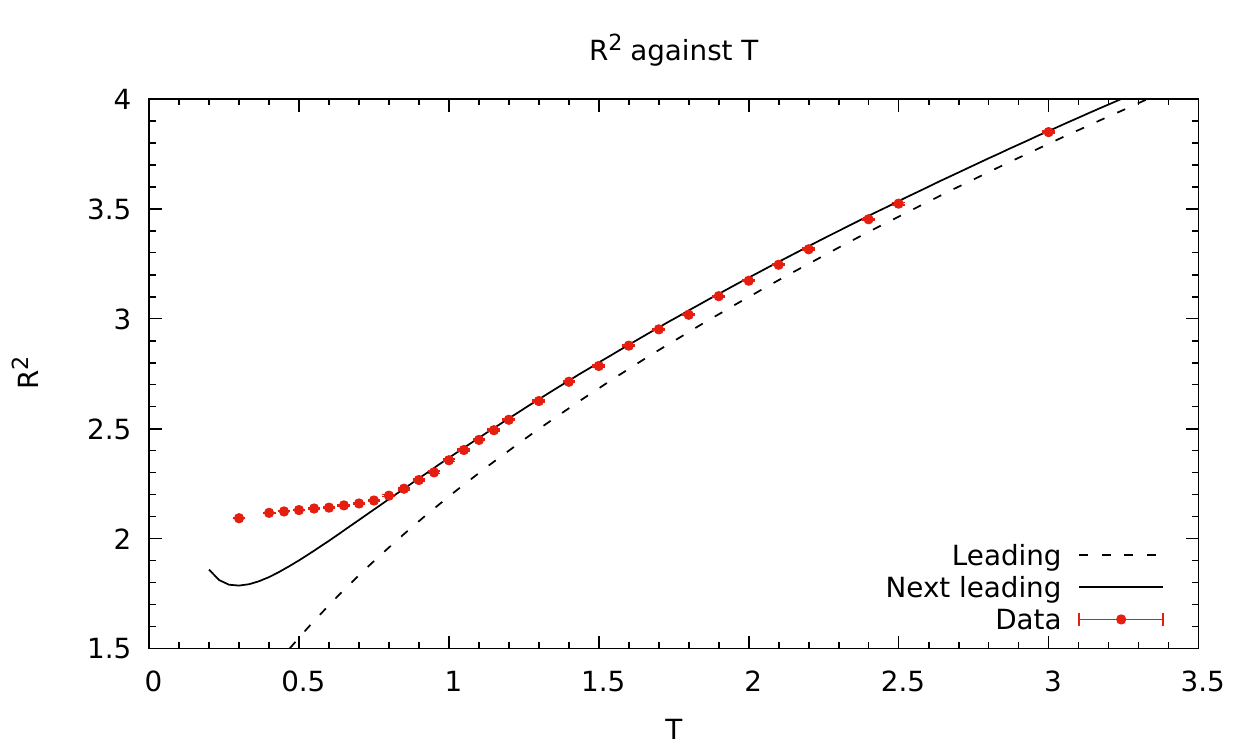}
\caption[The plot of extent of space $R^2$ against temperature $T$.]{The plot of extent of space $R^2$ against temperature $T$. The dotted and bold black curves show the leading and the next-to-leading order terms of the high temperature expansion (HTE).}
\label{fig4.8}
\end{figure}

In Fig. \ref{fig4.7} we show the plot of the normalized internal energy $\frac{E}{N^2}$ against temperature $T$. Our plot shows a smooth transition of $\frac{E}{N^2}$ across different phases but there should be a kink at $T_c$. This kink becomes visible for larger $N$, which is a clear indication of a second-order phase transition. In Ref. \cite{Filev_2016}, the authors simulated the same model for $N = 16$ and there we can clearly see a kink in the $\frac{E}{N^2}$ plot at $T_c$. In Fig. \ref{fig4.8} we show the plot of the extent of space $R^2$ against temperature $T$. We can see that all our plots are in excellent agreement with the high temperature expansion (HTE) predictions.

% Chapter5 -----------------------------------------------------------------------------------------------------------------------------
 % How to give reference to the Appendix
% Footnote hyperlink goes to page 1??

%%%%%%
\chapter{IKKT Matrix Model}\addcontentsline{}{×}{×}
%%%%%%

In this chapter we discuss the numerical simulations of the bosonic IKKT Model. We also discuss the simulations of the model after the inclusion of fermions. We also discuss the simulations of the full IKKT Model.

%%%%%% 
\section{Bosonic IKKT Matrix Model}
%%%%%%

This model is obtained by dimensionally reducing the Euclidean Yang-Mills action with gauge group $SU(N)$ from $9+1$ dimensions to $0+0$ dimensions. The Euclidean action of the model is given by
\begin{align}
S_{\rm E} = - \frac{1}{4 g^2} \Tr ([X^i, X^j]^2),
\end{align}
where $X^i$ are $N \times N$ traceless Hermitian matrices, $i, j$ varies from $1, 2, \cdots, 10$ and $g$ is the Yang-Mills coupling in $0+0$ dimensions. We can replace $g$ with $\lambda$ the 't Hooft coupling, which follows the relation $\lambda = N g^2$. The action then takes the form 
\begin{align}
S_{\rm E} = - \frac{N}{4 \lambda} \Tr ([X^i, X^j]^2).
\end{align}

Now by rescaling the fields we can absorb $\lambda$ into the field variables. If we rescale the fields by $X^i \rightarrow \lambda^{1/4} X^i$, then $\lambda$ will be absorbed in the field variables and the action simplifies to
\begin{align}
S_{\rm E} = - \frac{N}{4} \Tr ([X^i, X^j]^2). 
\label{eq5.3}
\end{align}

In the above action all the field variables are dimensionless. The action is gauge invariant under the following gauge transformation
\begin{align}
X^i \longrightarrow V X^i V^\dagger,
\end{align}
where $V \in SU(N)$. The partition function of this model is given by
\begin{align}
Z = \int \mathcal{D}X \,\, \exp{- S_{\rm E}}.
\end{align}

Apart from this gauge symmetry, the action Eq. \eqref{eq5.3} also has $SO(10)$ symmetry. That is, if we change $X^i$ by the transformation
\begin{align}
X^i \longrightarrow O_{i k} X^k,
\end{align}
where $O \in SO(10)$, the action remains invariant. Changing the fields with this transformation, the action becomes
\begin{align}
S_{\rm E} &= - \frac{N}{4} \Tr ([O_{i k} X^k, O_{j l} X^l] [O_{i m} X^m, O_{j n} X^n]) \nonumber \\
&= - \frac{N}{4} O_{i k} O_{i m} O_{j l} O_{j n} \Tr ([X^k, X^l] [X^m, X^n]) \nonumber \\
&= - \frac{N}{4} \delta_{k m} \delta_{l n} \Tr ([X^k, X^l] [X^m, X^n]) \qquad \textrm{ using } O_{i k} O_{i m} = \delta_{k m} \nonumber \\
&= - \frac{N}{4} \Tr ([X^k,X^l]^2).
\end{align}

Thus the action remains invariant under this transformation telling us that $SO(10)$ is a symmetry of this system. Putting this theory on the lattice is a trivial task as this model is $0$-dimensional, and thus everything is happening on a single lattice site. The action Eq. \eqref{eq5.3} itself represents the action of the continuum theory as well as of the lattice theory.

%%%%%%
\subsection{HMC for Bosonic IKKT Matrix Model}
%%%%%%

We will now introduce conjugate momenta for each of the field variable. The momenta conjugate to $X^i$ is $P^i$, where $P^i$ is an $N \times N$ traceless Hermitian matrix. The Hamiltonian for the molecular dynamics part of the HMC is given by
\begin{align}
H = \frac{1}{2} \sum \limits_{i=1}^{10} \Tr({P^i}^2) - \frac{N}{4} \sum \limits_{i, j = 1}^{10} \Tr ([X^i, X^j]^2). 
\label{eq5.8}
\end{align}

Hamilton's equations are given by
\begin{align}
\dot{X}_{rs}^i = \frac{\partial H}{\partial P_{sr}^i} = P_{rs}^i, \qquad {\dot{P}_{rs}^i} = - \frac{\partial H}{\partial X_{sr}^i},
\end{align}
where the dot represents time derivative with respect to the fictitious time $\tau$. 

The forces can be calculated using the Hamiltonian, Eq. \eqref{eq5.8}, and are given by
\begin{align}
- \frac{\partial H}{\partial X_{sr}^i} = N \sum \limits_{\substack{j=1 \\ j \neq i}}^{10} [X^j, [X^i, X^j]]_{rs}.
\end{align}

%%%%%%
\subsection{Observables}
\label{sec5.1.2}
%%%%%%

The observables we are interested in the IKKT model are the {\it extent of spacetime} $R^2$ and eigenvalues of the {\it moment of inertia tensor} $I_{\mu \nu}$. The definitions of the observables are given below

\begin{itemize}
\item Extent of spacetime $R^2$
\begin{align}
\left\langle R^2 \right\rangle = \left\langle \frac{1}{N} \Tr({X^i}^2) \right\rangle.
\end{align}
\item Eigenvalues of moment of inertia tensor $I_{\mu \nu}$
\begin{align}
I_{\mu \nu} = \frac{1}{N} \Tr(X^{\mu} X^{\nu}).
\end{align}
\end{itemize}
Eigenvalues of the moment of inertia tensor are $\lambda_i$, $i = 1, 2, \cdots, 10$.

%%%%%%
\subsection{Simulation Details and Results}
%%%%%%

We simulated this model for different values of $N$. Simulating this model is quite easy compared to the BFSS model. Again to save time, we run the simulation for each $N$ on a single core of the computer. This way, we were able to run the simulations for all the values of $N$ simultaneously using multiple cores at the same time. Here we can see that the simulation depends on $N$ itself, so it will take different time for each $N$.

We performed the simulations for 10 different values of $N$ each running for 110K Monte Carlo sweeps. The results are provided below. Basically, our aim is to study how the system behaves at large $N$.

%%%%%%
\subsubsection{Eigenvalues of $I_{\mu \nu}$}
%%%%%%

\begin{figure}[!h]
\centering
\includegraphics[width = \textwidth]{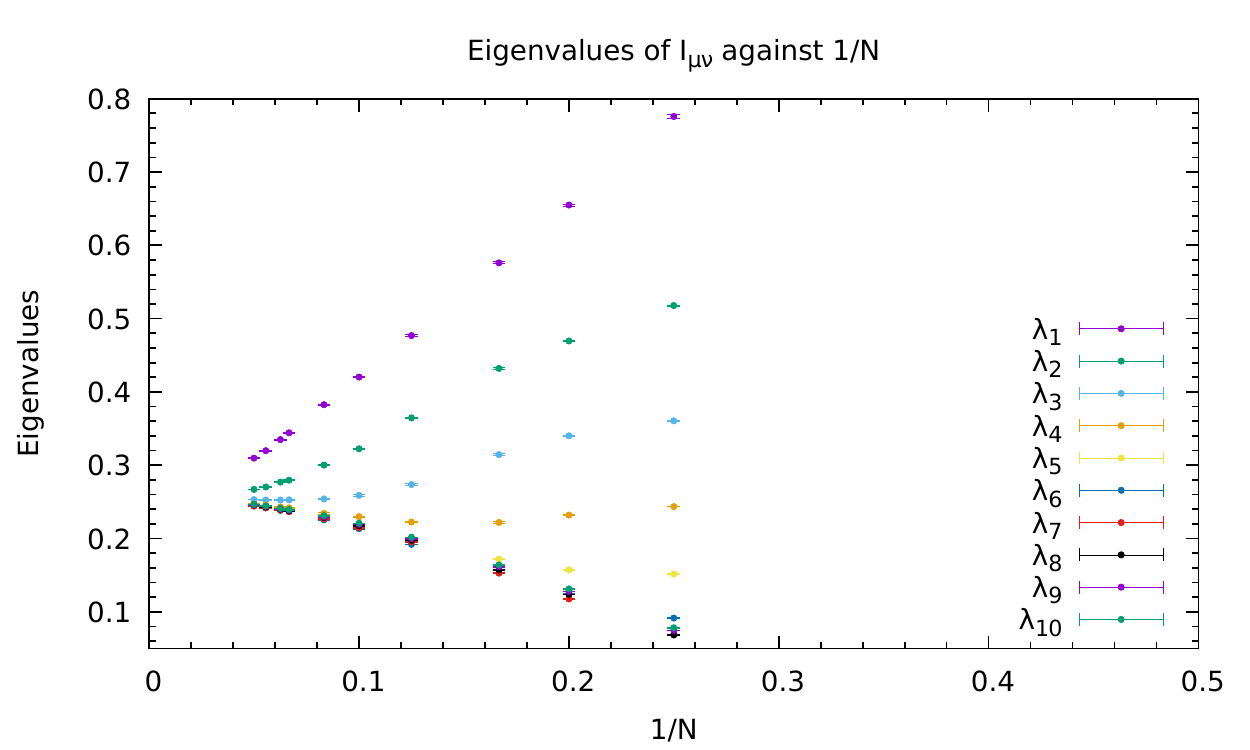}
\caption[Plot of the eigenvalues of $I_{\mu \nu}$ against $\frac{1}{N}$.]{Plot of the eigenvalues of $I_{\mu \nu}$ against $\frac{1}{N}$. We can see that the eigenvalues are converging towards a single point for large $N$.}
\label{fig5.2}
\end{figure}

Figure \ref{fig5.2} shows the plot of the eigenvalues of $I_{\mu \nu}$ against $\frac{1}{N}$. We can see in the plot that the eigenvalues are converging towards a single point. This shows that we do not encounter $SO(10)$ spontaneous symmetry breaking (SSB) in the limit $N \rightarrow \infty$. This same conclusion was also drawn in the Ref. \cite{Hotta_1999}. However, they used different criteria to arrive at this conclusion. 

%%%%%%
\subsubsection{Extent of spacetime $R^2$}
%%%%%%

\begin{figure}[!h]
\centering
\includegraphics[width = \textwidth]{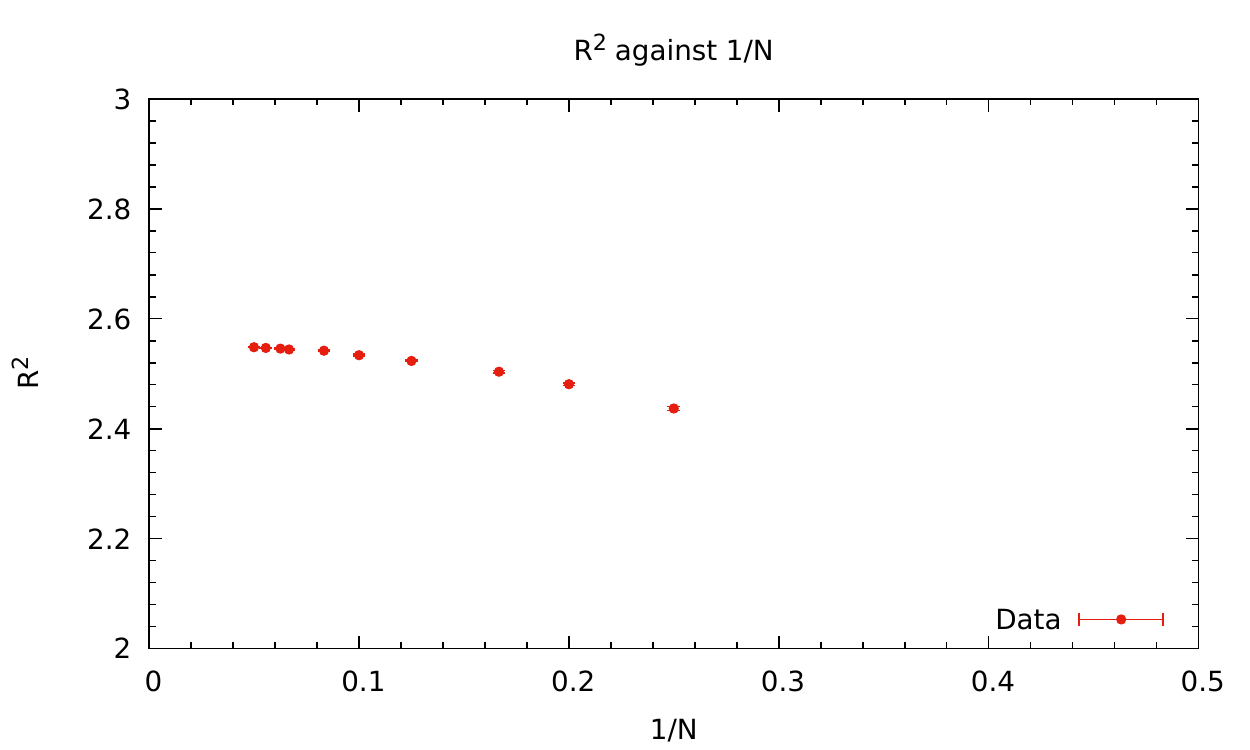}
\caption[Extent of spacetime $R^2$ against $\frac{1}{N}$.]{Extent of spacetime $R^2$ against $\frac{1}{N}$.}
\label{fig5.1}
\end{figure}

In Fig. \ref{fig5.1} we show the plot of the extent of spacetime $R^2$ against $\frac{1}{N}$. 

We can also arrive at the same conclusion using the extent of spacetime. But here we calculate $R_i^2 = \frac{1}{N} \Tr({X^i}^2)$ and this time we do not have a sum over repeated indices. If we plot these $R_i$'s, then we will see that all of them converges to the same value in the limit $N \rightarrow \infty$. This will also indicate that there is no SSB of $SO(10)$ in the limit $N \rightarrow \infty$. If there is SSB of $SO(10)$, then some $d$ values of $R_i$ will have smaller values as compared to the remaining ones in the limit $N \rightarrow \infty$. This way we will conclude that the $SO(10)$ will spontaneously break to $SO(d) \times SO(10-d)$.

%%%%%%
\section{Full IKKT Matrix Model}
%%%%%%

We will now add fermions in our theory and study the full IKKT matrix model. This model is obtained by dimensionally reducing the $\mathcal{N} =1$ super Yang-Mills model in $9+1$ dimensions to $0+0$ dimensions or dimensionally reducing the BFSS model to $0+0$ dimensions. The Euclidean action is given by
\begin{align}
S_{\rm E} = - \frac{N}{4 \lambda} \Tr ([X^i, X^j]^2) - \frac{N}{2 \lambda} \psi_{\alpha} (C_9 \gamma^i)_{\alpha \beta} [X^i, \psi_{\beta}],
\end{align}
where $i, j = 1, 2, \cdots, 10$; $\alpha, \beta = 1, 2, \cdots, 16$; $X^i$ are $N \times N$ traceless Hermitian matrices; $\lambda$ is the 't Hooft coupling; $\psi$ is a sixteen component Majorana fermion with each component of the fermion being an $N \times N$ traceless Hermitian matrix; $C_9$ is the Euclidean charge conjugation matrix in the nine dimensions; and $\gamma^i$ are the Euclidean gamma matrices in nine dimensions. Fermionic part of the action is obtained by taking a particular choice for the Majorana-Weyl fermion. The details are given in Ref. \cite{Filev_2016}. The choice of gamma matrices are such that the charge conjugation matrix becomes identity matrix in this representation. The details of the choice of gamma matrices are given in the Ref. \cite{Ambj_rn_2000}. In this representation of gamma matrices the action simplifies to
\begin{align}
S_{\rm E} = - \frac{N}{4 \lambda} \Tr ([X^i, X^j]^2) - \frac{N}{2 \lambda} \Tr( \psi_{\alpha} \gamma^i_{\alpha \beta} [X^i, \psi_{\beta}]).
\end{align}

The action is invariant under the following gauge transformation
\begin{align}
X^i \longrightarrow & V X^i V^{\dagger}, \nonumber \\
\psi_{\alpha} \longrightarrow & V \psi_{\alpha} V^{\dagger},
\end{align}
where $V \in SU(N)$. Apart from the gauge symmetry, the action also remains invariant under $SO(10)$ symmetry as $X^i$ transforms as a vector and $\psi$ transforms as a Majorana-Weyl spinor under this transformation. The partition function of this model is given by
\begin{align}
Z = \int \mathcal{D}X \,\, \mathcal{D} \psi \,\, \exp{-S_{\rm E}}. 
\label{eq5.16}
\end{align} 

We can integrate out the fermions from this partition function. For that, we first need to form the fermion matrix. We can simplify the fermionic part of the action to get the fermion matrix. First, we will decompose the fields with the help of the $SU(N)$ generators.
\begin{align}
X^i =& X^i_a T_a, \nonumber \\
\psi_{\alpha} =& \psi_{\alpha}^{a} T_a.
\end{align}

Substituting this in the fermionic part of the action, we get
\begin{align}
S_{\rm f} =& - \frac{N}{2 \lambda} \Tr( \psi_{\alpha}^{a} T_a \gamma^j_{\alpha \beta} [X^j_b T_b, \psi_{\beta}^{c} T_c] ) \nonumber \\
=& - \frac{N}{2 \lambda} \psi_{\alpha}^{a} \gamma^j_{\alpha \beta} X^j_b \psi_{\beta}^{c} \Tr( T_a [ T_b, T_c]) \nonumber \\
=& - \frac{N}{2 \lambda} \psi_{\alpha}^{a} \gamma^j_{\alpha \beta} X^j_b \psi_{\beta}^{c} ( i f_{bcd}) \Tr( T_a T_d) \qquad \textrm{ using } [T_b, T_c] = i f_{bcd} T_d \nonumber \\
=& \,\, \frac{i N}{2 \lambda} \psi_{\alpha}^{a} \gamma^j_{\alpha \beta} X^j_b \psi_{\beta}^{c} f_{bca} \qquad \textrm{ using } \Tr(T_a T_b) = - \delta_{ab} \nonumber \\
=& \,\, \psi_{\alpha}^{a} \left(- \frac{i N}{2 \lambda} \gamma^j_{\alpha \beta} X^j_c f_{abc} \right) \psi_{\beta}^{b} \qquad \textrm{ using totally antisymmetric property of } f_{abc} \nonumber \\
=& \,\, \psi_{\alpha}^{a} \,\, \mathcal{M}_{\alpha a, \beta b} \,\, \psi_{\beta}^{b},
\end{align}
where 
\begin{align}
\mathcal{M}_{\alpha a, \beta b} = - \frac{i N}{2 \lambda} \gamma^j_{\alpha \beta} X^j_c f_{abc} = \frac{N}{2 \lambda} \gamma^j_{\alpha \beta} \Tr( X^j [T_a,T_b] ),
\end{align}
with $a, b, c = 1, 2, \cdots, N^2-1$ and $\alpha, \beta = 1, 2, \cdots, 16$. The fermion matrix $\mathcal{M}_{\alpha a, \beta b}$ is a $16(N^2 - 1) \times 16(N^2 - 1)$ size matrix. Now using the result that
\begin{align}
\int d\psi \,\, \exp{- \psi^{T} A \psi} = \textrm{Pf}(A) = \det(A)^{1/2} \qquad \textrm{where } A \textrm{  is an antisymmetric matrix}
\end{align}
we can integrate out the fermions from the partition function Eq. \eqref{eq5.16}. Integrating out the fermions from the partition function we get
\begin{align}
Z = \int \mathcal{D}X \,\, \exp{-S_{\rm b}} \,\, \textrm{Pf}( \mathcal{M} ), 
\end{align} 
where $S_{\rm b}$ is the bosonic part of the action. $\textrm{Pf}(\mathcal{M})$ for this model comes to be a complex number. Since $\textrm{Pf}(\mathcal{M})$ is a complex number so we cannot treat $\exp{-S_{\rm E}}$ as the probability distribution. This is known as the {\it sign problem}. There are methods to deal with this problem by using other techniques like {\it Lefschetz thimbles} and {\it complex Langevin with stochastic quantization}.

One another technique is to consider only the absolute part of the Pfaffian in the partition function and monitor the phase of the Pfaffian during the simulations. If the phase of the Pfaffian does not vary much, then the Monte Carlo simulations can be trusted, but if it varies too much than we cannot trust the Monte Carlo simulations. We are going to use this technique for our simulation. So our partition function becomes
\begin{align}
Z = \int \mathcal{D}X \,\, \exp{-S_{\rm b}} \,\, \left| \textrm{Pf}(\mathcal{M}) \right|. 
\end{align} 

For calculation of $\textrm{Pf}(\mathcal{M}) = \det(\mathcal{M})^{1/2 }$ our best algorithm provides the time complexity of $\mathcal{O}(N^6)$. So calculating this determinant is quite a computationally heavy task. So we do not calculate this determinant directly, rather we use a rational approximation to approximate this determinant within certain error for our simulation purpose. The details are provided in the next section.

%%%%%%
\subsection{RHMC and Fermionic Forces} 
%%%%%%

We can write the absolute value of the Pfaffian as
\begin{align}
\left| \textrm{Pf}(\mathcal{M}) \right| = \det(\mathcal{M}^{\dagger} \, \mathcal{M})^{1/4},
\end{align} 
and using the result
\begin{align}
\int d\psi \, d\psi^{\dagger} \,\, \exp{- \psi^{\dagger} A \psi } \propto \frac{1}{\det(A)} \qquad \textrm{where  } A = A^{\dagger}
\end{align}
we can write
\begin{align}
\det(\mathcal{M}^{\dagger} \, \mathcal{M})^{1/4} \propto \int d\xi \, d\xi^{\dagger} \,\, \exp{- \xi^{\dagger} \, (\mathcal{M}^{\dagger} \, \mathcal{M})^{-1/4} \, \xi }.
\end{align}

So the partition function becomes
\begin{align}
Z \propto \int \mathcal{D}X \, d\xi \, d\xi^{\dagger} \,\, \exp{-S_{\rm b}[X] - S_{\rm ps f} },
\end{align}
where
\begin{align}
S_{\rm ps \, f} = \xi^{\dagger} \, (\mathcal{M}^{\dagger} \, \mathcal{M})^{-1/4} \, \xi,
\end{align}
with $\xi$ being a $16(N^2-1)$-dimensional complex vector known as the pseudo-fermion fields. $S_{\rm ps \, f}$ is known as  the pseudo-fermion action. First we note that if we define
\begin{align}
\eta = (\mathcal{M}^{\dagger} \, \mathcal{M})^{-1/8} \xi 
\label{eq5.28}
\end{align}
then the pseudo-fermion action becomes $S_{\rm ps \, f} = \eta^{\dagger} \, \eta$. Then we can simply take $\eta$ randomly from the Gaussian distribution and using Eq. \eqref{eq5.28} we get the value of $\xi$. But the main question is how to calculate $(\mathcal{M}^{\dagger} \, \mathcal{M})^{-1/8}$. For this we use the rational approximation to approximate this and after that we can calculate $\xi$. The idea is to approximate the rational exponent of the $\mathcal{M}^{\dagger} \, \mathcal{M}$ with the partial sum
\begin{align}
( \mathcal{M}^{\dagger} \, \mathcal{M} )^d = \alpha_0 + \sum \limits_{i=1}^p \alpha_i ( \mathcal{M}^\dagger \, \mathcal{M} + \beta_{i} \, \mathbb{1} )^{-1},
\end{align}
where $d$ is the rational exponent, which we require for our calculation purpose and $p$ depends on the accuracy of the approximation. If we want more accurate results then we will use a large value for $p$. {\it Remez algorithm} is used to obtain $\alpha_0, \, \alpha_i$ and $\beta_i$. The detailed theory of rational approximation is given in Ref. \cite{Ydri:2015zba} and the program of the Remez algorithm is on {\it Github} as an open source code - see Ref. \cite{Remez}. To calculate $\xi$ we require $d = 1/8$. Then, we use the rational approximation to calculate $\xi$ from $\eta$. We call this algorithm the RHMC algorithm since the rational approximation is used to approximate $(\mathcal{M}^{\dagger} \, \mathcal{M})^{d}$, and HMC is used to generate the states distributed according to the probability distribution.

The momentum conjugate to $X^i$ is $P^i$, where $P^i$ is an $N \times N$ traceless Hermitian matrix. We do not introduce momentum variable for pseudo-fermion fields as they are already distributed according to the required probability distribution. So the Hamiltonian of the system is given by
\begin{align}
H = \frac{1}{2} \Tr({P^i}^2) - \frac{N}{4 \lambda} \Tr ([X^i, X^j]^2) + \xi^{\dagger} \, (\mathcal{M}^{\dagger} \, \mathcal{M})^{-1/4} \, \xi. 
\label{eq5.30}
\end{align}

Thus we require two rational approximations, one for $d = 1/8$ and other for $d = -1/4$. So
\begin{align}
(\mathcal{M}^{\dagger} \, \mathcal{M})^{1/8} = \alpha_0 + \sum \limits_{i=1}^p \alpha_i ( \mathcal{M}^{\dagger} \, \mathcal{M} + \beta_i \, \mathbb{1} )^{-1} \\
(\mathcal{M}^{\dagger} \, \mathcal{M})^{-1/4} = a_0 + \sum \limits_{i=1}^q a_i ( \mathcal{M}^{\dagger} \, \mathcal{M} + b_i \, \mathbb{1} )^{-1}.
\end{align} 

Substituting the value of $(\mathcal{M}^{\dagger} \, \mathcal{M})^{-1/4}$ back in Eq. \eqref{eq5.30} we get
\begin{align}
H = \frac{1}{2} \Tr({P^i}^2) - \frac{N}{4 \lambda} \Tr ([X^i, X^j]^2) + \xi^{\dagger} \, \left( a_0 \xi + \sum \limits_{i=1}^q a_i ( \mathcal{M}^{\dagger} \, \mathcal{M} + b_i \, \mathbb{1} )^{-1} \xi \right). \label{eq5.33}
\end{align}

Hamilton's equations are given by
\begin{align}
\dot{X}_{rs}^i = \frac{\partial H}{\partial P_{sr}^i} = P_{rs}^i, \qquad {\dot{P}_{rs}^i} = - \frac{\partial H}{\partial X_{sr}^i}
\end{align}
where the dot represents time derivative with respect to the fictitious time $\tau$. 

The forces can be calculated using the Hamiltonian, Eq. \eqref{eq5.33}, and are given by
\begin{align}
- \frac{\partial H}{\partial X_{sr}^i} = N \sum \limits_{\substack{j=1 \\ j \neq i}}^{10} [X^j, [X^i, X^j]]_{rs} - \frac{\partial S_{\rm ps \, f}}{\partial X_{sr}^i},
\end{align}
where
\begin{align}
- \frac{\partial S_{\rm ps \, f}}{\partial X_{sr}^i} = \sum \limits_{i=1}^{q} a_i \, \xi^{\dagger} ( \mathcal{M}^{\dagger} \, \mathcal{M} + b_i \, \mathbb{1} )^{-1} \,\, \frac{\partial (\mathcal{M}^{\dagger} \, \mathcal{M})}{\partial X_{sr}^i} \,\, ( \mathcal{M}^{\dagger} \, \mathcal{M} + b_i \, \mathbb{1} )^{-1} \xi.
\end{align}

We can calculate $\dfrac{\partial (\mathcal{M}^{\dagger} \, \mathcal{M})}{\partial X_{sr}^i}$ using the expression of $\mathcal{M}$. This expression turns out to be
\begin{align}
\frac{\partial (\mathcal{M}^{\dagger} \, \mathcal{M})_{a \alpha, b \beta}}{\partial X_{sr}^i} = \left( \frac{N}{2 \lambda} \right)^2 & \Bigg[ \Big\{ (\gamma^i_{\mu \alpha}) [T_c, T_a]_{sr} \Big\}^{*} (\gamma^j_{\mu \beta}) \Tr(X^j \, [T_c, T_b]) \Bigg. \nonumber \\
 & \Bigg. + \, \Big\{ (\gamma^j_{\mu \alpha}) \Tr( X^j \, [T_c, T_a] ) \Big\}^{*} (\gamma^i_{\mu \beta}) [T_c, T_b]_{rs}  \Bigg].
\end{align}

Previously we mentioned that we can obtain $\xi$ from $\eta$. Calculations leading to this statement are provided below.
\begin{align}
\xi =& (\mathcal{M}^{\dagger} \, \mathcal{M})^{1/8} \eta, \nonumber \\
\xi =& \left( \alpha_0 + \sum \limits_{i=1}^{p} \alpha_{i} ( \mathcal{M}^{\dagger} \, \mathcal{M} + \beta_{i} \, \mathbb{1} )^{-1} \right) \eta. 
\label{eq5.38}
\end{align}

Now $( \mathcal{M}^{\dagger} \, \mathcal{M} + \beta_{i} \, \mathbb{1} )^{-1} \eta = f_i$ can be evaluated using the {\it Multimass Conjugate Gradient} (CG) method. The details of this is given in Ref. \cite{Ydri:2015zba}. As we have obtained $\xi$ so we can now evaluate the Hamiltonian and forces using $\xi$.

Below we briefly summarize the whole algorithm. 
\begin{itemize}
\item[]
\begin{itemize}
\item[{\bf Step 1:}] Take $\eta$ randomly from a Gaussian distribution $\mathcal{N}(0, 1)$. Since $\eta$ is a complex valued $16(N^2-1)$ dimensional vector both the real and imaginary parts of each component will be taken from this Gaussian distribution.
\item[{\bf Step 2:}] Evaluate $\xi$ using Eq. \eqref{eq5.38}, and use Multimass CG for evaluating $( \mathcal{M}^{\dagger} \, \mathcal{M} + \beta_{i} \, \mathbb{1} )^{-1} \eta = f_i$. 
\item[{\bf Step 3:}] Apply HMC to this system. Whenever we need to evaluate equations like $( \mathcal{M}^{\dagger} \, \mathcal{M} + b_{i} \, \mathbb{1} )^{-1} \xi = h_i$ use Multimass CG. We can see that we require this for both the Hamiltonian and force evaluations.
\item[{\bf Step 4:}] Accept/reject the state according to the Metropolis test. Repeat Steps 1 to 4 until we have enough data for evaluating the observables. We also need to monitor the phase of the Pfaffian along this.
\end{itemize}
\end{itemize}

%%%%%%
\subsection{Observables}
%%%%%%

The observables we are interested in are the same observables we computed in the bosonic IKKT model. (See Sec. \ref{sec5.1.2} for details.)

\begin{figure}[!t]
\centering
\includegraphics[width=\textwidth]{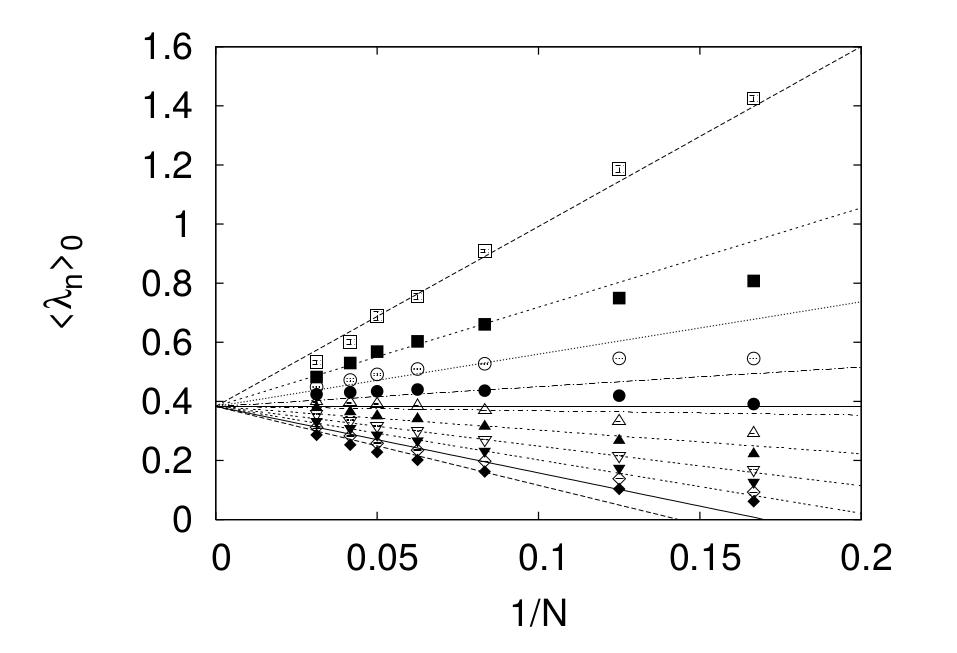}
\caption[Plot of the eigenvalues of $I_{\mu \nu}$ against $1/N$ for phase quenched IKKT model.]{Plot of the eigenvalues of $I_{\mu \nu}$ against $1/N$ for the phase quenched IKKT model. From the figure it is clear that all the eigenvalues converge to a single point. This figure is taken from the Ref. \cite{anagnostopoulos2015monte}.}
\label{fig5.3}
\end{figure}

%%%%%%
\subsection{Results}
%%%%%%

Due to lack of time during the final phase of the project, we were not able to produce the results for the full IKKT model. We are expecting the results provided in the Ref. \cite{anagnostopoulos2015monte}. Fig. \ref{fig5.3} shows the result given in the Ref. \cite{anagnostopoulos2015monte}. In the limit $N \rightarrow \infty$ eigenvalues converge to a single point, which clearly shows that there is no $SO(10)$ SSB in the phase quenched IKKT model.

Recent work on this model using complex Langevin algorithm showed that there is SSB of $SO(10)$ to $SO(3) \times SO(7)$. (See Ref. \cite{anagnostopoulos2020complex}.) In their paper, they further concluded that $SO(7)$ symmetry is also broken.

% Chapter6 -----------------------------------------------------------------------------------------------------------------------------
 % How to give reference to the Appendix
% Footnote hyperlink goes to page 1??

%%%%%%
\chapter{Conclusion and Future Work}\addcontentsline{}{×}{×}
%%%%%%

The goal of this thesis is to study the bosonic BFSS model and the IKKT model using Monte Carlo simulations.

In the BFSS model we used Polyakov loop as an order parameter to investigate the large-$N$ behaviour of this model at different temperatures. We clearly observed the confinement-deconfinement phase transition in the quenched form of this model. From the energy vs temperature plot given in Ref. \cite{Filev_2016} it is clear that the phase transition is of second order. All our results are in excellent agreement with the results produced by other authors.
 
In the bosonic IKKT model we studied the spontaneous symmetry breaking (SSB) of $SO(10)$ symmetry using the eigenvalues of the moment of inertia tensor and found that the system does not undergo SSB in the  bosonic model. This directly suggests that dynamical compactification of extra dimensions is not possible in the bosonic IKKT model. We also tried to study the phase-quenched IKKT model but did not get enough time simulate it. In the future, we can study the phase-quenched IKKT model using RHMC algorithm and full IKKT model using complex Langevin with stochastic quantization. These studies have been  carried out in Refs. \cite{Ambj_rn_2000, anagnostopoulos2020complex}.
%-------------------------------------------------------------------------------------------------------------------------------------
\appendix
%%%%%%
\chapter{Dimensional Reduction}\label{appen1}
%%%%%%

This appendix provides an introduction on how to obtain the action of certain theories by dimensionally reducing from higher dimensions to lower dimensions using Kaluza-Klein compactification. We will also talk about Wick rotation, using which one can go from the Lorentzian field theory to the Euclidean one and vice-versa.

We will discuss how to dimensionally reduce the $\mathcal{N} = 1$ super Yang-Mills theory in $9+1$ dimensions to $0+1$ dimensions and to $0+0$ dimension. In between we will also discuss the process of Wick rotation. The metric is given by $\eta_{\mu \nu} = \textrm{diag}(1, -1, -1, \cdots, -1)$. The action of the $\mathcal{N} = 1$ super Yang-Mills theory in $9+1$ dimensions with gauge group $SU(N)$ is
\begin{align}
S = \int d^{10}x \, \left\lbrace -\frac{1}{2 g^2_{10}} \, \Tr( F_{\mu \nu} F^{\mu \nu} + 2 i \bar{\Psi}_{\alpha} \gamma^{\mu}_{\alpha \beta} D_{\mu} \Psi_{\beta} ) \right\rbrace, 
\label{eqA.1}
\end{align}
where the integral for the temporal direction is from $0$ to $t_0$. We have
\begin{align}
D_\mu \Psi_\beta =& \partial_\mu \Psi - i [A_\mu, \Psi_\beta], \\
F_{\mu \nu} =& \partial_{[\mu} A_{\nu]} - i [A_\mu, A_\nu],
\end{align}
where $\mu, \nu = 0, 1, 2, \cdots, 9$; $\alpha, \beta = 1, 2, \cdots, 32$; $g_{10}$ is the coupling constant in $9+1$ dimensions; $\Psi$ is the Majorana-Weyl fermion in $9+1$ dimensions with 32 components and each component is an $N \times N$ traceless Hermitian matrix. $\Psi$ is in the adjoint representation of $SU(N)$, which is clear from the form of the covariant derivative $D_\mu$. The above action, Eq. \ref{eqA.1}, is invariant under the following gauge transformations
\begin{align}
A_\mu(x) \rightarrow & V(x) \, ( A_\mu(x) + i \partial_\mu) \, V^{\dagger}(x), \\
\Psi_\alpha(x) \rightarrow & V(x) \, \Psi_\alpha(x) \, V^{\dagger}(x).
\end{align}

The path integral for this system is given by
\begin{align}
Z = \int \mathcal{D} A_\mu \,\, \mathcal{D} \Psi \, \exp{i S}.
\end{align}

Now we will apply the Kaluza-Klein compactification to dimensionally reduce this theory to lower dimensions. The idea of the Kaluza-Klein compactification is not to treat the dimension as of infinite length but rather compactify the dimension on the circle by introducing periodicity in this length dimension. For example, suppose there is only one spatial dimension and the objects repeat in this dimension after every $2 \pi R$ distance. So it is clear that if we are close enough we will not be able to distinguish it from a line of infinite length but if we move far enough than we can clearly see that it is a circle of finite circumference. So if we want to compactify one dimension then we need to identify points after every $2 \pi R$ distance and then take the limit $R \rightarrow 0$. In the limit $R \rightarrow 0$ with finite energy system, all the fields become independent of the coordinate of that dimension - see Ref. \cite{zwiebach2004first}. The derivative in that direction no longer makes sense and all the derivative terms from the action disappears and component of the vector along that direction becomes a scalar.

Let us summarize how things change after compactification. Let us say we want to compactify along direction $x_9$. Then
\begin{align}
\phi(x_0, x_1, \cdots, x_9) \rightarrow & \,\, \phi(x_0, x_1, \cdots, x_8), \\ 
\partial_9 \phi(x) \rightarrow & \,\, 0, \\
A^9(x_0, x_1, \cdots, x_9) \rightarrow & \,\, X_9(x_0, x_1, \cdots, x_8) \quad \textrm{(Scalar field)}.
\end{align} 

Now let us compactify the direction $x_9$ on $\mathbf{S}^1$ of radius $R$ for the action Eq. \ref{eqA.1}. So $A^9(x)$ will be changed to $X_9(x)$ (Scalar field) and $\partial_9 \cdot$ will vanish. So $F_{9 \nu}$ and $D_9 \Psi_\beta$ will become
\begin{align}
F_{9 \nu} =& \partial_\nu X_9 + i[ X_9, A_{\nu}] \qquad \textrm{ where } \nu = 0, 1, \cdots, 8, \nonumber \\
F_{9 \nu} =&  D_\nu X_9, \qquad F^{9 \nu} = - D^{\nu} X_9, \\
D_9 \Psi_\beta =& i [X_9, \Psi_{\beta}].
\end{align} 

Substituting these back into the action we get
\begin{align}
S =& \int dx_9 \, \int d^9x \, \Big\{ -\frac{1}{2 g^2_{10}} \, \Tr \Big( F_{\mu \nu} F^{\mu \nu} + 2 F_{9 \nu} F^{9 \nu} + 2 i \bar{\Psi}_{\alpha} \gamma^{\mu}_{\alpha \beta} D_{\mu} \Psi_{\beta} \nonumber \\
& ~~~~~~ - 2 \bar{\Psi}_{\alpha} \gamma^{9}_{\alpha \beta} [X_9, \Psi_{\beta}] \Big) \Big\} \nonumber \\
=& \int dx_9 \, \int d^9x \, \Big\{ -\frac{1}{2 g^2_{10}} \, \Tr \Big( F_{\mu \nu} F^{\mu \nu} - 2 D_\mu X_9 D^\mu X_9 + 2 i \bar{\Psi}_{\alpha} \gamma^\mu_{\alpha \beta} D_\mu \Psi_\beta \nonumber \\
& ~~~~~~ - 2 \bar{\Psi}_\alpha \gamma^9_{\alpha \beta} [X_9, \Psi_\beta] ) \Big\} \nonumber \\
=& 2 \pi R \, \int d^9 x \, \Big\{ -\frac{1}{2 g^2_{10}} \, \Tr \Big( F_{\mu \nu} F^{\mu \nu} - 2 D_\mu X_9 D^\mu X_9 + 2 i \bar{\Psi}_\alpha \gamma^\mu_{\alpha \beta} D_\mu \Psi_\beta \nonumber \\
& ~~~~~~ - 2 \bar{\Psi}_\alpha \gamma^9_{\alpha \beta} [X_9, \Psi_\beta] \Big) \Big\}.
\end{align}

This gives
\begin{align}
S_9 =& \int d^9 x \, \Big\{ -\frac{1}{2 g^2_9} \, \Tr \Big( F_{\mu \nu} F^{\mu \nu} - 2 D_\mu X_9 D^\mu X_9 + 2 i \bar{\Psi}_\alpha \gamma^\mu_{\alpha \beta} D_\mu \Psi_\beta \nonumber \\
& ~~~~~~ - 2 \bar{\Psi}_\alpha \gamma^9_{\alpha \beta} [X_9, \Psi_\beta] \Big) \Big\}, 
\label{eqA.11}
\end{align}
where 
\begin{equation}
\frac{1}{2 g_9^2} = \frac{\pi R}{g^2_{10}},
\end{equation}
and $\mu, \, \nu = \, 0, 1, \cdots, 8$. Equation \eqref{eqA.11} is the action of the theory after compactifying $x_9$ direction. Further, we notice that the coupling constant also changes. Now we will compactify the $x_8$ direction on $\mathbf{S}^1$ of radius $R$. The following changes take place in the action
\begin{align}
F_{8 \nu} =&  D_\nu X_8, \quad F^{8 \nu} = - D^\nu X_8 \qquad  \textrm{ where } \nu = 0, 1, \cdots, 7, \\
D_8 X_9 D^8 X_9 =& [X_8, X_9]^2, \\
D_8 \Psi_\beta =& i [X_8, \Psi_\beta], \\
\frac{1}{2 g_8^2} =& \frac{\pi R}{g^2_{9}}.
\end{align}

The action becomes 
\begin{align}
S_8 =  \int d^8 x \, & \Bigg\{ -\frac{1}{2 g^2_8} \,  \Tr \Bigg( F_{\mu \nu} F^{\mu \nu} - 2 \sum \limits_{i = 8}^9 D_\mu X_i D^\mu X_i - 2 \sum \limits_{\substack{i, j = 8 \\ j > i}}^9 [X_i, X_j]^2 \Bigg. \Bigg. \nonumber \\
& \Bigg. \Bigg. \qquad \qquad \qquad + 2 i \bar{\Psi}_\alpha \gamma^\mu_{\alpha \beta} D_\mu \Psi_\beta - 2 \sum \limits_{i = 8}^9 \bar{\Psi}_\alpha \gamma^i_{\alpha \beta} [X_i, \Psi_\beta] \Bigg) \Bigg\},
\end{align}
where $\mu, \, \nu = \, 0, 1, \cdots, 7$. Now we can clearly see the pattern here. If we compactify one more spatial direction, say $x_7$, then we have
\begin{align}
\sum \limits_{i = 8}^9 D_\mu X_i D^\mu X_i & \to \sum \limits_{i = 7}^9 D_\mu X_i D^\mu X_i, \\
\sum \limits_{\substack{i, j = 8 \\ j > i}}^9 [X_i, X_j]^2 & \to \sum \limits_{\substack{i, j = 7 \\ j > i}}^9 [X_i, X_j]^2, \\
\sum \limits_{i = 8}^9 \bar{\Psi}_\alpha \gamma^i_{\alpha \beta} [X_i, \Psi_\beta] & \to \sum \limits_{i = 7}^9 \bar{\Psi}_\alpha \gamma^i_{\alpha \beta} [X_i, \Psi_\beta].
\end{align}

Applying this till we reach $0+1$ dimensions, the theory gets dimensionally reduced to a theory in $0+1$ dimensions. The action of this theory is given by
\begin{align}
S_1 = & -\frac{1}{2 g^2_{1}} \int \limits_{0}^{t_0} dt \,  \Tr \Bigg( - 2 (D_t X_i)^2 - [X_i, X_j]^2 + 2 i \bar{\Psi}_\alpha \gamma^0_{\alpha \beta} D_t \Psi_\beta - 2 \bar{\Psi}_\alpha \gamma^i_{\alpha \beta} [X_i, \Psi_\beta] \Bigg) \nonumber \\
= & \frac{1}{g^2} \int \limits_{0}^{t_0} dt \,  \Tr \Bigg( \frac{1}{2}(D_t X_i)^2 + \frac{1}{4} [X_i, X_j]^2 - \frac{i}{2} \bar{\Psi}_\alpha \gamma^0_{\alpha \beta} D_t \Psi_{\beta} + \frac{1}{2} \bar{\Psi}_\alpha \gamma^i_{\alpha \beta} [X_i, \Psi_\beta] \Bigg),
\end{align}
where $\frac{2}{g_1^2} = \frac{1}{g^2}$, $i, j \, = \, 1, 2, \cdots, 9 $ and repeated indices are summed over. Since $\Psi$ is a Majorana fermion so we can use $\bar{\Psi} = \Psi^{T} C_{10}$. Then the action becomes
\begin{align}
S = & \frac{1}{2 g^2} \int \limits_0^{t_0} dt \,  \Tr \Bigg( (D_t X_i)^2 + \frac{1}{2} [X_i, X_j]^2 -  i \Psi^{T}_\alpha C_{10} \gamma^0_{\alpha \beta} D_t \Psi_\beta + \Psi^T_\alpha C_{10} \gamma^i_{\alpha \beta} [X_i, \Psi_\beta] \Bigg). 
\label{eqA.22}
\end{align}
Equation \eqref{eqA.22} represents the action of the BFSS matrix model (see Ref. \cite{Filev_2016}).

Now we use Wick rotation to go from the Lorentzian action to the Euclidean action. Under Wick rotation the following changes take place
\begin{align}
t \to & \, \tau = i t, \\
dt \to & \, d\tau = i dt, \\
\partial_t \to & \, \partial_{\tau} = -i \partial_t, \\
A(t) \to & \, A(\tau) = -i A(t).
\end{align}

Path integral changes to the partition function with $\beta = i t_0 $ and the action changes to the Euclidean action. We take the representation of gamma matrices and $C_{10}$ apart from $\gamma^0$ given in Ref. \cite{Filev_2016}. We take the following choice for the Majorana-Weyl fermion and $\gamma^0$
\begin{align}
\Psi =& \psi \otimes 
\begin{pmatrix}
1 \\
0
\end{pmatrix}, \\
\gamma^0 =& \Gamma^{10} \otimes \sigma_2.
\end{align}

Taking these things in account, the action and the path integral becomes
\begin{align}
Z = \int \mathcal{D} A_\tau \,\, \mathcal{D} X \,\, \mathcal{D} \Psi \, \exp{- S_{\rm E}}. 
\end{align}

\begin{align}
S_{\rm E} = \frac{1}{2 g^2} \int \limits_0^\beta d\tau \, \Tr( (D_\tau X_i)^2 - \frac{1}{2} [X_i, X_j]^2 - i \psi^T_\alpha C_9 \Gamma^{10}_{\alpha \beta} D_\tau \psi_\beta - \psi^T_\alpha C_9 \Gamma^i_{\alpha \beta} [X_i, \psi_\beta] ), 
\label{eqA.32}
\end{align}
where $D_\tau = \partial_\tau - i [A(\tau), \cdot]$; $\alpha, \, \beta = 1, 2, \cdots, 16$; $i, \, j = 1, 2, \cdots, 9$; $C_9$ is the 9-dimensional Euclidean charge conjugation matrix; $\Gamma^i$ are the Euclidean gamma matrices in 9 dimensions. Bosonic fields follow periodic boundary conditions, i.e., $A(\tau + \beta) = A(\tau)$ and $X_i(\tau + \beta) = X_i(\tau)$, and fermionic fields follow anti-periodic boundary conditions, i.e., $\psi(\tau + \beta) = - \psi(\tau)$. Eq. \eqref{eqA.32} represents the Euclidean action of the BFSS matrix model. This equation is used in Chapter 4.

If we dimensionally reduce the above model to $0+0$ dimensions, we will get the IKKT matrix model. The Euclidean action of IKKT matrix model is given by
\begin{align}
S_{\rm E} = -\frac{1}{4 g_0^2} \Tr([X_i, X_j]^2) - \frac{1}{2 g_0^2} \Tr( \psi^T_\alpha C_9 \Gamma^i_{\alpha \beta} [X_i, \psi_\beta] ),
\end{align}
where $g_0$ is the 0-dimensional Yang-Mills coupling,$i \, , j = 1, 2, \cdots, 10$. This equation is used in Chapter 5.

% \input appendix2.tex
% \input appendix3.tex
%----------------------------------------------------------------------------------------------------------------------------
%\end{flushleft}
% Bibliography ------------------------------------------------------------------------------------------------------------------------------
%unsrt

\bibliographystyle{these}\addcontentsline{toc}{chapter}{Bibliography}
\bibliography{Bib/5re}
%------------------------------------------------------------------------------------------------------------------------------
\end{document}